\documentclass[preprint2]{aastex7}

\newcommand{\HI}{\mbox{H\,{\sc i}}}

\newcommand{\oii}{\mbox{[O\,{\sc ii}]}}
\newcommand{\nii}{\mbox{[N\,{\sc ii}]}}

\usepackage{comment}
\usepackage{booktabs}
\usepackage{multirow}

\shorttitle{Jet–ISM Interactions in NGC~1316}

\begin{document}

\title{Insights into Jet-Induced Cloud Disruption in NGC 1316: ALMA Reveals a Spatially Extended Molecular Gas}

\author[orcid=0000-0003-3932-0952, gname=Kana,sname='Morokuma-Matsui']{Kana Morokuma-Matsui}
\affiliation{Institute of Astronomy, Graduate School of Science, The University of Tokyo, 2-21-1 Osawa, Mitaka, Tokyo 181-0015, Japan}
\email[show]{kanamoro@ioa.s.u-tokyo.ac.jp}  

\author[orcid=0000-0002-5104-6434, gname="Alexander Y.",sname=Wagner]{Alexander Y. Wagner}
\affiliation{Center for Computational Sciences, University of Tsukuba, 1-1-1 Tennodai, Tsukuba, Ibaraki 3058577, Japan}
\email{ayw@ccs.tsukuba.ac.jp}  

\author[orcid=0000-0002-9930-1844, gname=Filippo,sname=Maccagni]{Filippo M. Maccagni}
\affiliation{INAF – Osservatorio Astronomico di Cagliari, Via della Scienza 5, I-09047 Selargius, (CA), Italy}
\affiliation{Wits Centre for Astrophysics, School of Physics, University of the Witwatersrand, 1 Jan Smuts Avenue, 2000, Johannesburg, South Africa}
\email{}  


\author[0000-0002-3249-8224]{Lauranne Lanz}
\affil{Department of Physics, The College of New Jersey, 2000 Pennington Road, Ewing, NJ 08628, USA}
\email{}  

\author[orcid=0000-0002-8868-1255, gname=Fumiya,sname=Maeda]{Fumiya Maeda}
\affiliation{Research Center for Physics and Mathematics, Osaka Electro-Communication University, 18-8 Hatsucho, Neyagawa, Osaka 572-8530, Japan}
\email{}  

\author[orcid=0000-0002-8762-7863, gname=Jin,sname=Koda]{Jin Koda}
\affiliation{Department of Physics and Astronomy, Stony Brook University, Stony Brook, NY 11794-3800, USA}
\email{jin.koda@stonybrook.edu}

\author[orcid=0000-0002-0465-5421, gname=Akihiko,sname=Hirota]{Akihiko Hirota}
\affiliation{Joint ALMA Observatory, Alonso de Córdova 3107, Vitacura, Santiago 763-0355, Chile}
\affiliation{National Astronomical Observatory of Japan, 2-21-1 Osawa, Mitaka, Tokyo 181-8588, Japan}
\email{}

\author[orcid=0000-0003-0058-9719,sname='Fujita']{Yutaka Fujita}
\affiliation{Department of Physics, Graduate School of Science, Tokyo Metropolitan University, 1-1 Minami-Osawa, Hachioji-shi, Tokyo 192-0397, Japan}
\email[show]{y-fujita@tmu.ac.jp}

\author[orcid=0000-0002-4052-2394, gname=Kotaro,sname='Kohno']{Kotaro Kohno}
\affiliation{Institute of Astronomy, Graduate School of Science, The University of Tokyo, 2-21-1 Osawa, Mitaka, Tokyo 181-0015, Japan}
\email{kkohno@ioa.s.u-tokyo.ac.jp}

\author[orcid=0000-0001-7449-4814, gname=Tomoki,sname=Morokuma]{Tomoki Morokuma}
\affiliation{Astronomy Research Center, Chiba Institute of Technology, 2-17-1, Tsudanuma, Narashino, Chiba 275-0016, Japan}
\email{}  

\author[0000-0001-8416-7673, gname=Tsutomu, sname=Takeuchi]{Tsutomu T.~Takeuchi}
\affiliation{Division of Particle and Astrophysical Science, Nagoya University, Furo-cho, Chikusa-ku, Nagoya 464-8602, Japan}
\affiliation{The Research Center for Statistical Machine Learning, The Institute of Statistical Mathematics, 10-3 Midori-cho, Tachikawa, Tokyo 190-8562, Japan}
\email{tsutomu.takeuchi.ttt@gmail.com}

\author[orcid=0000-0002-6939-0372, gname=Kouichiro,sname=Nakanishi]{Kouichiro Nakanishi}
\affiliation{National Astronomical Observatory of Japan, 2-21-1 Osawa, Mitaka, Tokyo 181-8588, Japan}
\affiliation{Department of Astronomy, School of Science, The Graduate University for Advanced Studies (SOKENDAI), 2-21-1 Osawa, Mitaka, Tokyo 181-1855, Japan}
\email{}

\author[orcid=0000-0002-1639-1515, gname=Fumi,sname=‘Egusa’]{Fumi Egusa}
\affiliation{Institute of Astronomy, Graduate School of Science, The University of Tokyo, 2-21-1 Osawa, Mitaka, Tokyo 181-0015, Japan}
\email{fegusa@ioa.s.u-tokyo.ac.jp}

\author[orcid=0000-0001-6163-4726, gname=Kenji,sname=‘Bekki’]{Kenji Bekki}
\affiliation{ICRAR M468 The University of Western Australia, 35 Stirling Hwy, Crawley, Western Australia 6009, Australia}
\email{kenji.bekki@uwa.edu.au}

\author[0000-0002-8726-7685, gname=Daniel,sname='Espada']{Daniel Espada}
\affiliation{Departamento de F\'{i}sica Te\'{o}rica y del Cosmos, Campus de Fuentenueva, Edificio Mecenas, Universidad de Granada, E-18071, Granada, Spain}
\affiliation{Instituto Carlos I de F\'{i}sica Te\'{o}rica y Computacional, Facultad de Ciencias, E-18071, Granada, Spain}
\email{despada@ugr.es}

\author[0000-0003-4351-993X, gname=B\"{a}rbel,sname='Koribalski']{B\"{a}rbel Koribalski}
\affiliation{Australia Telescope National Facility, CSIRO, Space and Astronomy, P.O. Box 76, Epping, NSW 1710, Australia}
\affiliation{Western Sydney University, Locked Bag 1797, Penrith South DC, NSW 2751, Australia}
\email{Baerbel.Koribalski@csiro.au}

\author[0000-0002-6593-8820, gname=Jing,sname='Wang']{Jing Wang}
\affiliation{Kavli Institute for Astronomy and Astrophysics, Peking University, Beijing 100871, China}
\email{jwang_astro@pku.edu.cn}








\begin{abstract}

We present ALMA CO($J=1–0$) observations of a nearby radio galaxy NGC 1316 at a 100-pc resolution to investigate the impact of AGN jets on the molecular gas. The molecular gas exhibits complex spatial and kinematic distributions, with broad CO line widths ($>50$ km s$^{-1}$) observed in several regions. The interferometric CO flux is only 34\%–38\% compared to single-dish data, indicating a large fraction of spatially extended molecular gas, especially in the central regions. We identified 24 Giant Molecular Clouds Associations (GMAs) primarily within the ``NW Shell'' and the ``SE Blob''; these GMAs show velocity dispersions approximately twice as high as those in typical star-forming galaxies for their sizes. Analysis of archival ALMA CO($J=2–1$) and CO($J=3–2$) data reveals elevated line ratios ($R_{21} \sim 1$ and $R_{31} \sim 1$) in gas near the jet, whereas, away from the jet, typical values ($R_{21} \sim 0.7$, $R_{31} \sim 0.3$). A multi-wavelength comparison reveals a $\sim$5 kpc warm ionized gas shell that encompasses the molecular NW Shell. The observed energetics and bubble morphology are consistent with an expanding bubble model driven by the jet assuming a jet power of $1.6\times10^{43}$~erg~s$^{-1}$. We propose that the high extended gas fraction results from the destruction of molecular clouds due to interactions with the jet plasma. 
NGC~1316 may be a good example of jet-induced negative feedback through the ablation, dispersal, and rarification of dense molecular clouds through jet-ISM interactions.

\end{abstract}



\section{Introduction}


Understanding the mechanisms that suppress star formation is essential for comprehending galaxy evolution. The cosmic star formation rate (SFR) density peaked around 10 billion years ago and has been declining since then \citep[e.g.,][]{Madau:2014ff}. Moreover, comparisons between the dark matter mass function obtained from $\Lambda$-CDM $N$-body simulations and the observed stellar mass function of galaxies suggest a need for some mechanism to suppress the formation in massive galaxies \citep[e.g.,][]{Silk:2012ls}.

One of the primary mechanisms for suppressing star formation in massive galaxies is feedback associated to the active galactic nuclei (AGN) in the center of galaxies.
Although recent studies suggest that AGN feedback can potentially enhance star formation under certain conditions \citep[see e.g.,][for a review]{Harrison:2024aa}, it is predominantly treated as a key process for suppressing star formation in most cosmological galaxy formation simulations \citep[e.g.,][]{Vogelsberger:2014ak,Schaye:2015vc,Dave:2019ei}.
AGN feedback can be broadly classified into radio mode and quasar mode \citep[e.g.,][]{Fabian:2012us}. Conventionally, the former is interpreted as heating the circumgalactic gas ($\gtrsim$100 kpc) through relativistic jets, indirectly suppressing star formation by inhibiting cooling and accretion onto the central galaxy \citep[e.g.,][]{Croton:2006av,Bower:2006ty}. In contrast, the latter is thought to directly suppress star formation in galaxies by blowing away or heating the interstellar medium (ISM) ($\lesssim$10 kpc) through the strong radiation of quasars \citep[e.g.,][]{Silk:1998jh}.
Recent theoretical predictions and observations have shown that expanding radio jets on (sub-)kpc scales can also have a huge impact onto the surrounding galactic medium \citep[e.g.,][]{Mandal:2021jv,Venturi:2021to}. In these early phases of jet propagation through the ISM of galaxies, the jets efficiently transfer energy and momentum to the ISM, ablating clouds, driving shock-ionized outflows and bubbles, and thereby directly affect the properties of the ISM and star-formation activity of the host galaxies \citep[e.g.,][]{Alexander:2025xp}.
Jet-driven feedback on galaxy scales ($\lesssim 10-20$ kpc) has been extensively studied through numerical simulations \citep{Sutherland:2007sf,Wagner:2011dt,Wagner:2012lz,Bieri:2016un,Mukherjee:2016wi,Mukherjee:2018my,Cielo:2018bv,Mandal:2021jv,Dutta:2024gd,Mukherjee:2025oq}.

The impact of jets on the ISM and star formation activity is difficult to generalize due to its dependence on many parameters of both the ISM and jets. Numerical simulations of the interaction between clumpy gas disks and jets \cite[e.g.,][]{Mandal:2021jv} show that while jets can compress some molecular clouds, locally increasing their SFRs, the velocity dispersion also increases while the binding energy decreases, leading to a decrease in the star formation efficiency (SFE) per unit mass of molecular clouds. 
The time evolution of the SFR depends on the inclination angle of the relativistic jet relative to the gas disk, with jets closer to the disk leading to a larger decrease in the overall SFR and a longer time to recover. 
Furthermore, the overall outcome of the jet-ISM interaction arises from a complex combination of all these factors, including the jet power, jet age, inclination angle relative to the disk, and the properties of the clumpy ISM such as the size of the molecular clouds \citep[e.g.,][and references therein]{Wagner:2012lz,Mukherjee:2018my,Mukherjee:2025oq}.
To verify this scenario, observations at the molecular cloud scale of various radio galaxies are essential. 
However, prior to the advent of the Atacama Large Millimeter/submillimeter Array (ALMA), probing the molecular ISM at cloud-scale resolutions was technically challenging, primarily due to the large distances of these systems and the performance constraints of earlier radio telescopes.

Detailed observations of cold molecular gas with modern facilities like ALMA have steadily revealed the complex interplay between jets and the ISM in radio galaxies across a vast range of scales.
Recent studies have targeted numerous systems, ranging from nearby sources such as NGC~5128 \citep[$D=3.42$~Mpc,][]{Salome:2016zx,Salome:2017xi,Salome:2019df}, NGC~4258 \citep[$D=7.2$~Mpc][]{Ogle:2014rf} NGC~4579 \citep[$D=21$~Mpc,][]{Ogle:2024eq}, NGC~3100 \citep[$D=38$~Mpc,][]{Ruffa:2019ul,Ruffa:2019hk,Ruffa:2020tj,Ruffa:2022zo}, IC~5063 \citep[$D=47.9$~Mpc,][]{Morganti:2015hx,Dasyra:2016jl,Oosterloo:2017oc}, NGC~3393 \citep[$D=52$~Mpc,][]{Finlez:2018zr}, NGC~6328 \citep[$D=63$~Mpc,][]{Papachristou:2023ch}, NGC~1167 \citep[$D=70$~Mpc,][]{Fabbiano:2022ow,Murthy:2022tb,Murthy:2025om}, and NGC~1275 \citep[$D=76$~Mpc,][]{Oosterloo:2024fr}, to more distant galaxies like The Teacup \citep[$z=0.085$][]{Audibert:2023gv}, 3C~326 \citep[$z=0.089$,][]{Nesvadba:2010lt}, PKS~1549-79 \citep[$z=0.1525$,][]{Oosterloo:2019kf}, 3C~273 \citep[$z=0.158$,][]{Husemann:2019kq}, PKS~0023-26 \citep[$z=0.32$,][]{Morganti:2021dg,Oosterloo:2025zg}, and, IRAS~00183-111 \citep[$z=0.328$][]{Ruffa:2026jf}.
The resulting physical effects are diverse, spanning the acceleration of outflows and the injection of turbulence to the localized compression and heating of the molecular phase \citep[see][Tables A1 and A2, for a recent summary of observations]{Mukherjee:2025oq}.

In the study of radio galaxies, cloud-scale spatial resolution has been achieved primarily through the ALMA observations of NGC~5128 by \citet{Salome:2017xi}. At a resolution of 23.8~pc, they revealed the detailed properties of molecular gas within a filament located 13.5~kpc from the galactic center along the northeast jet. Their results showed that the interferometric flux was only one-fifth of the single-dish value, indicating that the majority of the gas resides in a diffuse phase. Furthermore, although the 140 identified molecular clouds exhibit sizes and line widths comparable to those in the Milky Way, their relatively high virial ratios suggest that they are gravitationally stable. Given the absence of star-forming regions even in clouds with lower virial ratios, the authors argued that these structures formed via external pressure from the jet rather than through self-gravity or stellar feedback.
Verifying whether these molecular gas properties are universal across other radio galaxies is of fundamental importance. Since the jet in NGC~5128 is oriented nearly perpendicular to the galactic disk \citep[e.g.,][]{Quillen:2010vq}, its interaction with the ISM is expected to be relatively weak \cite[e.g.,][]{Mukherjee:2018my}. Therefore, further investigations into systems with stronger jet-ISM interactions are essential to understand the full impact of the radio-mode AGN feedback. To this end, we explore the properties of molecular clouds in NGC~1316 and investigate whether the radio jet is responsible in shaping some of their properties.

This paper is structured as follows. In Section~\ref{sec:ngc1316}, we provide an overview of our target galaxy, NGC 1316. Section~\ref{sec:analysis} describes the ALMA observations and the subsequent data reduction procedures. We present the ALMA observational results in Section~\ref{sec:results}. Finally, in Section\ref{sec:discussions}, we discuss the nature of the jet-ISM interaction through a comparison with multi-wavelength archival data and theoretical models.

\section{NGC~1316, a.k.a. Fornax~A} \label{sec:ngc1316}

\begin{table*}
\begin{center}
\caption{Basic information of NGC~1316 \label{tab:ngc1316}}
\begin{tabular}{lcc}
\tableline\tableline
Coordinate & ($03^h 22^m 41.718^s$, $-37^\circ 12'29''.62$) & \cite{Shaya:1996nu} \\ 
Velocity & $1732\pm10$~km~s$^{-1}$$^{\dag}$ & \cite{Longhetti:1998gk} \\ 
Redshift & $0.00587$ & \cite{Sanders:2013vr} \\ 
Distance & $20.8\pm0.5$ Mpc & \cite{Cantiello:2013ul}\\
& ($1''\sim100$~pc) &\\
Stellar mass & $5.9\times10^{11}$ M$_\odot$$^{\ddag}$ & \cite{Duah-Asabere:2016ur}\\
SFR & $0.30-0.77$ M$_\odot$ yr$^{-1}$$^{\ddag}$ & \cite{Duah-Asabere:2016ur}\\
Dust mass & $2.0\times10^{7}$ M$_\odot$ & \cite{Draine:2007ja,Lanz:2010ll}\\
\HI~mass & $7\times10^{8}$ M$_\odot$ & \cite{Serra:2019ph}\\
H$_2$~mass & $5.5\times10^{8}$ M$_\odot$ & \cite{Morokuma-Matsui:2022xi}\\
Morph. & cD & \cite{Schweizer:1980gl}\\
Blackhole mass & $1.5\times10^8$ M$_\odot$ & \cite{Nowak:2008df}\\
Eddigton luminosity & $2.3\times10^{46}$ erg~s$^{-1}$ & \cite{Lanz:2010ll}\\
Bolometric luminosity & $2.4\times10^{42}$ erg~s$^{-1}$ & \cite{Lanz:2010ll}\\
Jet power & $1.6\times10^{43}$ erg~s$^{-1}$ & \cite{Maccagni:2020pq}, this study\\
\tableline
\end{tabular}
\end{center}
\tablecomments{
$\dag$ with \oii$\lambda3727$, kinematic-LSR systemic velocity in radio definition;
$\ddag$ assuming the Chabrier IMF \citep{Chabrier:2003ql} for the stellar mass and the Kroupa IMF \citep{Kroupa:2001ak} for SFR;
}
\end{table*}

\begin{figure*}[]
\begin{center}
\includegraphics[width=0.8\textwidth, bb=0 0 768 734]{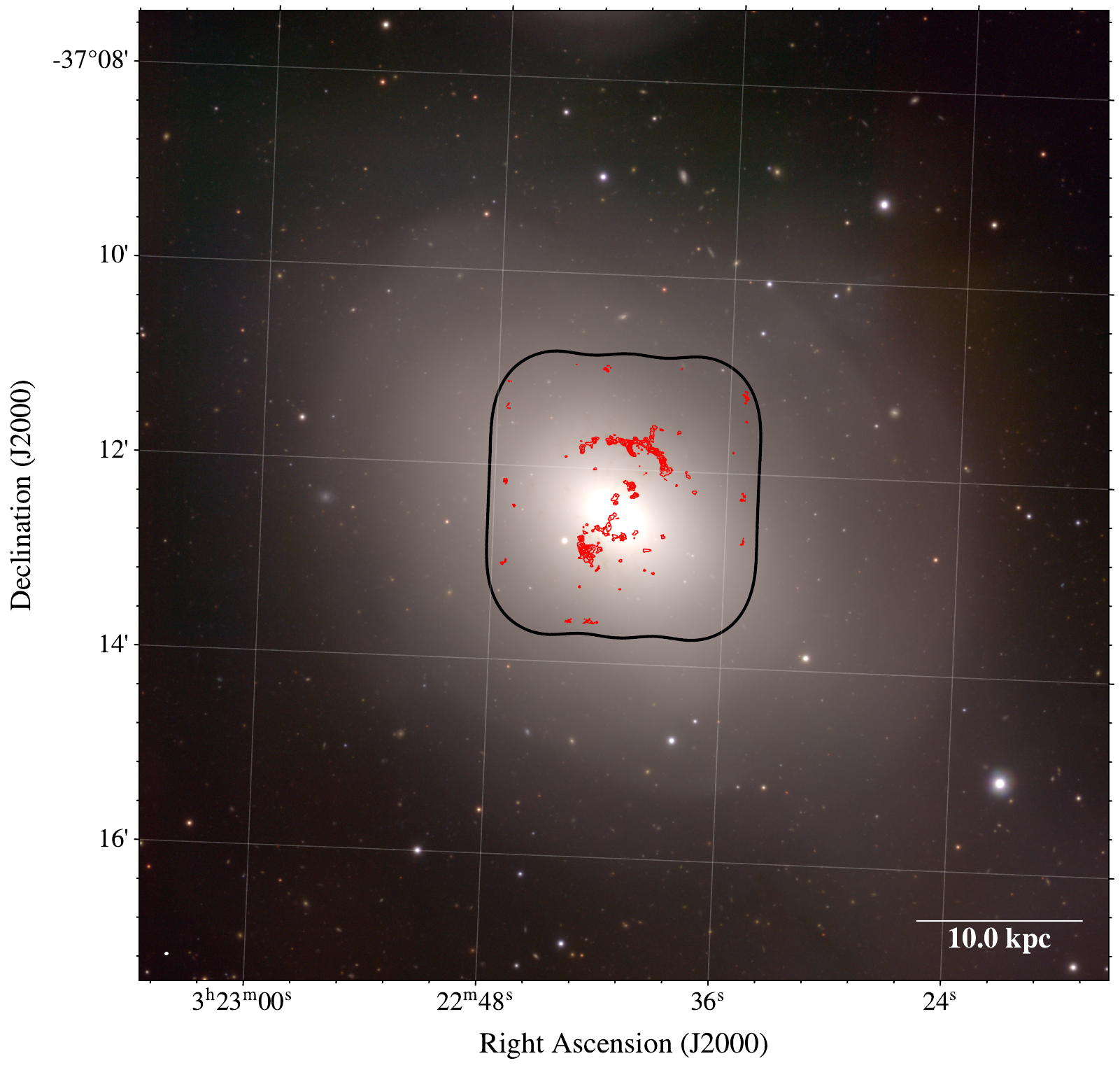}
\end{center}
\caption{
Three-color composite image of NGC~1316 from the Fornax Deep Survey \citep[FDS;][]{Iodice:2016tc,Iodice:2017pf}, using $g$-, $r$-, and $i$-band data obtained with the ESO VLT Survey Telescope. The ALMA observations were conducted to resolve the molecular gas properties within the central region of the galaxy. Red contours represent the CO($J=1-0$) emission map obtained with our ALMA observations. The solid black line indicates the field of view of the ALMA CO($J=1-0$) observations.
}
\label{fig:fds}
\end{figure*}

NGC~1316 (Fornax~A) is the central elliptical galaxy of the Fornax A group \citep[Figure~\ref{fig:fds},][]{Schweizer:1980gl} located at the distance of 20.8~Mpc \citep{Cantiello:2013ul}, characterized by its complex radio morphology including large-scale lobes exceeding 100 kpc and a 10 kpc-scale nuclear jet \citep[e.g.,][]{Ekers:1983nd,Geldzahler:1984rl,Fomalont:1989sh,Maccagni:2020pq}. The basic properties of NGC~1316 are summarized in Table~\ref{tab:ngc1316}. Radio spectral energy distribution (SED) analysis indicates that the large-scale lobes and nuclear jet are distinct episodes of AGN activity: the giant lobes were formed approximately 28 Myr ago, while the nuclear jet resulted from a subsequent burst 3 Myr ago \citep{Maccagni:2020pq}. Notably, \cite{Maccagni:2020pq} also suggested the presence of a third, even more recent episode of activity. This hypothesis has been further supported by the first successful VLBI detection in this system \citep{Paraschos:2024vg}, which provides direct evidence for the onset of a new active phase in the central engine.

The Fornax~A group resides within a dark matter halo of approximately $1.6 \times 10^{13} M_{\odot}$ \citep{Maddox:2019gl}, indicating a hot ($10^6$ K) intra-group medium (IGM). Indeed, recent \textit{SRG}/eROSITA all-sky survey observations have confirmed the presence of an extended hot gas halo associated with the Fornax~A group spanning at least $1^{\circ}$, as well as an apparent low-surface-brightness X-ray bridge connecting it to the main Fornax cluster beyond the virial radius \citep{Reiprich:2025cr}. Filaments of warm ionized gas and associated neutral hydrogen (\HI) have been detected within this IGM, likely products of runaway thermal instabilities where the hot plasma cools to a neutral state \citep[e.g.,][]{Nelson:2020nu,Kleiner:2021yg}. This cooling may be driven by Chaotic Cold Accretion (CCA), where jet-induced turbulence triggers the precipitation of cold gas onto the central galaxy \citep[e.g.,][]{Gaspari:2013rf,Maccagni:2021dz}. Simultaneously, the galaxy exhibits significant evidence of a complex merger history, including ripples, shells \citep{Schweizer:1980gl,Iodice:2017pf}, and a 150-kpc tidal tail containing \HI~gas stripped during past interactions \citep{Serra:2019ph}.

Despite its classification as an elliptical galaxy, i.e., a class typically poor in gas and dust, NGC~1316 hosts a remarkably diverse multiphase ISM. Observations have confirmed the coexistence of (1) hot ionized gas ($10^6$ K) visible in X-rays \citep{Kim:2003yb}; (2) warm ionized gas ($10^4$ K) identified via \nii~and H$\alpha$ emission \citep{Maccagni:2021dz}; (3) neutral atomic hydrogen ($100$ K) detected in \HI~\citep{Horellou:2001hb,Serra:2019ph}; (4) cold molecular gas ($10$ K) traced by CO emission \citep{Horellou:2001hb,Morokuma-Matsui:2019gv}; (5) dust  (tens of K) observed through mid- and far-infrared emission and optical/UV extinction \citep{Lanz:2010ll,Duah-Asabere:2016ur}.
Given its proximity and gas-rich circumnuclear environment, NGC~1316 provides an ideal laboratory for investigating the origins of these multiphase components. It further allows for a detailed assessment of mass fractions as determined by the interplay between the group-scale environment and mechanical/thermal feedback from radio jets.

\section{ALMA Observations and Data Analysis} \label{sec:analysis}


\begin{table*}
\begin{center}
\caption{Observations for the 12M-array and ACA data \label{tab:sbs}}
\begin{tabular}{lccccccc}
\tableline\tableline
SB name & Project ID& CO & Observation & $N_{\rm EB}^\dag$ & $N_{\rm ant.}$ & Baseline$^\ddag$ & MRS$^\mathsection$\\
 & & transition & date & & & [m] & [arcsec]\\
\tableline
NGC1316\_c\_03\_TP & 2017.1.00129.S & CO($J$=1-0) & 29 May--31 May, 2017 & $5$ & $3$ & $-$ & $-$\\
NGC1316\_a\_03\_7M & 2017.1.00129.S & CO($J$=1-0) & 16 Oct.--21 Dec., 2017 & $33$ & $11$ & $8.9-48.9$ & $72.1$\\
NGC1316\_a\_03\_TM2 & 2019.1.01845.S & CO($J$=1-0) & 12 Nov., 2019 & $1$ & $45$ & $15.1-500.2$ & $23.4$\\
 &&& 15 Nov., 2019 & $1$ & $47$ & $15.1-455.6$\\
NGC1316\_a\_03\_TM1 & 2019.1.01845.S & CO($J$=1-0) & 7 Apr., 2021 & $1$ & $40$ & $15.1-1400$ & $9.2$\\
 &&& 10 Apr., 2021 & $1$ & $42$ & $15.1-1400$\\
NGC\_1316\_a\_06\_7M & 2017.1.01140.S & CO($J$=2-1) & 8 Apr.--10 Sep., 2018 & $15$ & $12$ & $8.9-48.9$, & $32.1$\\
NGC\_1316\_a\_07\_7M & 2017.1.01140.S & CO($J$=3-2) & 26 May--14 Sep., 2018 & $19$ & $12$ & $8.9-48.9$ & $21.6$\\
\tableline
\end{tabular}
\end{center}
\tablecomments{
$^\dag$ Number of Execution Blocks (EB);
$^\ddag$ Apparent baseline (physical separation between antennas);
$^\mathsection$ Maximum Recoverable Scale, $\theta_{\rm MRS}\sim\frac{0.983\lambda}{L_5}$~radians, where $\lambda$ is the observing wavelength in meters, and $L_5$ is the 5th percentile of uv-distance in meters.
}
\end{table*}

\begin{table*}
\begin{center}
\caption{Spatial coverage and effective fields of view of the ALMA CO mosaic observations for NGC~1316.}
\label{tab:alma_fov_summary}
\begin{tabular}{llccccc}
\hline\hline
Line & Array & Project ID & \multicolumn{2}{c}{Mosaic Center} & \multicolumn{2}{c}{Effective FoV} \\
\cline{4-5} \cline{6-7}
 & & & R.A. (h:m:s) & Decl. ($^\circ$:$'$:$''$) & Width ($''$) & Height ($''$) \\
\hline
CO($J=1-0$) & ACA (7m + TP) & 2017.1.00129.S & 03:22:41.79 & $-37$:12:29.5 & 143 & 169 \\
CO($J=1-0$) & 12m Array     & 2019.1.01845.S & 03:22:41.33 & $-37$:12:16.2 & 111 & 132 \\
CO($J=2-1$) & ACA (7m)      & 2017.1.01140.S & 03:22:41.75 & $-37$:12:26.9 & 116 & 138 \\
CO($J=3-2$) & ACA (7m)      & 2017.1.01140.S & 03:22:41.62 & $-37$:12:05.4 &  77 &  82 \\
\hline
\end{tabular}
\end{center}
\tablecomments{The effective field of view (FoV) represents the total spatial extent of each mosaic, determined by combining the full span of the pointing centers with the primary beam FWHM at the corresponding line frequency.}
\end{table*}

For CO($J=1-0$), we used Morita Array or Atacema Compact Array (ACA) data (7-m array and total power array) from project 2017.1.00129.S (PI: K. Morokuma) and 12-m array data from project 2019.1.01845.S (PI: K. Morokuma).
The 7-m array data for CO($J=2-1$) and CO($J=3-2$) were obtained from project 2017.1.01140.S (PI: J. Kenney). 
The CO($J=1-0$) ACA data ($1$~kpc resolution) of NGC~1316 is already presented in \cite{Morokuma-Matsui:2019gv,Morokuma-Matsui:2022xi}.
Table~\ref{tab:sbs} summarizes the ALMA data used in this study, including the project IDs, target transitions, observation dates, and number of execution blocks. Furthermore, the detailed Field of View (FoV) parameters for each observation set, such as the phase center coordinates and the mosaic coverage, are comprehensively listed in Table~\ref{tab:alma_fov_summary}.

Calibration was performed using the standard CASA pipeline.
Specifically, the CO($J=1-0$) 12-m array data were calibrated with CASA version 6.2 and pipeline version 2021.2.0.128, while the CO($J=1-0$), CO($J=2-1$), and CO($J=3-2$) 7-m array data were calibrated with CASA version 5.4 and pipeline version Pipeline-CASA54-P1-B (42254M).
For both datasets, continuum subtraction was performed using {\tt uvcontsub}, employing data from velocity ranges where no CO emission was detected, prior to imaging.
Imaging was performed using {\tt CASA}'s {\tt tclean} task, employing Briggs weighting with a robust parameter of 0.5, which provides an optimal balance between spatial resolution and sensitivity.
The final data cubes for all three CO transitions were reconstructed with a consistent phase center of (RA, Dec) = (03h22m41.3000s, $-$37d12m15.985s). To properly sample the gas kinematics, the velocity channels for each cube were binned to a uniform width of 5.0 km~s$^{-1}$ in the LSRK frame, covering a range of 1300 to 2100 km~s$^{-1}$ across 160 channels.
For masking, we employed the '{\tt auto-multithresh}' option with parameters set to the recommended values from the Automasking Guide\footnote{\url{https://casaguides.nrao.edu/index.php/Automasking_Guide_CASA_6.6.1}}, specifically: {\tt sidelobethresh} of 2.0; {\tt noisethresh} of 4.25; {\tt minbeamfrac} of 0.3; {\tt lownoisethresh} of 1.5; {\tt negativethresh} of 0.0; {\tt stop} of 2.0 for CO($J=1-0$) data and {\tt sidelobethresh} of 1.25; {\tt noisethresh} of 5.0; {\tt minbeamfrac} of 0.1; {\tt lownoisethresh} of 2.0; {\tt negativethresh} of 0.0; {\tt stop} of 2.0 for CO($J=2-1$) and CO($J=3-2$) data.
The interferometric and single-dish CO($J=1-0$) data were combined using {\tt CASA}'s {\tt feather} task.
The transition-specific imaging parameters, including the individual cell sizes, total image pixels, resulting beam sizes, and achieved rms noise levels, are comprehensively summarized in Table~\ref{tab:co_data}.
The following describes the data analysis sequence, first for the high spatial resolution CO($J=1-0$) data, and then for the data analysis to calculate the CO line ratios.


\section{Results} \label{sec:results}

In the following sections, we present the primary imaging results for each CO emission line data set (Section~\ref{sec:mom}). To investigate how the molecular gas properties in NGC~1316 differ from those in typical galactic environments, we provide a detailed analysis based on: (1) the extended emission fraction of CO($J=1-0$) derived from a comparison between the interferometric and TP array data (Section~\ref{sec:ext}), (2) the identification and characterization of Giant Molecular Associations (GMAs)\footnote{Here, we adopt the term ``GMAs'' rather than ``Giant Molecular Clouds (GMCs)'' because the spatial resolution of our data corresponds to a physical scale of $\sim100$~pc, which is larger than the typical size of individual GMCs \citep[$\lesssim 40$~pc; e.g.,][]{Solomon:1987gx,Scoville:1987vp,Hirota:2024ij} and thus likely captures blended complexes of multiple molecular clouds.} using the CO($J=1-0$) data (Section~\ref{sec:gma}), and (3) a comparative analysis of different CO rotational transitions (Section~\ref{sec:co_ratios}).

\subsection{Moment maps}\label{sec:mom}


\begin{table*}
\begin{center}
\caption{Summary of CO data\label{tab:co_data}}
\begin{tabular}{lcccccc}
\tableline\tableline
CO & Beam$^\dag$ & Pixel Size & Image Size & $\Delta v^\ddag$ & rms \\
Transition & BMAJ$\times$BMIN, BPA & [arcsec] & [pixels] & [km~s$^{-1}$] & [mJy~beam$^{-1}$] \\
\tableline
CO($J=1-0$) & $1''.1 \times 0''.8, -83.9~{\rm deg}$ & $0.1$ & $1500 \times 1601$ & $5.0$ & $1.7$ \\
 & $(0.11~{\rm kpc} \times 81~{\rm pc})$ & $(10~{\rm pc})$ & $(15~{\rm kpc} \times 16~{\rm kpc})$ & & \\
CO($J=2-1$) & $6''.6 \times 4''.1, -87.6~{\rm deg}$ & $0.9$ & $200 \times 214$ & $5.0$ & $6\phantom{.0}$ \\
 & $(0.67~{\rm kpc} \times 0.41~{\rm kpc})$ & $(91~{\rm pc})$ & $(18~{\rm kpc} \times 19~{\rm kpc})$ & & \\
CO($J=3-2$) & $4''.3 \times 2''.9, -83.4~{\rm deg}$ & $0.6$ & $300 \times 320$ & $5.0$ & $11\phantom{.0}$ \\
 & $(0.43~{\rm kpc} \times 0.29~{\rm kpc})$ & $(61~{\rm pc})$ & $(18~{\rm kpc} \times 19~{\rm kpc})$ & & \\
\tableline
\end{tabular}
\end{center}
\tablecomments{
$^\dag$ BMAJ: Beam Major Axis, BMIN: Beam Minor Axis, BPA: Beam Position Angle;
$^\ddag$ The binned velocity resolution
}
\end{table*}

\begin{figure*}[]
\begin{center}
\includegraphics[width=\textwidth, bb=0 0 1352 1147]{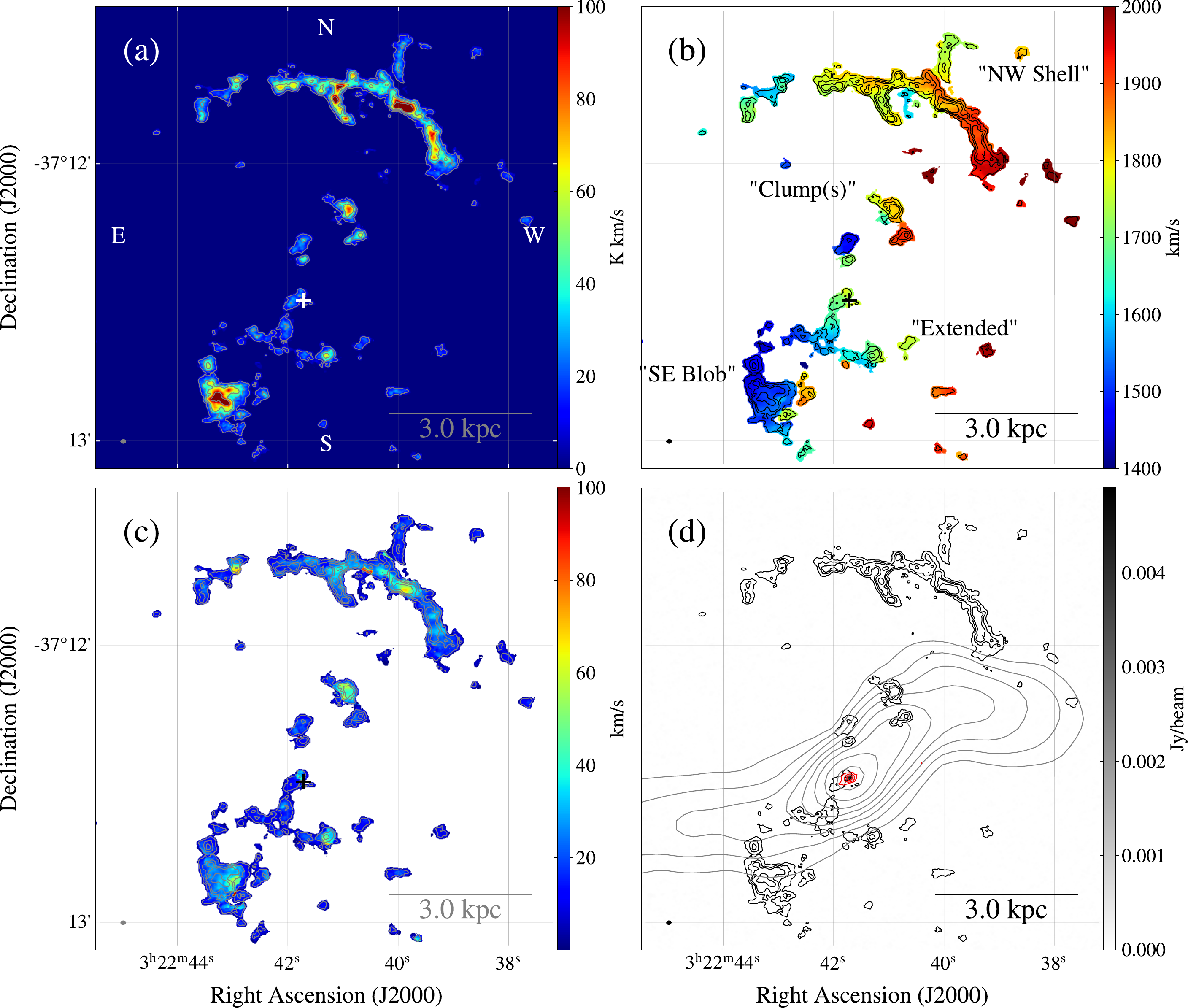}
\end{center}
\caption{
CO($J=1-0$) and 100 GHz continuum maps of NGC~1316. (a) Integrated intensity map in units of K~km~s$^{-1}$. (b) Intensity-weighted mean velocity map (moment 1) in km~s$^{-1}$. (c) Velocity dispersion map (moment 2) in km~s$^{-1}$. (d) 100 GHz continuum map in Jy~beam$^{-1}$. In all panels, the CO($J=1-0$) integrated intensity is shown as contours, with colors indicating grey in (a), black in (b), grey in (c), and black in (d). In panel (d), the red and grey contours represent the 100-GHz continuum from ALMA and the 1.4 GHz continuum from MeerKAT, respectively. Note that the beam size for the MeerKAT 1.4 GHz continuum is $6.8'' \times 5.8''$ with a position angle of $-62.5^\circ$. The galactic center is indicated by a cross. The synthesized beam is indicated in the bottom-left corner of each panel. North is up and east is to the left.
}
\label{fig:mom_co10}
\end{figure*}

\begin{figure}[]
\begin{center}
\includegraphics[width=0.5\textwidth, bb=0 0 404 389]{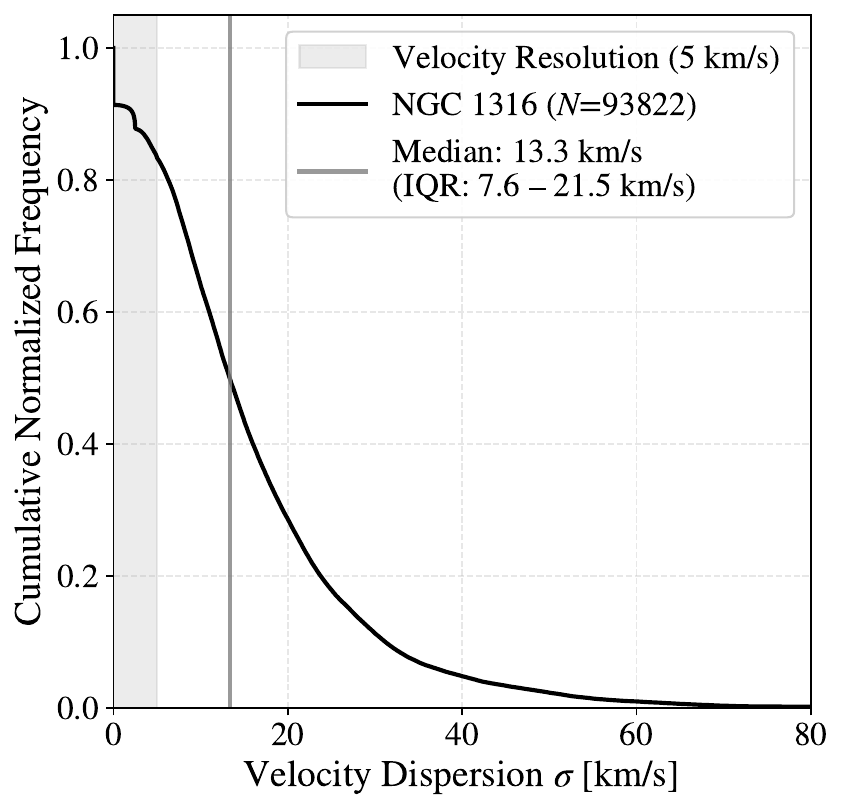}
\end{center}
\caption{
Cumulative frequency distribution of the velocity dispersion data shown in Figure~\ref{fig:mom_co10}. The median value ($13.3\text{ km s}^{-1}$) is indicated by the solid grey line. The velocity resolution of the data is $5\text{ km s}^{-1}$, represented by the grey-shaded region. While the majority of the distribution lies below $20\text{ km s}^{-1}$, there are some regions with dispersions exceeding $50\text{ km s}^{-1}$.
}
\label{fig:mom2_histo}
\end{figure}

Figure~\ref{fig:mom_co10} presents the CO($J=1-0$) integrated intensity, velocity, and velocity dispersion maps, along with the Band-3 continuum image. 
These moment maps were generated using the Source Finding Application 2 ({\tt SoFiA-2}; \citealt{Serra:2015ix, Westmeier:2021kk}). {\tt SoFiA-2} is a specialized pipeline designed for the automated detection of sources in 3D spectral-line data cubes; here, we employed the Smooth-and-Threshold (S+T) algorithm within {\tt SoFiA-2} to effectively extract signals across multiple spatial and velocity scales and construct a source mask. The specific parameters for noise scaling, the S+C finder, and the linker were set as summarized in Table~\ref{tab:sofia_parameters} in the Appendix.
Based on this constructed source mask, the integrated intensity map was computed using all data within the mask, while the velocity and velocity dispersion maps were calculated by restricting the integration to channels with positive flux densities within the mask.
The resultant beam size is $\sim1.0''$ ($\sim 100$~pc)\footnote{Although the original proposal aimed for a spatial resolution of 50~pc, the required extended antenna configurations could not be fully executed due to the constraints during the COVID-19 pandemic, resulting in a final resolution of 100~pc.}, with a pixel scale of $0.1''$ ($\sim 10$~pc) and a binned velocity resolution of 5~km~s$^{-1}$. The $1\sigma$ sensitivity per velocity channel is $1.7 \times 10^{-3}$~Jy~beam$^{-1}$ ($\sim 0.18$~K), corresponding to a molecular gas mass surface density of $3.9$~$M_\odot$~pc$^{-2}$ assuming $\alpha_{\rm CO} = 4.36$~$M_\odot$~pc$^{-2}$~(K~km~s$^{-1}$)$^{-1}$.
Because $\alpha_{\rm CO}$ is primarily a function of metallicity \citep{Bolatto:2013ou} and NGC~1316 is characterized by an approximately solar metallicity \citep{Goudfrooij:2001ya}, the standard Galactic disk value is a physically appropriate choice for this system.

In \cite{Morokuma-Matsui:2019gv}, the characteristic structures were named, from north to south, the ``NW Shell'', the ``Clump'', the ``Extended'', and the ``SE Blob'' structures, all of which appeared spatially connected at a resolution of about 1~kpc. In contrast, data with a 100~pc resolution reveal that each of these is a spatially and kinematically independent structure; furthermore, there are spatially and kinematically independent substructures within each of them.

The cumulative frequency distribution of the velocity dispersion (Figure~\ref{fig:mom2_histo}) shows that most regions exhibit values below $20\text{ km s}^{-1}$, with a median of $13.3\text{ km s}^{-1}$ and an interquartile range (IQR) of $7.6\text{--}21.5\text{ km s}^{-1}$. However, dispersions exceeding $50\text{ km s}^{-1}$ are observed in localized areas. While some of these high-dispersion regions in Figure~\ref{fig:mom_co10}(c) likely result from the superposition of multiple independent velocity components, several key features, including the high-intensity northeastern part of the NW Shell, the Clump(s), the region immediately north of the galactic center, and the eastern edge of the Extended structure, appear as single components or multiple components within a shared kinematic envelope, at least at the current velocity resolution. In contrast, the large velocity dispersion observed in the previous 1-kpc resolution data \citep{Morokuma-Matsui:2019gv} at the western edge of the ``SE Blob'' is now clearly shown to result from the projection of distinct velocity components within a single beam. Additionally, the CO emission in the nucleus is slightly offset by $\sim 1''$ ($\sim 100$~pc) from the 100~GHz continuum peak. While continuum emission is also visible to the west of the center, it does not constitute a significant detection on its own. This western feature is only considered a tentative detection when viewed in conjunction with the Band-6 and Band-7 data presented later.


\begin{figure*}[]
\begin{center}
\includegraphics[width=\textwidth, bb=0 0 1767 1518]{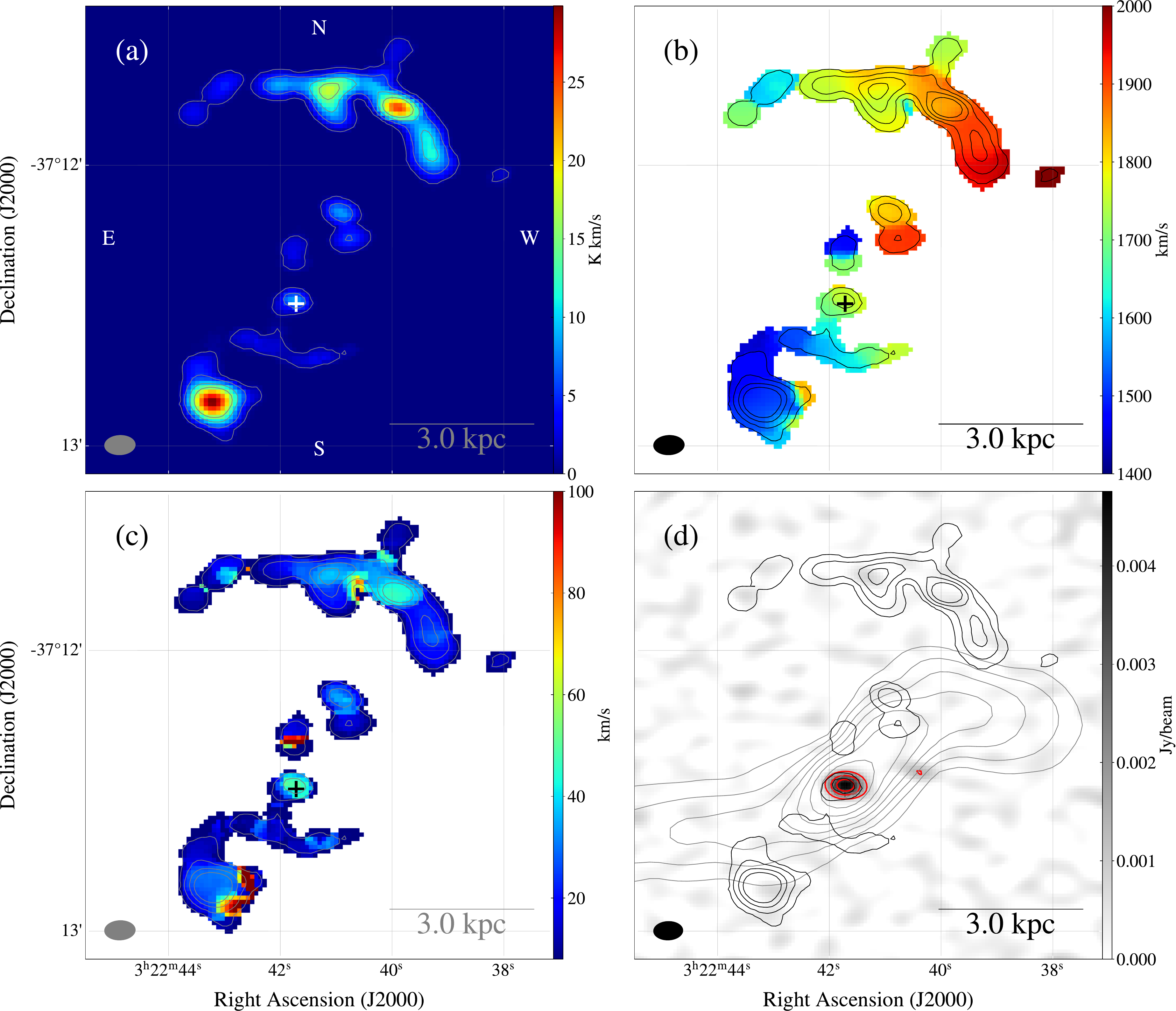}
\end{center}
\caption{
Same as Figure~\ref{fig:mom_co10}, but for the CO($J=2-1$) emission and the 230~GHz continuum. All panels and overlaying contours correspond to the CO($J=2-1$) transition.
}
\label{fig:mom_co21}
\end{figure*}

\begin{figure*}[]
\begin{center}
\includegraphics[width=\textwidth, bb=0 0 2638 2283]{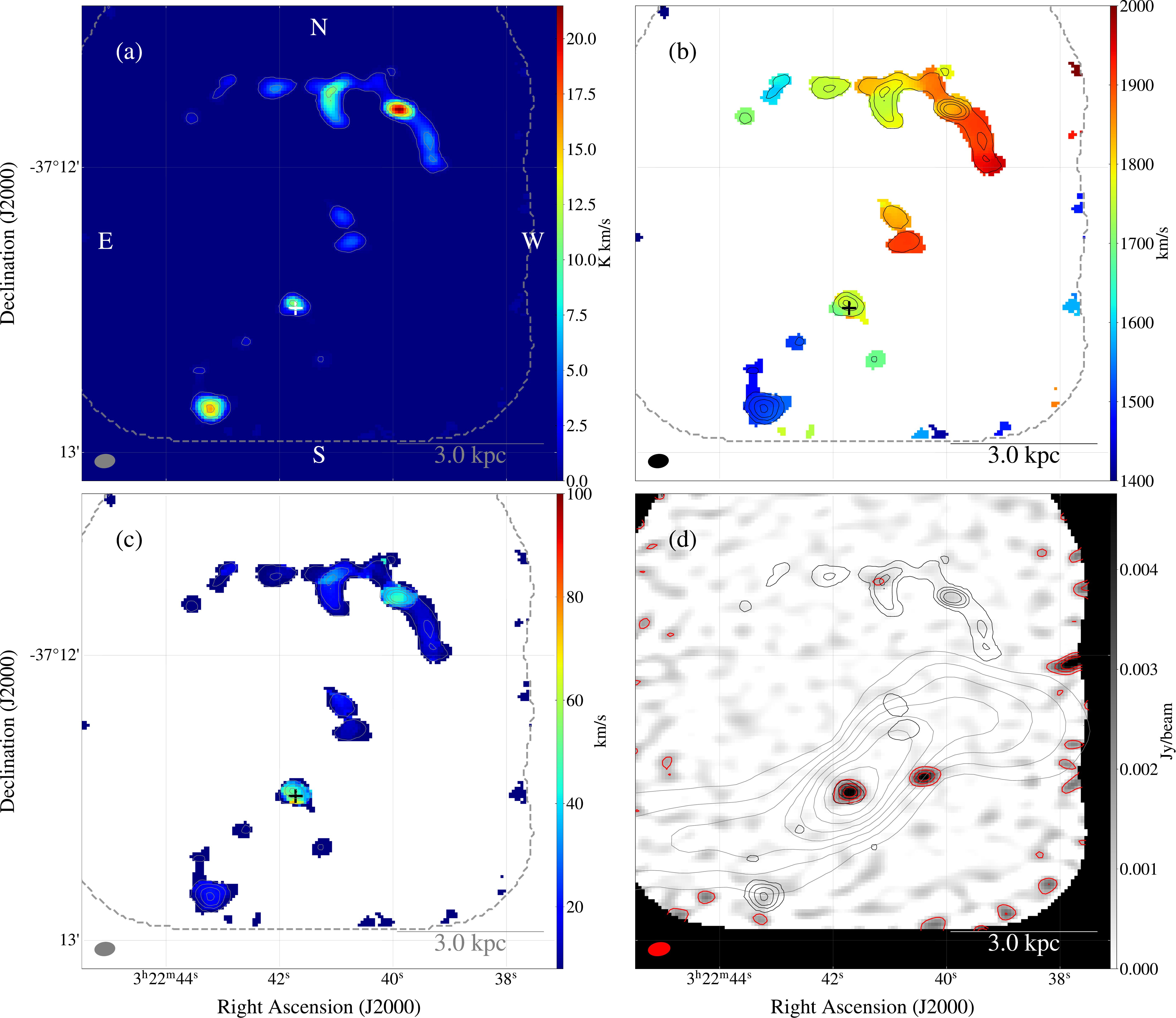}
\end{center}
\caption{
Same as Figure~\ref{fig:mom_co10}, but for the CO($J=3-2$) emission and the 330~GHz continuum. All panels and overlaying contours correspond to the CO($J=3-2$) transition. The grey dashed lines in panels (a)–(c) indicate the field of view (FoV) of the ALMA observations.
}
\label{fig:mom_co32}
\end{figure*}

Figures~\ref{fig:mom_co21} and \ref{fig:mom_co32} present the integrated intensity, velocity, and velocity dispersion maps, along with the 230~GHz and 330~GHz continuum emission, for the CO($J=2-1$) and CO($J=3-2$) transitions, respectively. It should be noted that the field of view of CO($J=3-2$) map is narrower than those of the CO($J=1-0$) and CO($J=2-1$) maps.
Consequently, the ``SE Blob'' structure is located near the very edge of the observed field of view of the CO($J=3-2$) data, resulting in increased noise contamination in this region. As anticipated, the derived velocity fields and velocity dispersions of CO($J=2-1$) and CO($J=3-2$) are consistent with those observed in the CO($J=1-0$) data.
The basic properties of three CO transition data are summarized in Table~\ref{tab:co_data}.

Regarding the continuum emission, both the central source and the secondary western source are clearly detected in both Band 6 and Band 7 data. We find no corresponding CO line emission associated with the western continuum source.
To characterize the emission mechanism of the western continuum source, we derived its spectral index $\alpha$, defined by $S_\nu \propto \nu^\alpha$. The continuum flux densities at each band were measured by fitting a 2D Gaussian component using the {\tt CASA} task {\tt imfit}. The source was robustly detected at 230 and 330 GHz, yielding flux densities of $1.20 \pm 0.31$~mJy and $4.88 \pm 0.48$~mJy, respectively. At 100 GHz, the source remained undetected, providing a $3\sigma$ upper limit of $0.06$~mJy. From these measurements, we obtain a steep spectral index of $\alpha_{230-330} \sim 4.0$, which is consistent with the lower limit derived from the 100 GHz non-detection ($\alpha_{100-230} > 3.7$). This value is highly indicative of thermal emission from optically thin cold dust in the Rayleigh--Jeans regime, where $\alpha = 2 + \beta$ with a dust emissivity index of $\beta \sim 2$ \citep[e.g.,][]{Chapin:2009cr, Planck-Collaboration:2011fx}. Given its compact morphology and the lack of a clear counterpart in other wavelengths, this source is likely a distant background galaxy (e.g., a submillimeter galaxy) rather than a structure associated with NGC~1316. Further investigation into this continuum source is beyond the scope of the current work.

\subsection{Extended molecular gas fraction}\label{sec:ext}

\begin{table}
\begin{center}
\caption{CO($J=1-0$) flux ratios among the 12-m, 7-m, and TP arrays \label{tab:fdiffuse}}
\begin{tabular}{lc}
\tableline\tableline
 & Ratio\\
\tableline
7M/TP & $0.34\pm0.08$\\
12M/TP & $0.38\pm0.10$\\
12M/7M & $1.12\pm0.19$\\
\tableline
\end{tabular}
\end{center}
\end{table}

\begin{figure}[]
\begin{center}
\includegraphics[width=0.5\textwidth, bb=0 0 573 556]{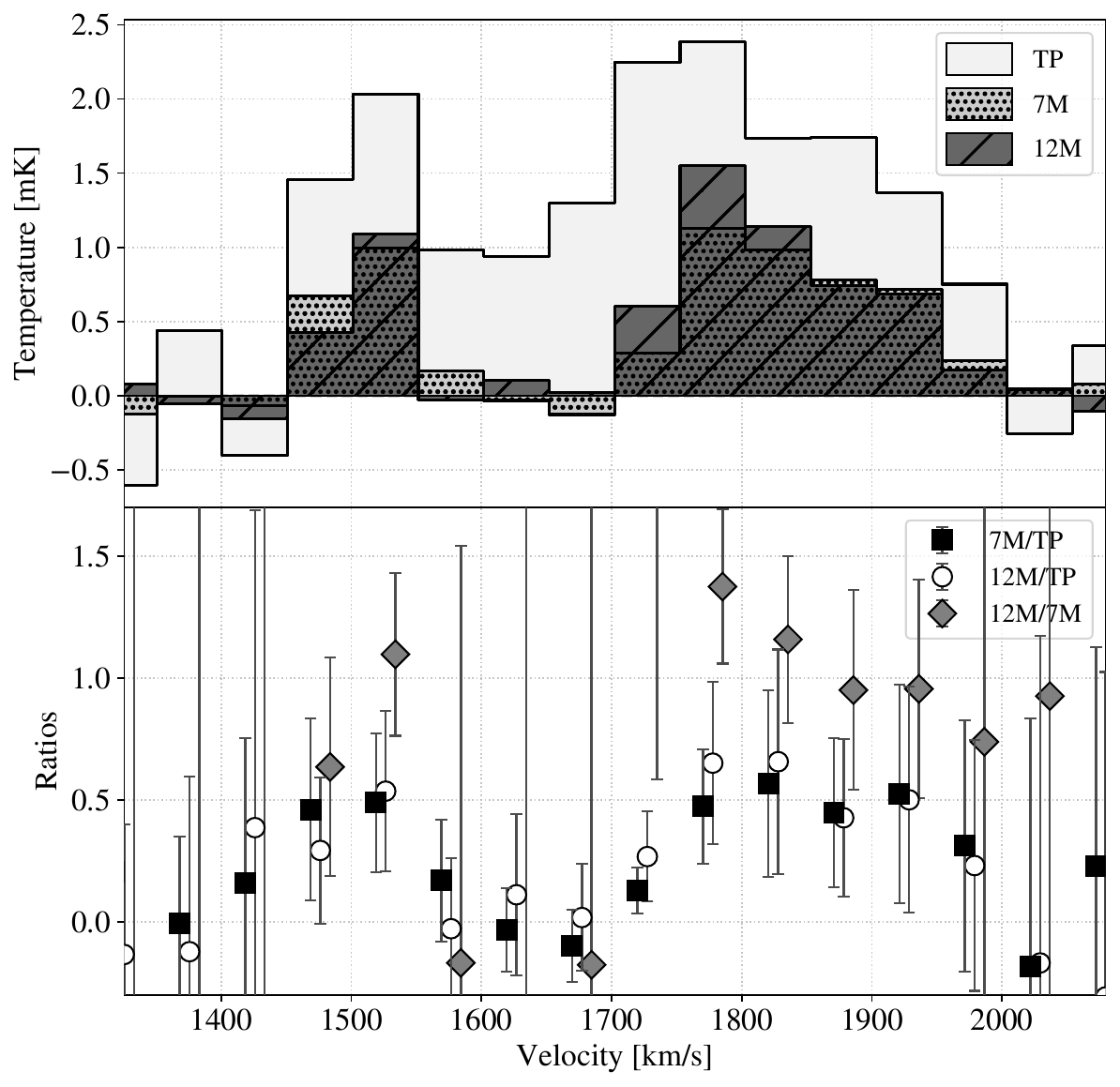}
\end{center}
\caption{
Top panel: Comparison of the CO($J=1-0$) spectra integrated over the entire galaxy obtained with the TP, 7M, and 12M arrays, smoothed to a velocity resolution of $50\text{ km s}^{-1}$. The flux was integrated over a $150'' \times 160''$ rectangular region centered at $(\alpha, \delta) = (03^{\text{h}}22^{\text{m}}41.300^{\text{s}}, -37^{\circ}12'15.985'')$, covering the full extent of the galactic emission. The spectra are distinguished by their hatching patterns: TP (no hatching), 7M (dotted pattern), and 12M (hatched).
Bottom panel: Brightness temperature ratios for each velocity channel: 7M/TP (filled black squares), 12M/TP (filled white circles), and 12M/7M (filled grey diamonds). Error bars were derived from the RMS noise of each individual spectrum. The data points for each ratio are slightly shifted along the velocity axis (by 7.5~km~s$^{-1}$ or 15~\% of the velocity resolution) to avoid overlapping of the error bars. The 7M and 12M intensities are negligible in the velocity range of $1550-1700\text{ km s}^{-1}$. Even in other channels where emission is detected, the intensities detected by the 7M and 12M arrays remain less than half of the TP intensity.
}
\label{fig:dfiffuse_totalspec}
\end{figure}

\begin{figure*}[]
\begin{center}
\includegraphics[width=0.329\textwidth, bb=0 0 909 715]{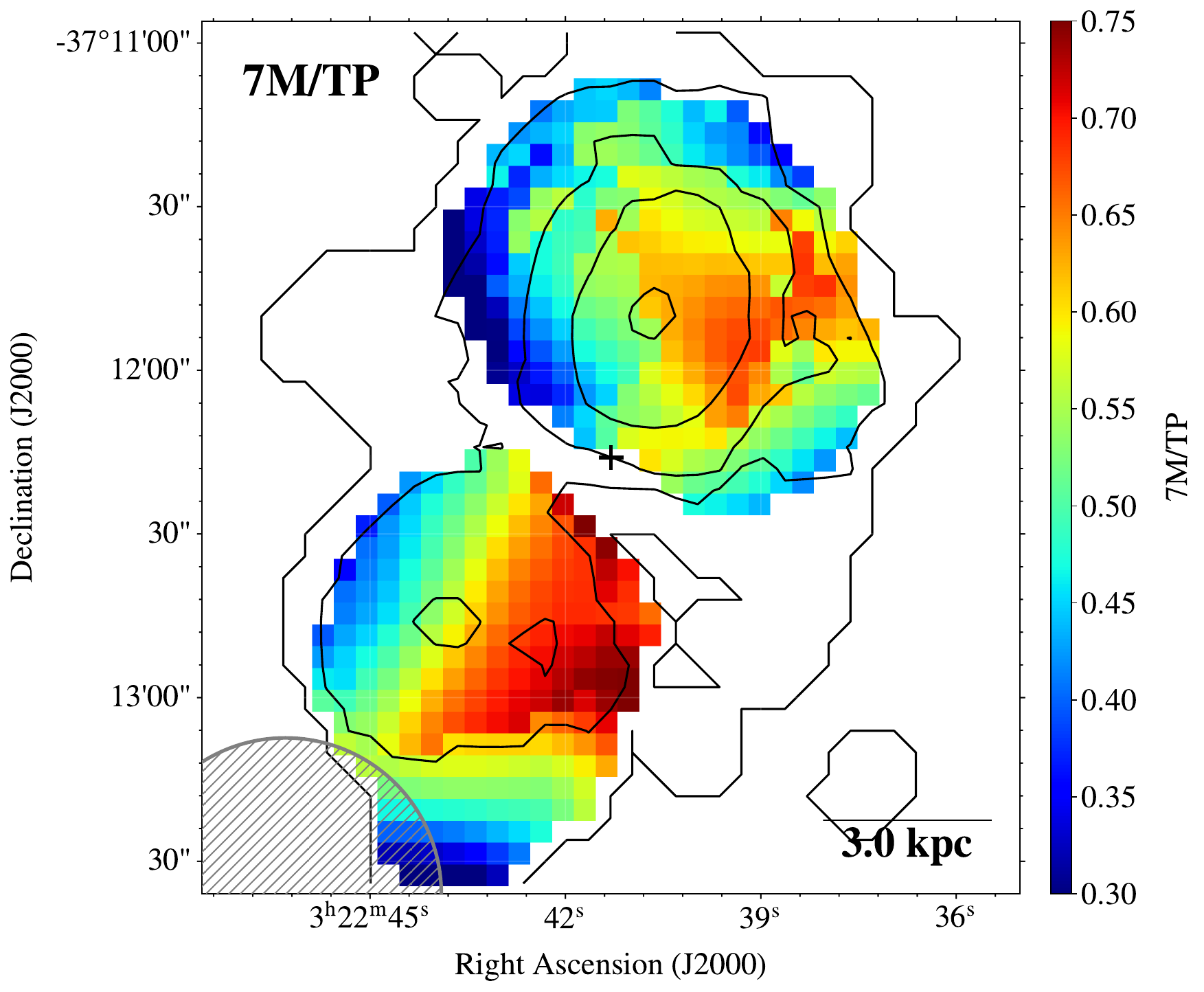}
\includegraphics[width=0.329\textwidth, bb=0 0 909 715]{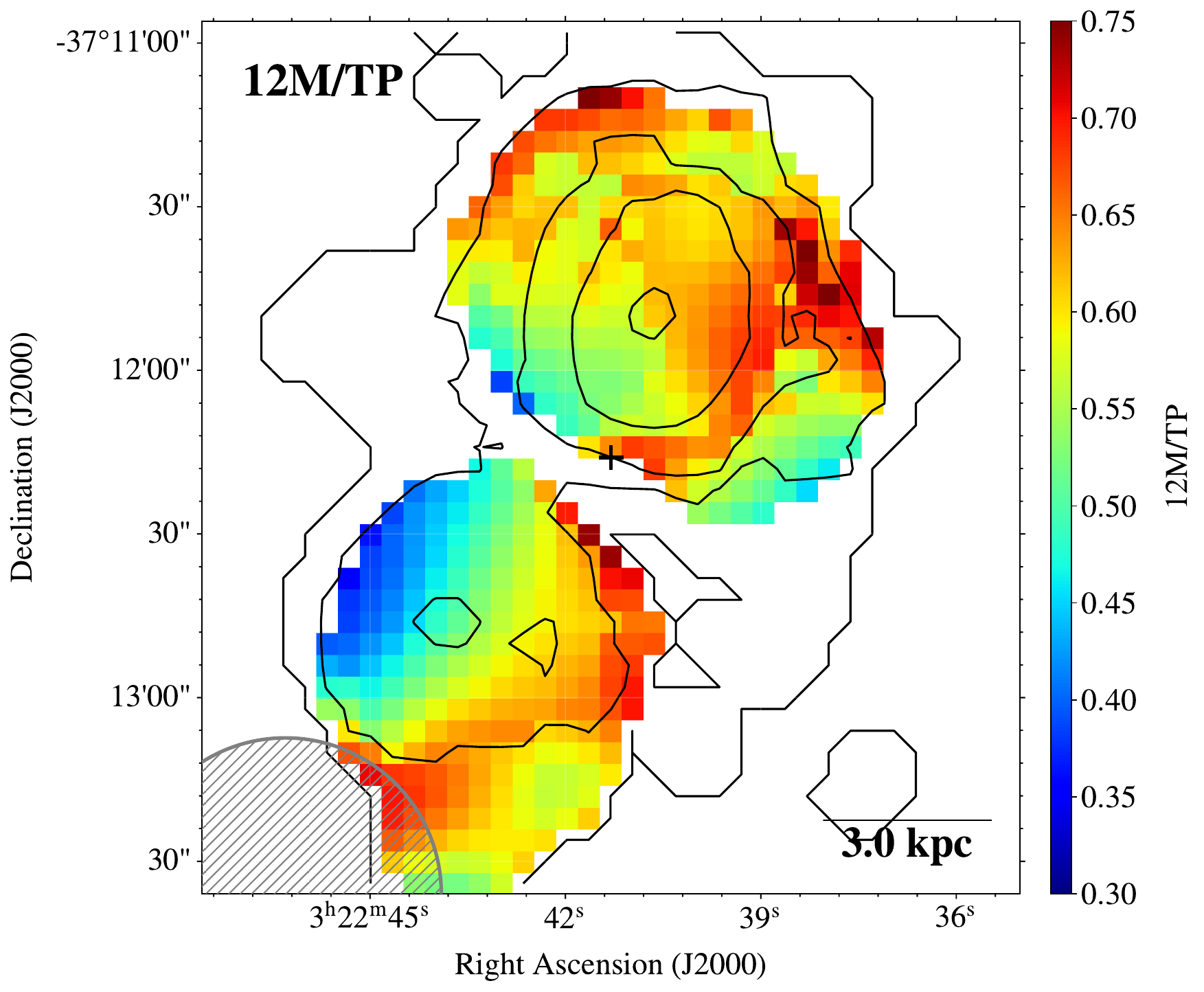}
\includegraphics[width=0.329\textwidth, bb=0 0 909 715]{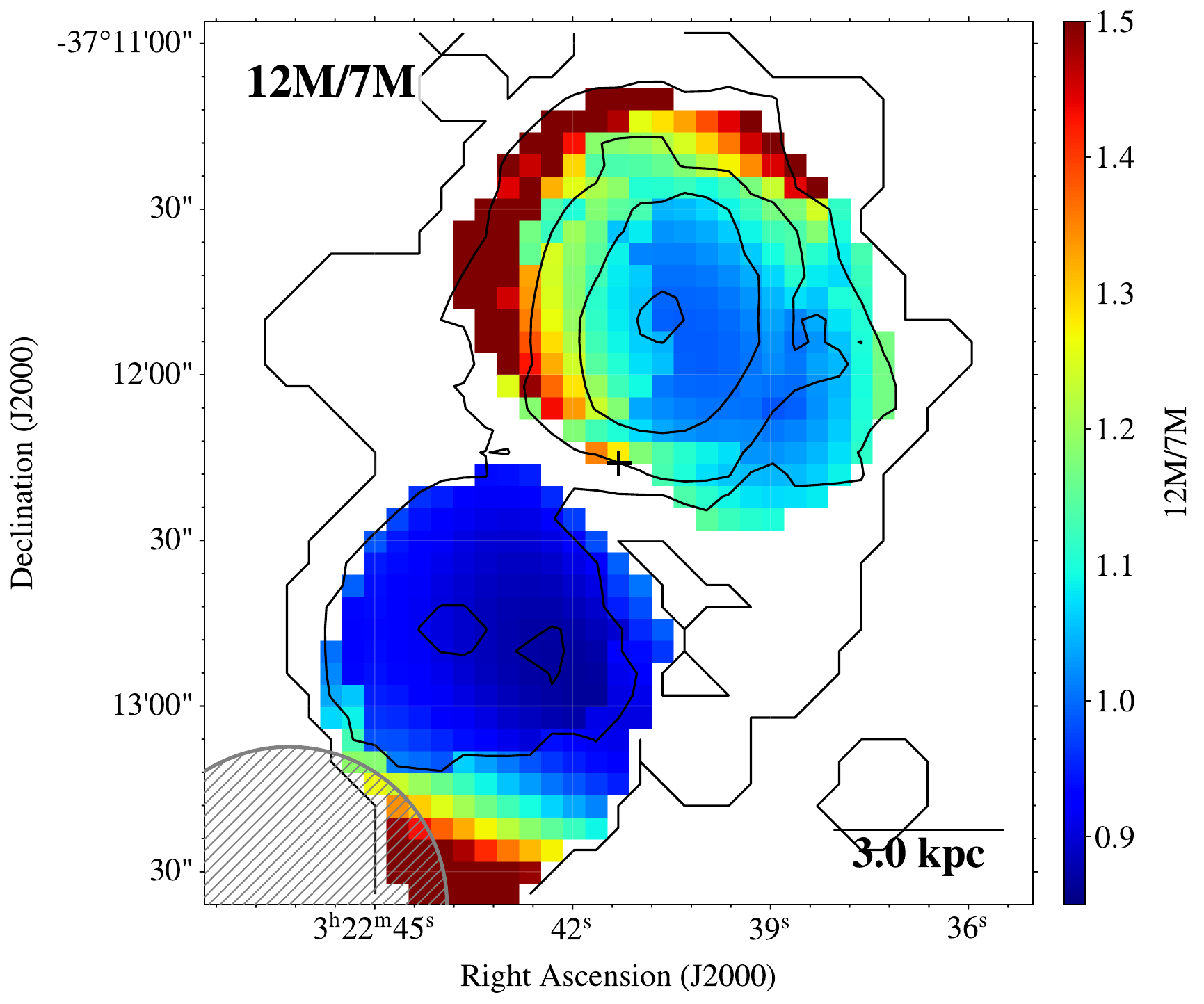}
\end{center}
\caption{
CO($J=1-0$) integrated intensity ratio maps derived from 12-m, 7-m, and TP array moment-0 data.
From left to right, panels show the 7M/TP, 12M/TP, and 12M/7M ratios.
The black contour represents the integrated intensity map from the TP data.
The grey-hatched circle in the lower-left corner indicates the beam size.
}
\label{fig:dfiffuse}
\end{figure*}

To estimate the extended gas fraction, we individually image the 12-m, 7-m, and TP array CO($J=1-0$) data.
The 12-m and 7-m array configurations provide a maximum recoverable scale (MRS) of approximately $23''$ (2.4~kpc) and $72''$ (7.2~kpc), respectively. Consequently, any molecular gas structures spatially extended beyond these scales are filtered out by the interferometric observations, allowing us to quantify the fraction of the extended molecular phase by comparing these data with the TP observations.

For comparison, the 12-m and 7-m datasets were first smoothed to the TP-map beam size ($57''.2$) using the {\tt CASA} task {\tt imsmooth}. Subsequently, these data were spatially and spectrally regridded to match the pixel increments ({\tt CDELT}1, 2, and 3) and the total number of pixels ({\tt NAXIS}1, 2, and 3) of the TP-array data cube using the {\tt imrebin} and {\tt imregrid} tasks, ensuring a consistent grid shape across all datasets.
To construct the integrated intensity ratio maps, we generated CO emission masks for each dataset using {\tt SoFiA 2} \citep{Serra:2015ix,Westmeier:2021kk}. Then, we created a common mask consisting only of spaxels where CO was detected in both datasets being compared. For instance, when calculating the 7-m/TP ratio, we restricted the analysis to spaxels with detections in both the 7-m and TP data. These common masks were then applied to the respective data cubes to produce the integrated intensities used for the final ratio maps (Figure~\ref{fig:dfiffuse}).

The global CO($J=1-0$) spectra of NGC~1316 obtained with the 12-m, 7-m, and TP arrays, along with their 7-m/TP, 12-m/TP, and 12-m/7-m ratios, are presented in Figure~\ref{fig:dfiffuse_totalspec}.
To extract the global spectrum of the galaxy, we utilized the {\tt CASA} task {\tt specflux}. The flux was integrated over a rectangular region centered at $(\alpha, \delta) = (03^{\text{h}}22^{\text{m}}41.300^{\text{s}}, -37^{\circ}12'15.985'')$ with a field of view of $150'' \times 160''$, covering the entire emission of the galaxy.
Table~\ref{tab:fdiffuse} shows the integrated intensity ratios derived from these spectra.


Our comparison reveals that the 12-m and 7-m arrays miss more than 60\% of the TP flux in NGC~1316. This fraction is notably higher than the $\sim$50\% typically filtered out in normal star-forming galaxies, where such missing flux is generally attributed to extended structures on scales exceeding a lower limit of only 1.0--1.3 kpc \citep[e.g.,][]{Pety:2013tw,Caldu-Primo:2015qg}. Considering that normal galaxies typically lose half their flux at the kiloparsec scale, it is remarkable that NGC~1316 misses over 60\% of its emission even at the 7.2~kpc scale (the MRS of the 7-m array). This suggests that a dominant portion of the molecular gas in NGC~1316 is distributed over exceptionally vast spatial scales, distinguishing its gas distribution from the much more compact structures seen in typical star-forming galaxies.

Notably, flux within the velocity range of $\sim$1,550--1,700~km~s$^{-1}$, corresponding to the central region between the NW shell and SE blob, is almost entirely undetected (Figures~\ref{fig:dfiffuse_totalspec}, \ref{fig:12mtp_v}, and \ref{fig:7mtp_v}). Even in regions where CO emission is detected with the 12-m or 7-m arrays, a spatial variation in the ratios is observed, with higher values present on the galaxy's west side (Figure~\ref{fig:dfiffuse}). While the 12-m/7-m flux ratio is consistent with unity within the uncertainties, it may be slightly higher than unity. This trend may be partly attributed to the presence of ``negative bowls'', extended negative sidelobes often observed in ACA data \citep{Leroy:2021ux}, which can artificially suppress the flux in the 7-m-only maps.

\subsection{Giant Molecular Clouds Associations}\label{sec:gma}

\begin{figure*}[]
\begin{center}
\includegraphics[width=0.32\textwidth, bb=0 0 343 318]{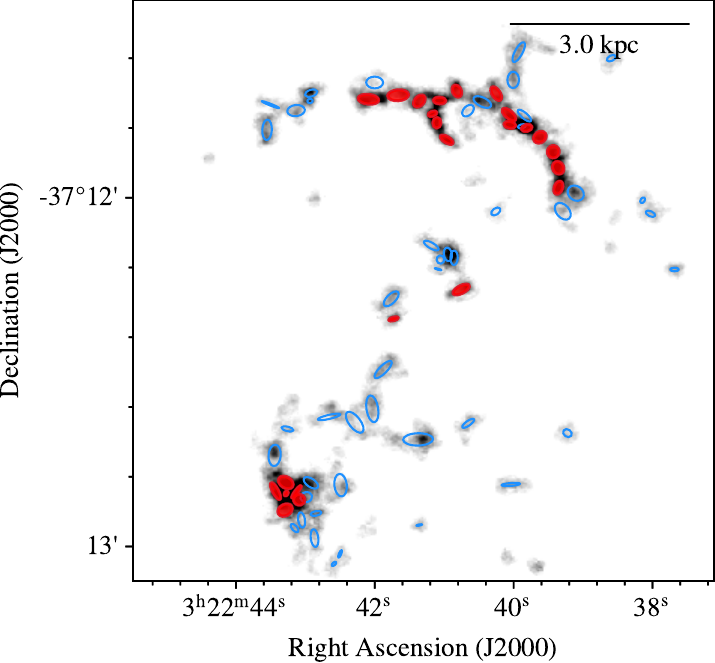}
\includegraphics[width=0.32\textwidth, bb=0 0 531 491]{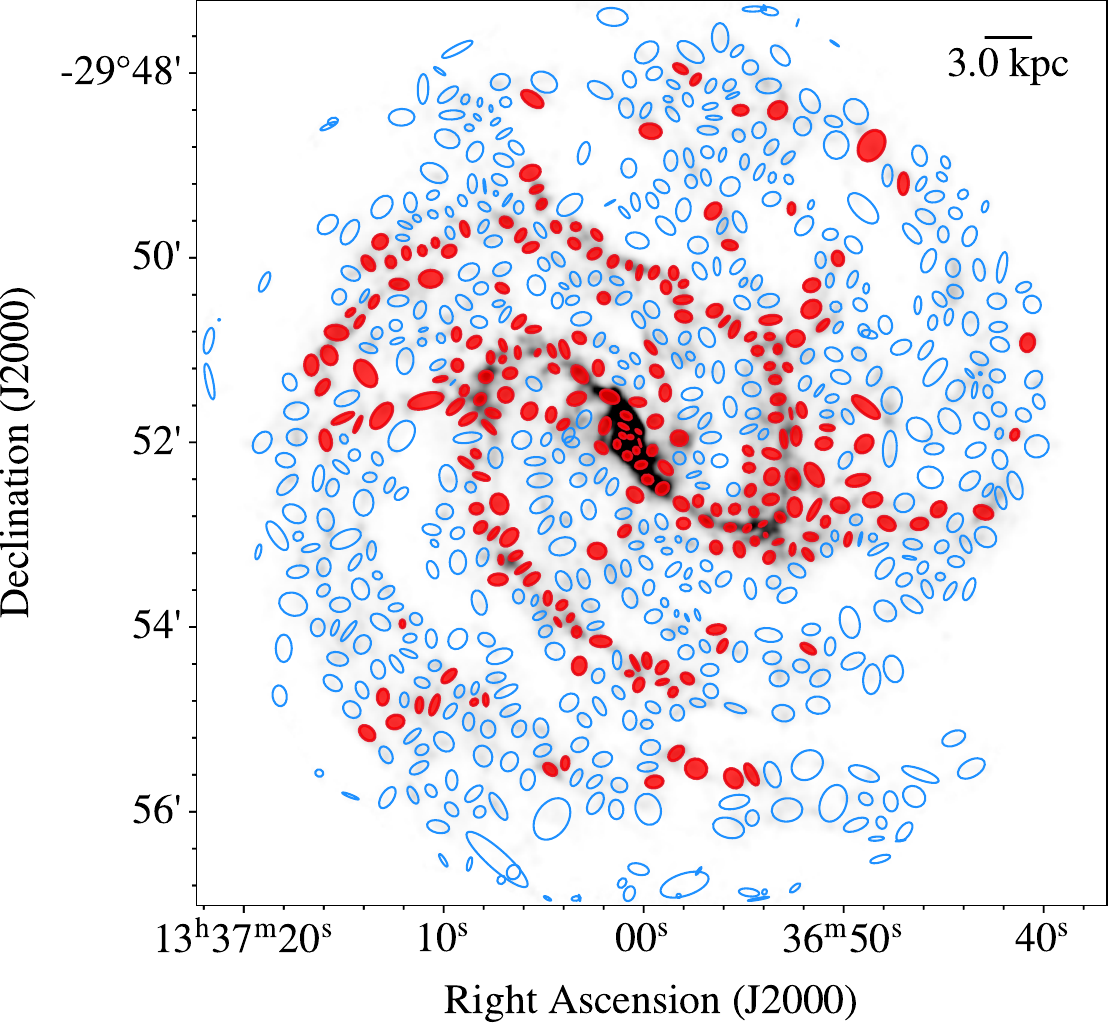}
\includegraphics[width=0.32\textwidth, bb=0 0 531 491]{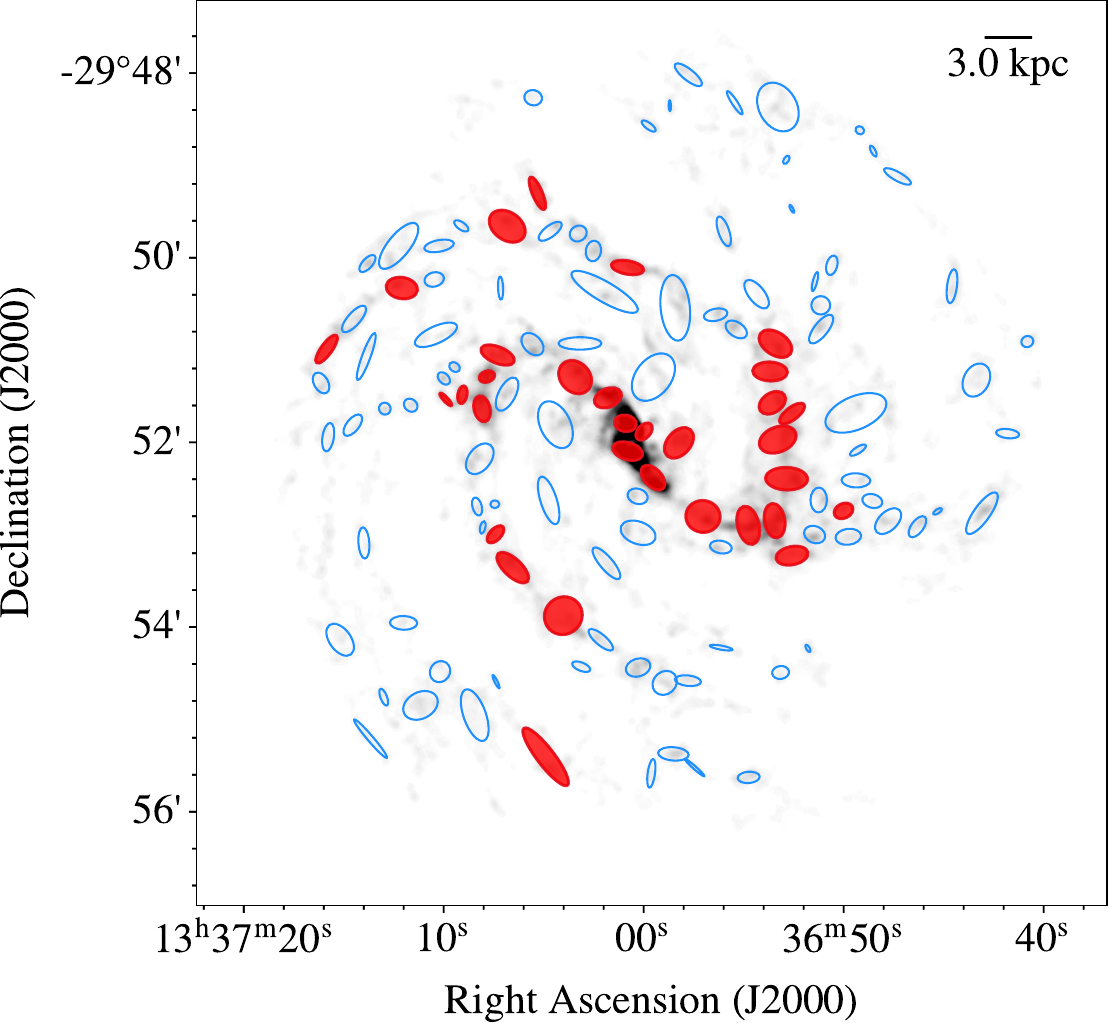}
\end{center}
\caption{
Spatial distribution of GMAs identified by {\tt PYCPROPS} in NGC~1316 (left), M~83 (middle), and a version of M~83 simulated to match the sensitivity of the NGC~1316 data (right). Red markers denote GMAs with S/N $\ge$ 6. In NGC 1316, GMAs are predominantly concentrated in the NW Shell and the SE Blob, whereas in M~83, they are primarily distributed along the spiral arms. Because the original M~83 data has an S/N ten times higher than that of NGC 1316, the emission-detection threshold used in SoFiA to generate the input mask for the GMA search was made ten times more stringent for M83. It is evident from this comparison that this adjustment results in the identification of larger GMA sizes.
}
\label{fig:pycprops}
\end{figure*}

\begin{figure*}[]
\begin{center}
\includegraphics[width=\textwidth, bb=0 0 1432 496]{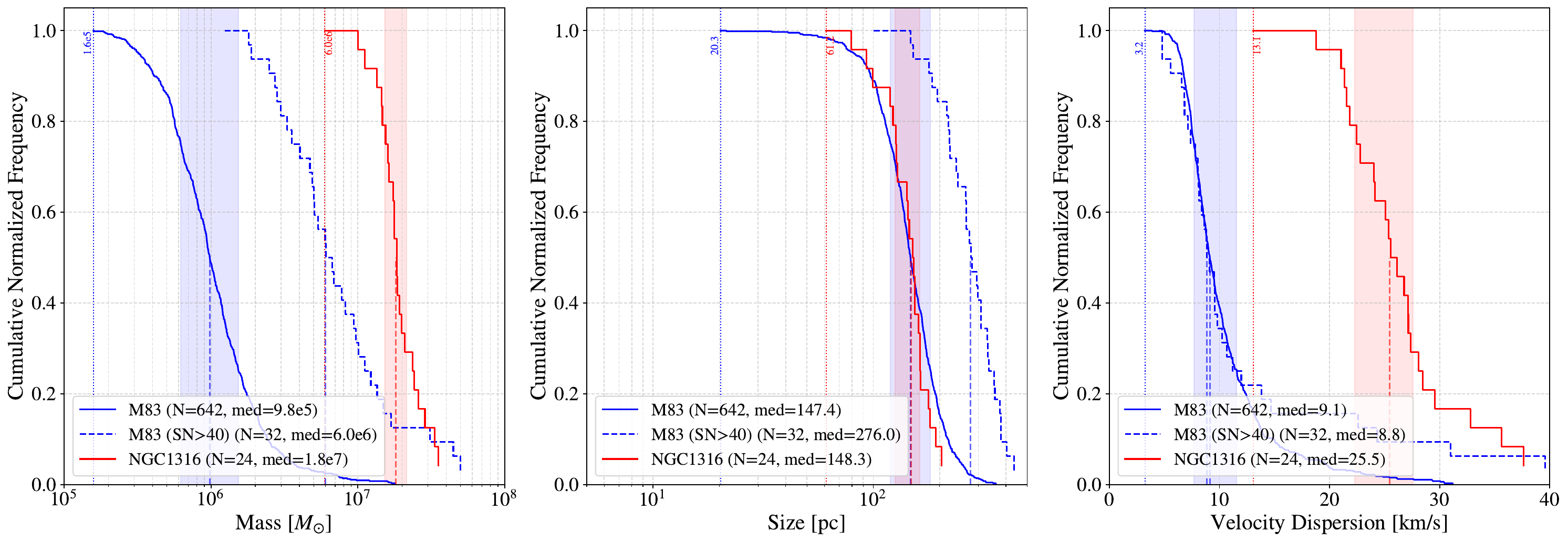}
\end{center}
\caption{
Comparison of GMA properties between NGC 1316 (red) and M83 (blue). From left to right, the panels show the distributions of GMA mass, size (pc), and velocity dispersion ($\text{km s}^{-1}$). Solid and dashed lines represent the results using ${\rm S/N} > 4$ and ${\rm S/N} > 40$ thresholds for cloud identification, respectively. For each distribution, the vertical dashed lines indicate the median values, while the surrounding semi-transparent shaded regions represent the interquartile range (IQR). The vertical dotted lines indicate the minimum mass, size, and velocity dispersion values for the GMAs identified in each galaxy.
}
\label{fig:gma_comparison}
\end{figure*}

\begin{figure}[]
\begin{center}
\includegraphics[width=0.5\textwidth, bb=0 0 495 496]{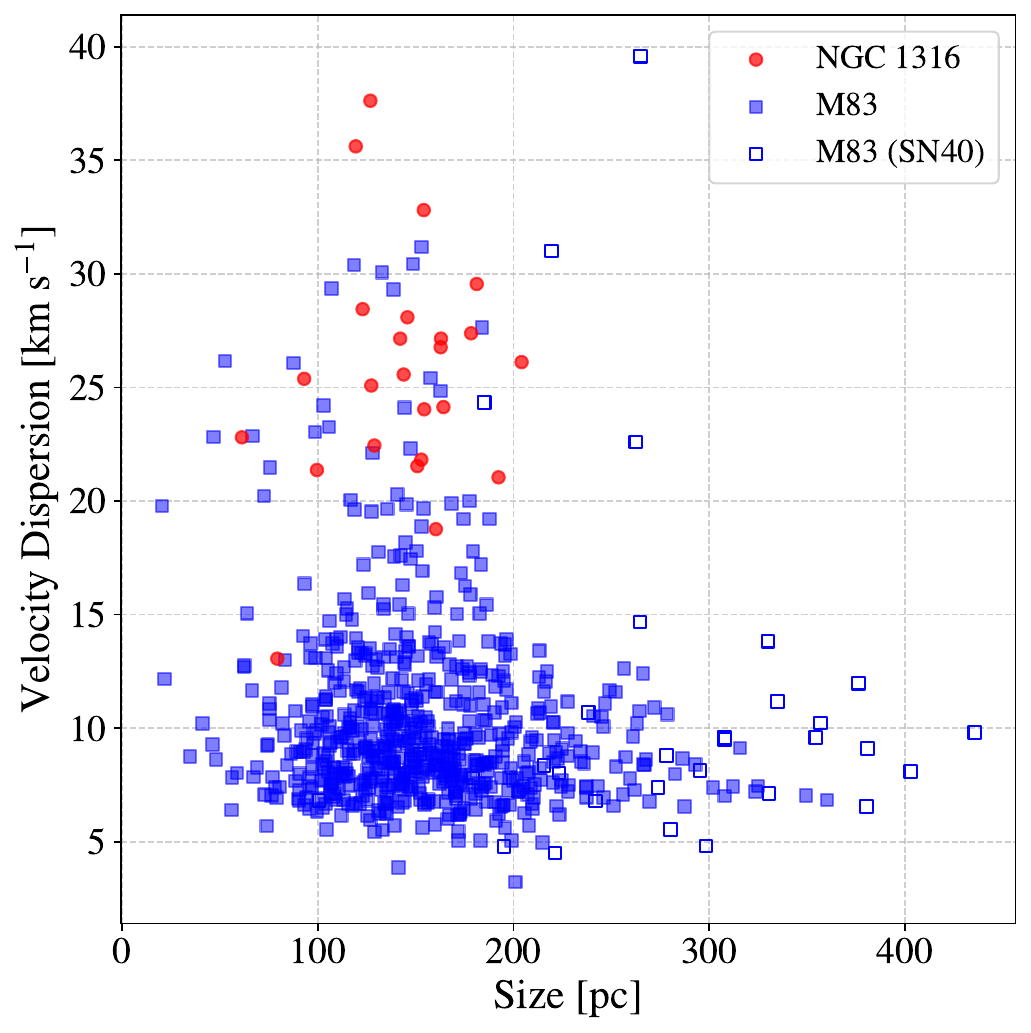}
\end{center}
\caption{
Size--velocity dispersion relation for GMAs in NGC~1316 (red circles) and M~83 (blue filled squares) identified by {\tt PYCPROPS}. The blue open squares denote ``sensitivity-degraded'' M~83 data, where the {\tt SoFiA2} S/N threshold was increased to 40 to mimic the tenfold lower sensitivity of the NGC~1316 data. While the majority of M~83 GMAs cluster at $\sigma < 15$~km~s$^{-1}$ (with a few reaching $\sim$30~km~s$^{-1}$), NGC~1316 shows a systematic shift toward higher dispersions, characterized by a notable absence of the low-dispersion population seen in M~83. All analyses are based on data with a spatial resolution of $\sim$100~pc and a velocity resolution of 5~km~s$^{-1}$.}
\label{fig:gma_size_vdisp}
\end{figure}

We investigate the properties of Giant Molecular Cloud Associations (GMAs) using the CO($J=1-0$) data. We identified the GMAs and derived the physical properties of the identified GMAs using {\tt PYCPROPS} \citep{Rosolowsky:2006uk,Rosolowsky:2021qo} basically with its default configuration. Specifically, the local RMS noise ($\sigma$) was estimated from the Median Absolute Deviation (MAD) of the data cube, which provides a robust measure of noise that is insensitive to the presence of real astronomical emission. The dendrogram pruning threshold ($\Delta$), which defines the minimum contrast required for a substructure to be identified as an independent entity distinct from neighboring structures, was set to $2\sigma$ to ensure the significance of the identified structures.
This conservative value was chosen to effectively filter out spurious noise fluctuations while retaining high completeness for faint but genuine structures, following the standard approach established in recent large-scale surveys \citep[e.g.,][]{Rosolowsky:2021qo}.
We adopted a CO-to-$\text{H}_2$ conversion factor of $\alpha_{\text{CO}} = 4.36 \, M_\odot \, (\text{K km s}^{-1} \text{pc}^2)^{-1}$, as justified in Section~\ref{sec:mom}. The cloud radii were calculated by applying a scaling factor of 1.91 to the RMS spatial size ($R = 1.91 \sigma_{\text{rms}}$), and all properties were extrapolated to account for sensitivity limits using the package’s default settings. No bootstrap resampling or channel correlation corrections were applied.

To facilitate this analysis, we utilized the input mask files created by {\tt SoFiA-2} (as described in Section~\ref{sec:mom}) to restrict the {\tt PYCPROPS} analysis to regions with CO emission. Using this masked approach, we successfully measured the size and velocity dispersion for 69 GMAs (Figure~\ref{fig:pycprops}). Among these, the numbers of GMAs with signal-to-noise ratios (S/N) $\ge 3, 6,$ and $10$ are 67, 24, and 4, respectively. The resulting mass, size, and velocity dispersion distributions for the GMAs in NGC~1316 are presented in Figure~\ref{fig:gma_comparison}.

While it would be ideal to compare the cloud properties in NGC~1316 directly with a comprehensive sample of similar radio galaxies with/without a prominent jet-ISM interaction, high-resolution datasets that simultaneously achieve comparable spatial and velocity resolutions, as well as sensitivity to resolve individual GMAs/GMCs, are extremely scarce for radio galaxies. Therefore, to isolate how the molecular gas properties in a jet-perturbed environment deviate from a normal galactic environment, we utilize a well-studied star-forming galaxy as a crucial baseline.
Following the same procedure, we analyzed ALMA CO($J=1-0$) data of M~83 to provide a comparative baseline for a typical star-forming environment. M~83 is a nearby spiral galaxy located at a distance of 4.5~Mpc, with a total molecular gas mass of $2.6 \times 10^9~M_\odot$ \citep{Koda:2023cr}, which is approximately 4.6 times that of NGC~1316. Its stellar mass \citep[$2.5 \times 10^{10}$~M$\odot$;][]{Barnes:2014np}, SFR \citep[$5.2$~M$\odot$~yr$^{-1}$;][]{Jarrett:2019vl}, and morphology closely resemble those of the Milky Way, making it an ideal control for investigating GMA properties in a ``normal'' galactic environment.

To ensure a fair comparison with the NGC 1316 dataset, the original M~83 data \citep[with an initial beam size of $2.09'' \times 1.68''$ ($\sim 46 \times 37$~pc)][]{Koda:2023cr} were smoothed to a $4.5''$ ($\sim 98$~pc) Gaussian beam to match the $\sim 100$~pc resolution of NGC~1316. The pixel size and velocity resolution were also matched to the NGC 1316 data at $0.5''$ ($\sim 10.9$~pc, corresponding to the $\sim 10$~pc pixel scale of NGC~1316) and 5~km~s$^{-1}$, respectively. The resulting $1\sigma$ sensitivity per velocity channel is $4.7 \times 10^{-3}$~Jy~beam$^{-1}$ ($\sim 2.1 \times 10^{-2}$~K), corresponding to a molecular gas surface density of $0.46$~$M_\odot$~pc$^{-2}$ assuming $\alpha_{\rm CO} = 4.36$~$M_\odot$~pc$^{-2}$~(K~km~s$^{-1}$)$^{-1}$. By applying the same analysis procedure to these two disparate systems with matched spatial, pixel, and spectral scales, we aim to isolate the distinct characteristics of GMAs in NGC~1316 that may be influenced by its radio jets. The results for M~83 are presented alongside the NGC~1316 data in Figures~\ref{fig:pycprops} and \ref{fig:gma_comparison}.

Focusing on GMAs detected with ${\rm S/N} > 6$, we examine and compare their masses, sizes, and velocity dispersions between NGC~1316 and M~83 in Figures~\ref{fig:gma_comparison} and \ref{fig:gma_size_vdisp}.
In Figure~\ref{fig:gma_comparison}, solid lines represent M~83 (blue) and NGC~1316 (red). The blue dashed line shows GMAs in M~83 detected using an increased {\tt scfind.threshold} of 40 in {\tt SoFiA2} (compared to 4 for solid lines). This higher S/N threshold serves as a first-order test to illustrate the potential impact of the tenfold sensitivity difference between the two galaxies, while noting that this approach does not perfectly account for the differences in the spatial and spectral noise non-uniformity and actual sensitivity limits between the two datasets.

Compared to M~83, NGC~1316 exhibits a tendency toward larger GMA masses, which is driven by its higher velocity dispersions rather than differences in size, as the size distributions of the two galaxies are comparable. Figure~\ref{fig:gma_size_vdisp} illustrates the size--velocity dispersion relation for the identified GMAs. 
We find that the majority of M~83 GMAs are characterized by relatively low velocity dispersions of $\sigma < 15$~km~s$^{-1}$, although a few outliers reach approximately 30~km~s$^{-1}$. 
While the NGC~1316 GMAs appear to exhibit higher overall velocity dispersions, this trend is primarily driven by the absence of the low-dispersion population ($\sigma < 15$~km~s$^{-1}$) that is prevalent in M~83. 

The contrast in the size–velocity dispersion relation between NGC~1316 and M~83 likely reflects intrinsic physical differences in their GMA properties, rather than being a mere observational artifact of sensitivity. A comparison between the standard (filled square) and high-threshold (open square) results for M~83 suggests that reduced sensitivity primarily tends to inflate the estimated sizes of GMAs, while its impact on the overall velocity dispersion distribution is less pronounced. 
This ``size-inflation'' appears to arise from a combination of two effects: the loss of faint, small-scale clouds below the detection limit and the artificial blending of adjacent clouds into larger structures (source confusion).

As a result, GMAs identified in the sensitivity-degraded data become biased toward the most massive cloud complexes. Importantly, the observed trend in NGC~1316, where velocity dispersions are substantially enhanced relative to cloud size, is inconsistent with the bias seen in the sensitivity-degraded M~83 data.
While spatial blending can theoretically merge multiple velocity components along the line of sight and artificially inflate the velocity dispersion of a combined structure, as potentially seen in the single high-dispersion outlier in the degraded M~83 dataset, such an effect fails to reproduce the overall shift of the bulk population toward high velocity dispersions seen in NGC~1316. Nevertheless, we again acknowledge that this threshold-based control test is not entirely exhaustive. To definitively confirm whether the observed high velocity dispersions are truly intrinsic to the NGC~1316 environment, future deeper observations matching the high sensitivity of the M~83 dataset are required.
While we note the limitations of this threshold-based test, these results suggest that the high velocity dispersions in NGC~1316 are an intrinsic characteristic of its GMAs, distinguishing it from typical star-forming galaxies like M~83.

\subsection{$R_{21}$ and $R_{31}$ ratios}\label{sec:co_ratios}

\begin{figure*}[]
\begin{center}
\includegraphics[width=0.49\textwidth, bb=0 0 935 765]{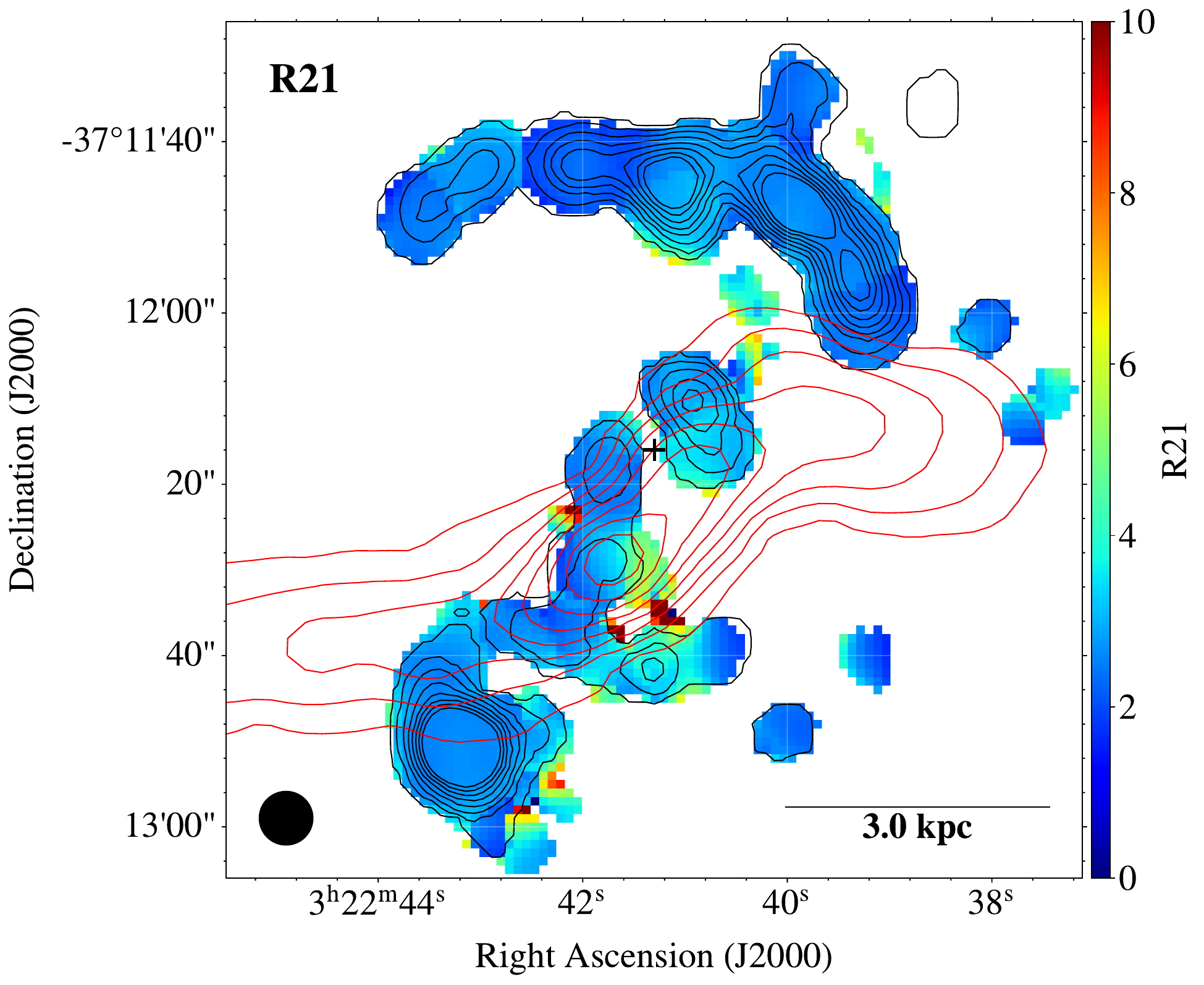}
\includegraphics[width=0.49\textwidth, bb=0 0 935 765]{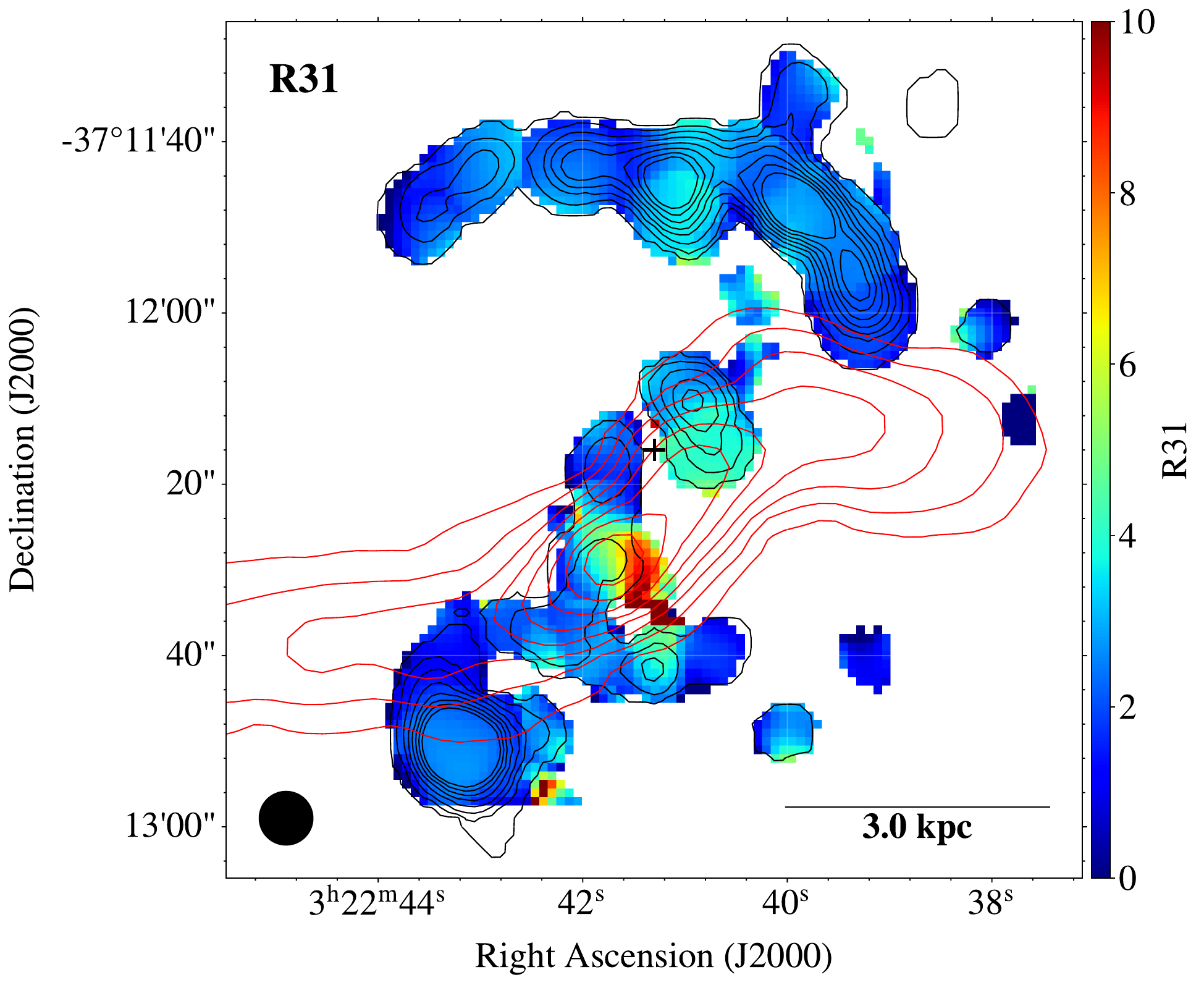}
\end{center}
\caption{
$R_{21}$ and $R_{31}$ maps using the peak intensity maps of each line in Jy units. The black contour indicates the CO($J=2-1$) integrated intensity map, and the red contour outlines the S-shaped nuclear jet. Elevated ratios are observed at several locations along the jet; in particular, the region to the west of the galactic center exhibits relatively high values, with $R_{21} \sim 4$ and $R_{31} \sim 9$.
}
\label{fig:coratios}
\end{figure*}

\begin{figure*}[]
\begin{center}
\includegraphics[width=0.49\textwidth, bb=0 0 533 532]{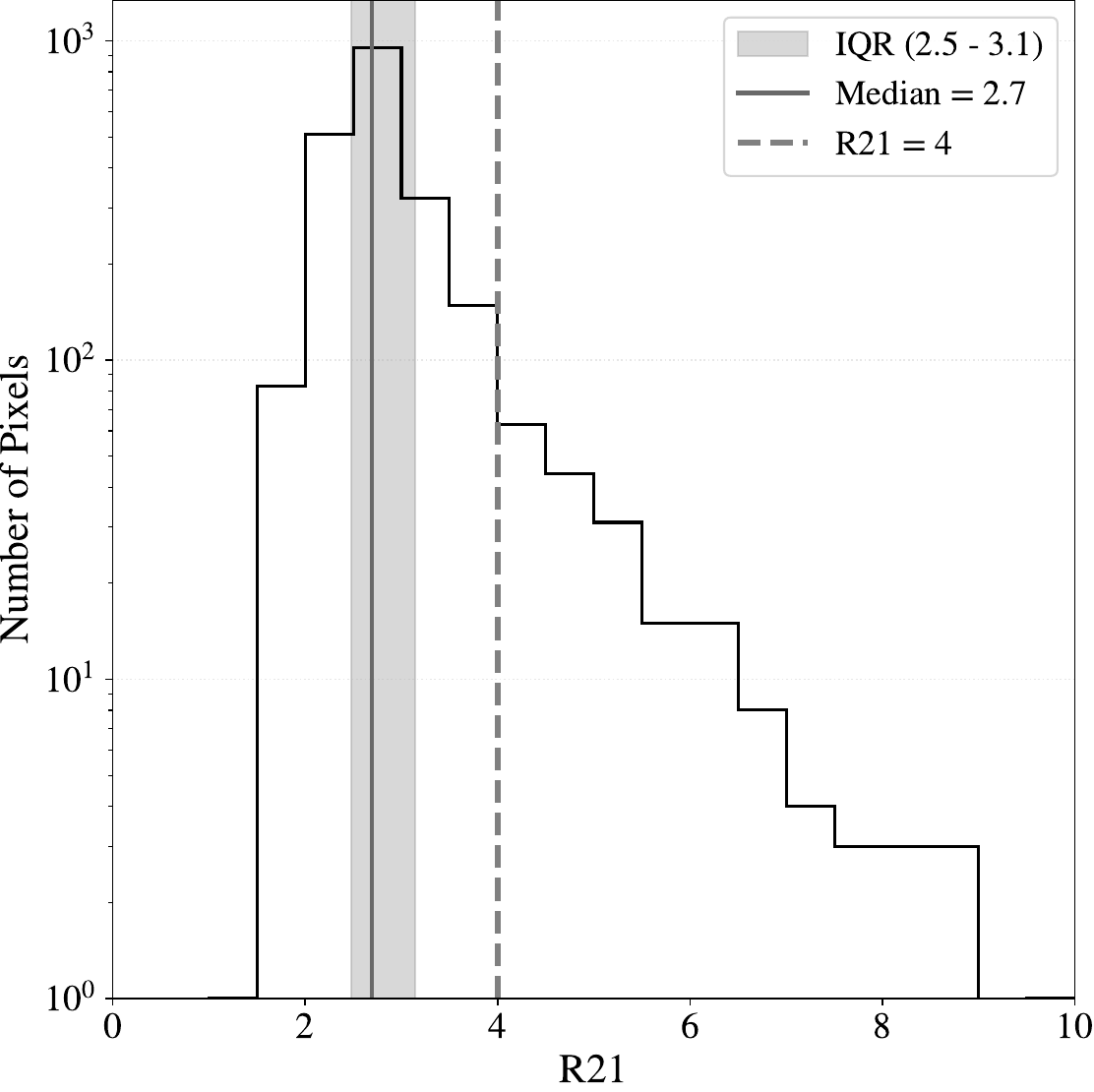}
\includegraphics[width=0.49\textwidth, bb=0 0 533 532]{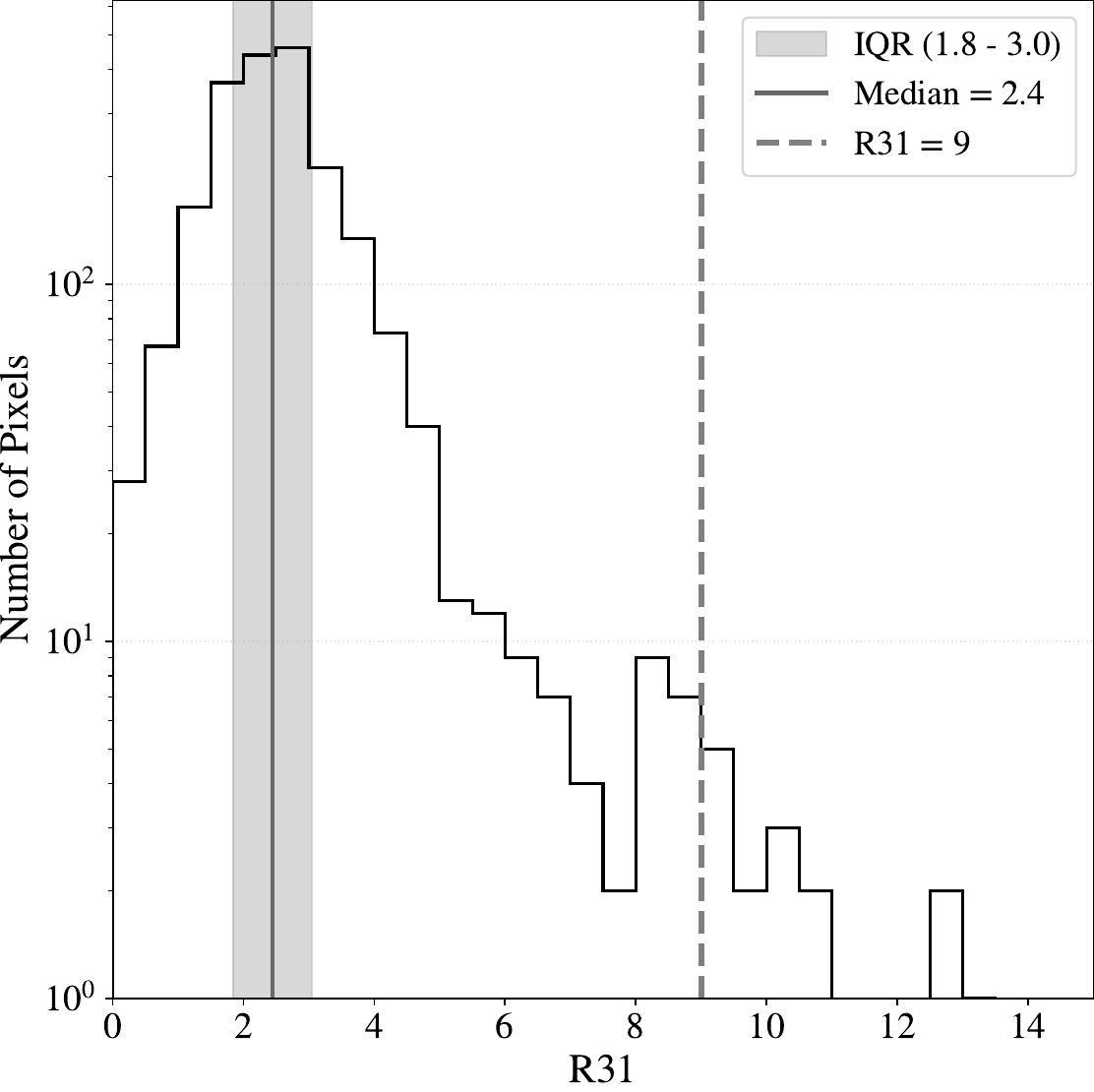}
\end{center}
\caption{
Histograms of $R_{21}$ and $R_{31}$ derived from the ratio maps shown in Figure~\ref{fig:coratios}. The ratios are calculated using flux units of Jy~beam$^{-1}$, with the vertical axis presented on a log scale. The median value for each distribution is indicated by a solid grey line, while the interquartile range (1st to 3rd quartile) is highlighted by the translucent grey shaded area. The grey dashed lines represent a ratio of unity ($R = 1$) in K units. While the ratios are generally comparable to those observed in typical star-forming galaxies, certain regions exhibit $R_{21}$ and $R_{31}$ values exceeding unity in K units.
}
\label{fig:coratios_histo}
\end{figure*}

\begin{figure*}[]
\begin{center}
\includegraphics[width=\textwidth, bb=0 0 1783 1495]{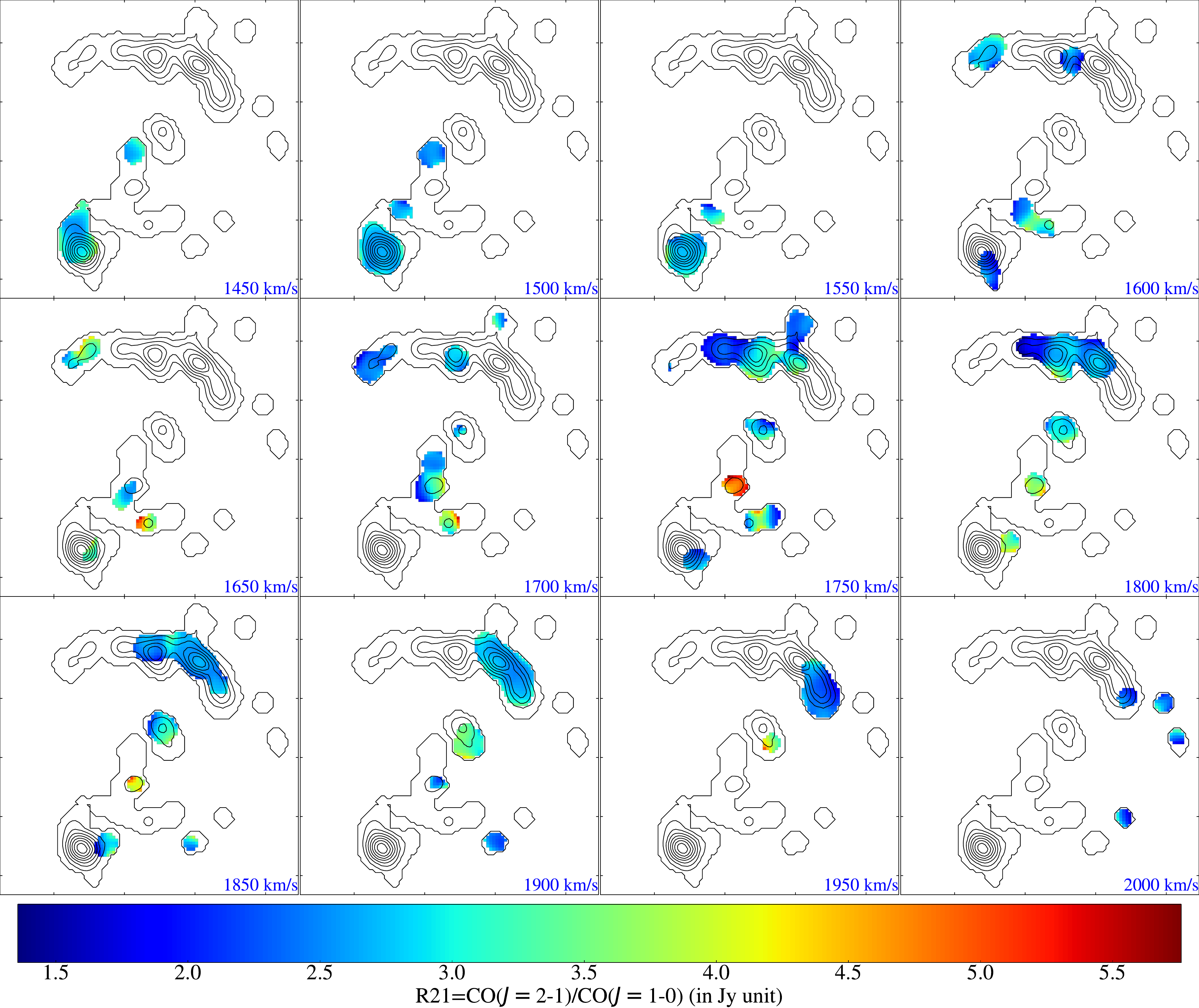}
\end{center}
\caption{
$R_{21}$ map at different velocity channels.
Contours show the CO($J=2–1$) integrated intensity.
Near the galactic center, an $R_{21}$ ratio exceeding 4 (in Jy units) is observed around a velocity of 1750~km~s$^{-1}$.
}
\label{fig:r21_v}
\end{figure*}

\begin{figure*}[]
\begin{center}
\includegraphics[width=\textwidth, bb=0 0 1783 1495]{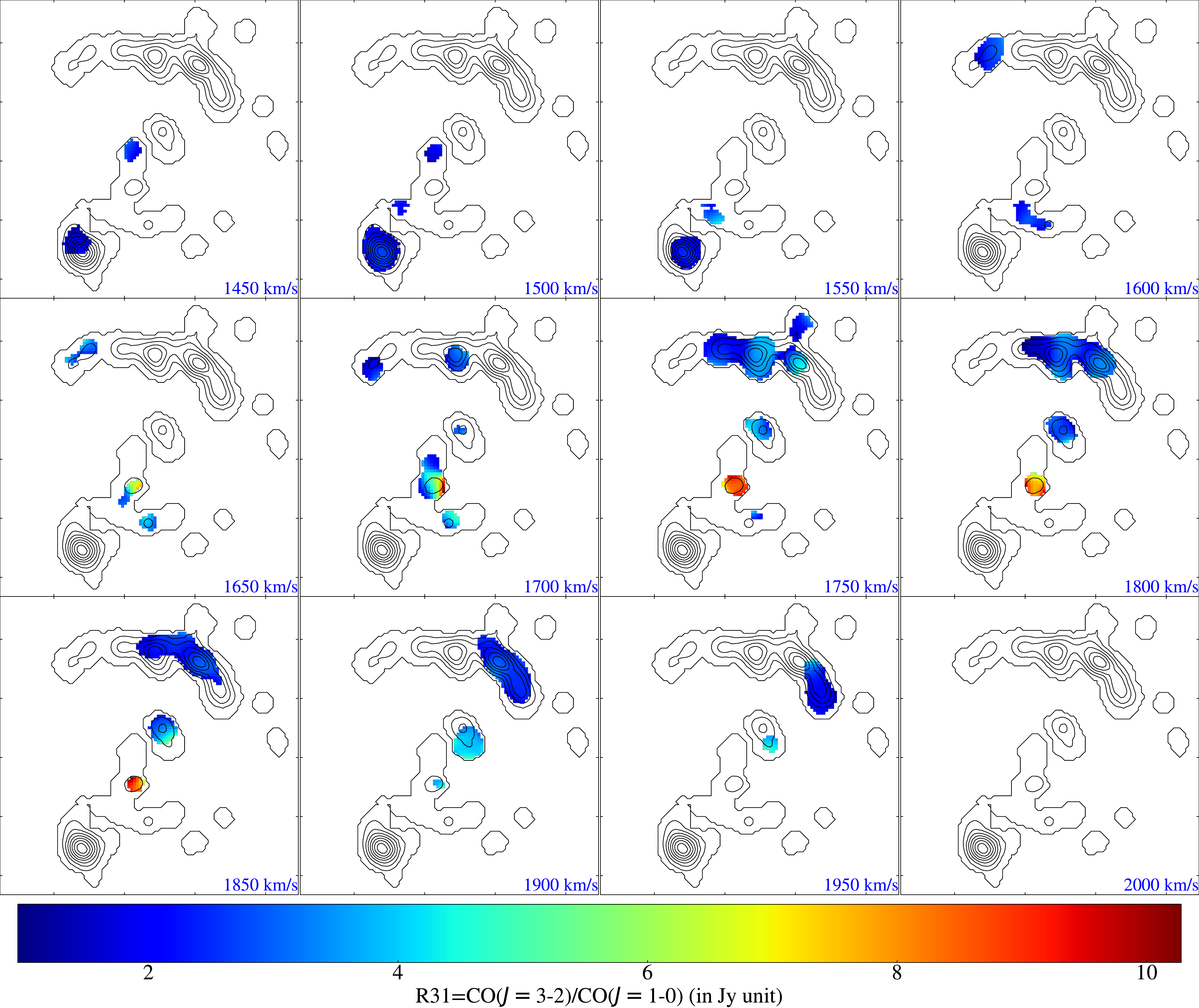}
\end{center}
\caption{
$R_{31}$ map at different velocity channels.
Contours show the CO($J=2–1$) integrated intensity.
Near the galactic center, an $R_{31}$ ratio exceeding 9 (in Jy units) is observed around a velocity of 1750~km~s$^{-1}$.
}
\label{fig:r31_v}
\end{figure*}

CO line ratios serve as a powerful diagnostic of the physical conditions of the molecular gas, as they primarily depend on the gas temperature and density, although these two parameters are often degenerate. In particular, previous studies of radio galaxies hosting jet-ISM interactions have frequently reported enhanced CO line ratios, suggesting that jet-induced shocks or turbulence can substantially alter the excitation state of the molecular gas \citep[e.g.,][]{Oosterloo:2017oc,Ruffa:2022zo,Ruffa:2026jf}.
In order to explore the excitation properties of CO emission in NGC~1316, we investigate the CO($J=3-2$) to CO($J=1-0$) ratio ($R_{31}$) and the CO($J=2-1$) to CO($J=1-0$) ratio ($R_{21}$).
To prepare data sensitive to the same spatial scales, we re-imaged the CO($J=3-2$), CO($J=2-1$), and CO($J=1-0$) data by matching their {\tt UVwave} range. 
The original {\tt UVwave} ranges were approximately $0-520~{\rm k}\lambda$ for CO($J=1-0$), $5-37~{\rm k}\lambda$ for CO($J=2-1$), and $8-54~{\rm k}\lambda$ for CO($J=3-2$). Therefore, we limited the {\tt UVwave} range of the data used to $8-37~{\rm k}\lambda$.

Imaging was performed using {\tt tclean} task, employing Briggs weighting with a robust parameter of 0.5.
Before calculating the CO line ratios, we ensured that all three datasets had identical spatial and spectral gridding, as well as identical beam sizes. Specifically, the data cubes were imaged with an image size of $180 \times 192$ pixels, a pixel scale of $1''.0$, and a spectral gridding of 18 channels with a channel width of 50~km~s$^{-1}$ covering a velocity range of 1300--2200~km~s$^{-1}$.
For masking, we employed the '{\tt auto-multithresh}' option with parameters set of: {\tt sidelobethresh} of 1.25; {\tt noisethresh} of 5.0; {\tt minbeamfrac} of 0.1; {\tt lownoisethresh} of 2.0; {\tt negativethresh} of 0.0; {\tt stop} of 2.0 for all the data.
The resulting beam sizes and data RMS levels for CO($J=1-0$), CO($J=2-1$), and CO($J=3-2$), all at a velocity resolution of 50~km~s$^{-1}$, were $4''.1 \times 3.8''$ at 3.76~mJy~beam$^{-1}$ (22 mK), $6''.3 \times 4''.0$ at 6.5~mJy beam$^{-1}$ (6.0 mK), and $4''.4 \times 3''.7$ at 12~mJy~beam$^{-1}$ (7.6~mK), respectively.
We unified the beam sizes of all data to $6''.4$ using {\tt imsmooth}, and aligned their gridding with {\tt imregrid}.

We used {\tt SoFiA2} to identify voxels where CO emission was detected. To construct the line ratio cubes, we generated a mask that includes only those voxels where emission was detected above the threshold in both the numerator and denominator CO transitions. The ratio cubes were then computed from these masked data cubes using the CASA task {\tt immath}.
The CO($J=3-2$) data's southern region suffers from reduced sensitivity near the edge of the observed field, leading to larger errors and a narrower detected emission area.

To account for the complex velocity structure in NGC~1316, the $R_{21}$ and $R_{31}$ ratio maps in Figure~\ref{fig:coratios} were derived from peak temperatures ($T_{\text{peak}}$; Moment~8). These peak temperature maps for each transition were constructed using the CASA task {\tt immoment} by applying the emission masks generated by {\tt SoFiA2}, and the final ratio maps were computed from these masked peak temperature maps. This approach ensures more physically meaningful excitation measurements in regions where multiple velocity components are superimposed, as simple integrated intensity ratios can obscure the distinct physical properties of overlapping gas clouds.


Figure~\ref{fig:coratios_histo} displays the pixel-value histograms for these ratio maps. It should be noted that these ratios are initially calculated using data in units of Jy~beam$^{-1}$. Because the conversion factor from flux density in Jy~beam$^{-1}$ to brightness temperature in Kelvin is inversely proportional to the square of the observation frequency ($\propto \nu^{-2}$), line ratios expressed in Kelvin units are systematically lower than those expressed in Jy units. For $R_{21}$, the resulting median and IQR are $2.7^{+0.4}_{-0.2}$ in Jy units, which corresponds to $0.68^{+0.05}_{-0.10}$ in Kelvin units. Similarly, the median and IQR for $R_{31}$ are $2.4 \pm 0.6$ in Jy units and $0.27 \pm 0.07$ in Kelvin units. These results are broadly consistent with the CO line ratios typically observed in normal star-forming galaxies, where $R_{21} \sim 0.7$ and $R_{31} \sim 0.3$ in Kelvin units \citep[e.g.,][]{Koda:2020mp, Koda:2025ho, Yajima:2021fh, Leroy:2022fx}.

However, several localized regions exhibit notably higher ratios, with both $R_{21}$ and $R_{31}$ reaching approximately 4 to 5 in Jy units. These regions include the eastern and central parts of the ``NW Shell'', the western side of the ``Clump'', and a portion of the ``Extended'' structure. Intriguingly, these high-ratio areas generally correspond to regions characterized by large CO velocity dispersions, as shown in Figure~\ref{fig:mom_co10}.

Furthermore, a distinct region located west of the galactic center exhibits relatively high ratios, specifically $R_{21} \sim 4$ and $R_{31} \sim 9$ in Jy units. These values are consistent with the expected ratios under conditions of local thermodynamic equilibrium (LTE). This specific location, and the physical implications of such elevated ratios that approach the theoretical LTE limit, will be discussed in detail in Section~\ref{sec:discussions}.

Given the complex velocity field exhibited by the molecular gas in NGC 1316, we also investigated the $R_{21}$ and $R_{31}$ channel maps (Figures~\ref{fig:r21_v} and \ref{fig:r31_v}) at a velocity resolution of $50\ \text{km s$^{-1}$}$. These figures demonstrate that, irrespective of the spatial position, the enhanced $R_{21}$ and $R_{31}$ ratios are primarily confined to the velocity range associated with the emission near the galactic center ($\sim1650-1900$~km~s$^{-1}$)

\subsection{Summary of Results}\label{sec:sum_results}

Our findings are summarized as follows:
\begin{itemize}
    \item \textbf{High-resolution Imaging and Kinematics (Section~\ref{sec:mom}):} Our $100\text{-pc}$ resolution CO($J=1-0$) data reveal complex sub-structures within the previously identified Shell, Clump, Extended, and Blob components. The kinematics are also complex, revealing intricate sub-structures that cannot be explained by simple circular rotation. While the velocity dispersion is generally $\lesssim 20\text{ km s}^{-1}$, it exceeds $50\text{ km s}^{-1}$ in several localized regions.
    
    \item \textbf{Extended CO Emission (Section~\ref{sec:ext}):} A comparison between interferometric and single-dish data indicates that $60\text{--}70\%$ of the total CO($J=1-0$) flux is missing from the interferometric maps. This missing flux is particularly pronounced near the galactic center, where emission was almost entirely filtered out by the interferometers.
    
    \item \textbf{GMA Identification and Characterization (Section~\ref{sec:gma}):} Using {\tt PYCPROPS}, we identified 24 GMAs. A comparison with M~83 at similar resolution shows that while GMA sizes are comparable ($\sim 148\text{ pc}$), those in NGC~1316 are significantly more massive ($1.8 \times 10^7$ vs. $9.8 \times 10^5\ M_\odot$) and exhibit much higher velocity dispersions ($25.5$ vs. $9.1\text{ km s}^{-1}$). Notably, GMAs are predominantly located in the NW Shell and SE Blob, while remaining largely absent near the nucleus.
    
    \item \textbf{CO Excitation (Section~\ref{sec:co_ratios}):} Most regions exhibit line ratios typical of normal star-forming environments ($R_{21} \sim 0.7$, $R_{31} \sim 0.3$ in K units). However, elevated ratios ($R_{21} \gtrsim 1.0$, $R_{31} \gtrsim 0.6$ in K units) were observed in the galactic center and within high-dispersion areas of the ``NW Shell'', ``Clump'' and ``Extended'' structures, suggesting different physical conditions in these components.
\end{itemize}

\section{Discussions} \label{sec:discussions}

The ISM kinematics of NGC 1316 have been shown to be inconsistent with a simple rotating disk model \citep[e.g.,][]{Morokuma-Matsui:2019gv,Maccagni:2021dz}.
The stellar body of this galaxy exhibits prominent tidal tails and loops, which provide strong morphological evidence of one or more merger events that occurred relatively recently, approximately 1--3~Gyr ago \citep[e.g.,][]{Schweizer:1980gl,Goudfrooij:2001ya,Serra:2019ph}.
\cite{Maccagni:2021dz} conducted a detailed kinematic analysis using multi-wavelength data, including ionized gas data from VLT/MUSE, neutral atomic hydrogen (\HI) gas from MeerKAT, and kpc-scale molecular gas from ALMA. Their findings revealed the presence of both inflowing and outflowing gas components, which were interpreted within the framework of CCA \citep[e.g.,][]{Gaspari:2013rf,Gaspari:2015bx,Gaspari:2017fb}. CCA is a mechanism proposed to efficiently feed cold gas toward galactic centers, primarily used to explain the cooling flows in Brightest Cluster Galaxies (BCGs), where runaway thermal instability in the turbulent hot gas halo leads to rapid cooling and subsequent infall.

Our analysis demonstrates that the molecular gas in NGC~1316 is characterized by a higher fraction of extended components (with a spatial extent of $>7.2$~kpc) compared to typical star-forming galaxies, particularly in the vicinity of the galactic center. Furthermore, the GMAs detected at larger galactocentric distances exhibit higher velocity dispersions than typical star-forming galaxies. We also identified localized regions where high velocity dispersion correlates with elevated $R_{21}$ and $R_{31}$ line ratios. Based on these findings, the following sections discuss the impact of relativistic jets on the ISM. Specifically, we compare our ALMA results with multi-wavelength data and analyze the gas properties through a comparison with numerical simulations of jet-ISM interactions and physical models of supernova remnants (SNRs), which provide a baseline for understanding the energetic feedback on the molecular gas.


\subsection{Comparison with other wavelengths data}

\begin{figure*}[]
\begin{center}
\includegraphics[width=1.0\textwidth, bb=0 0 1743 1227]{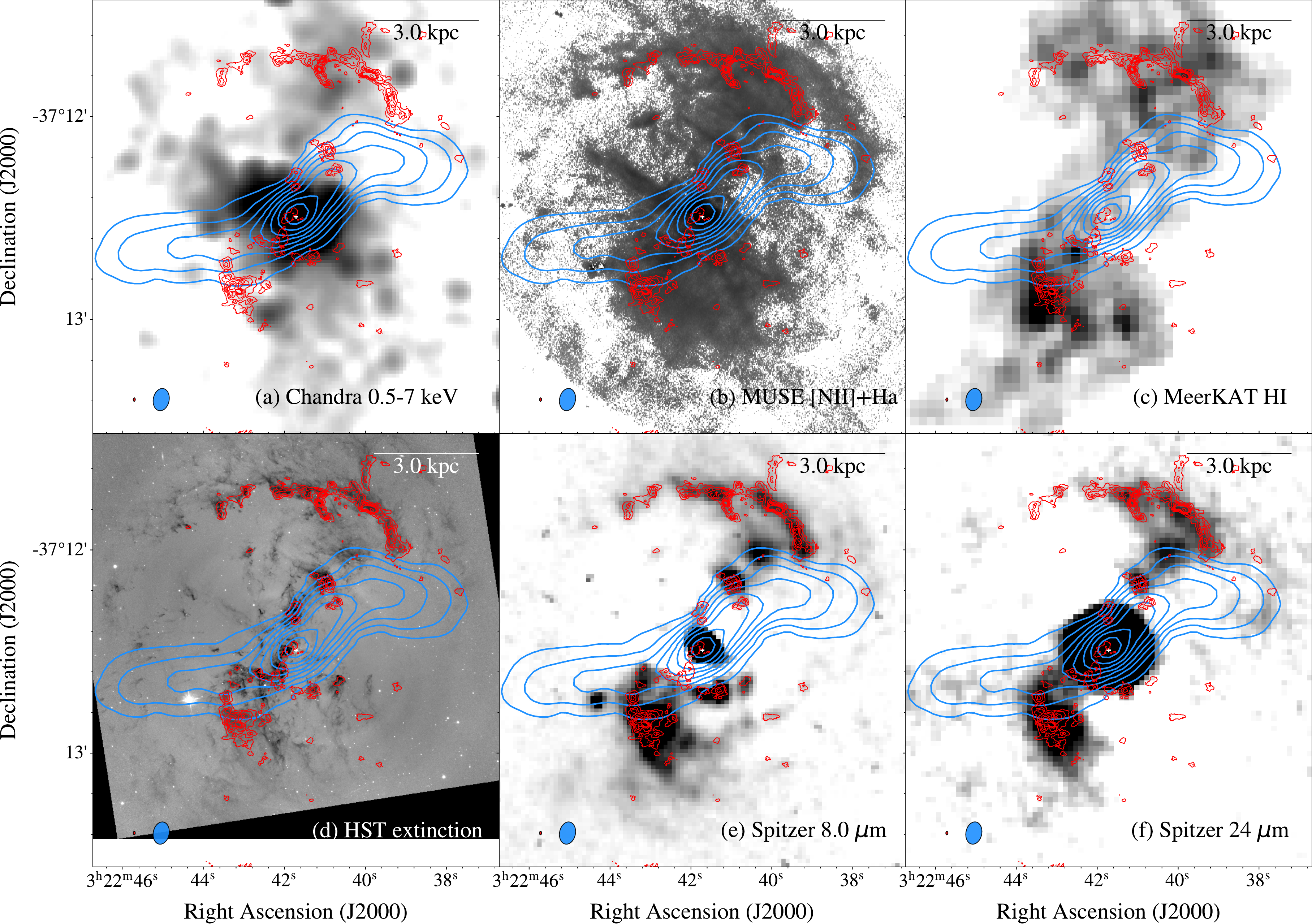}
\end{center}
\caption{
Comparison of ALMA CO($J$=1-0) integrated intensity (red solid contours) with multi-wavelength observations (grayscale). The S-shaped nuclear jet observed with MeerKAT at $1.4\ \text{GHz}$ is indicated by a blue solid contours in all panels. The filled ellipses at the bottom left of each panel represent the restoring beams for the CO (red) and Jet (blue) observations, both outlined with black edges. The greyscale images are: (a) Chandra X-ray map, (b) MUSE H$\alpha+$\nii~map, (c) MeerKAT~\HI~map, (d) HST extinction map, (e) stellar-subtracted Spitzer $8.0\ \mu\text{m}$ map, and (f) stellar-subtracted Spitzer $24\ \mu\text{m}$ map.
}
\label{fig:comparison_multiwavelengths}
\end{figure*}

\begin{figure*}[]
\begin{center}
\includegraphics[width=0.9\textwidth, bb=0 0 623 798]{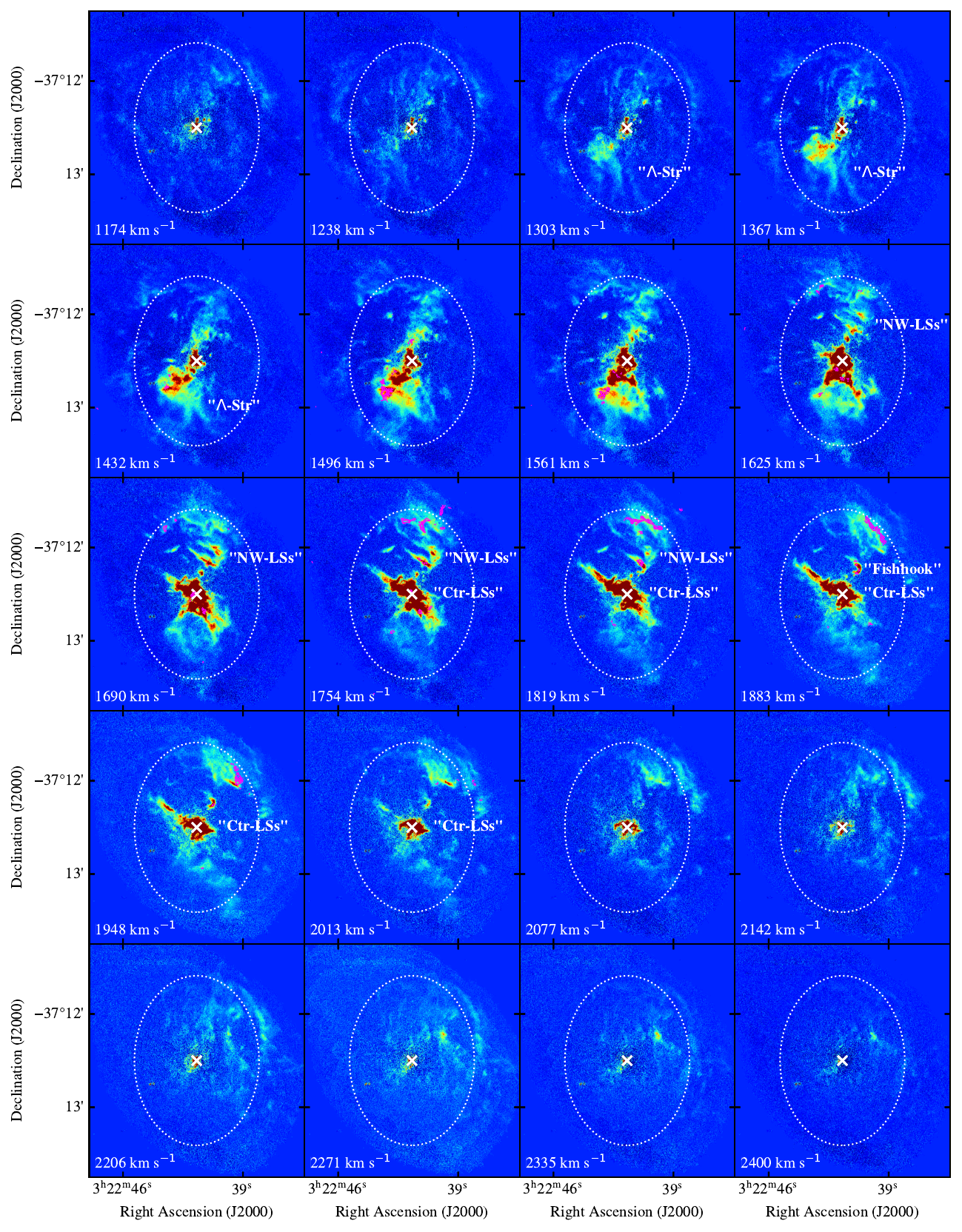}
\end{center}
\caption{
Velocity channel maps of MUSE \nii$\lambda$6583 (color) and ALMA CO($J=1-0$) (magenta contours). The white dashed circle indicates the characteristic shell-like structure of the WIG, with a radius of approximately 5.5~kpc. Panels are ordered by velocity from 1174 to 2400~km~s$^{-1}$. Partial spatial overlap is observed between the CO and \nii-bright regions. The \nii~kinematics reveal a shell structure originating in the east at low velocities, bifurcating into northern and southern components at intermediate velocities, and converging in the west at high velocities. The northern shell is prominent, while the southern part is more diffuse. Labeled WIG structures include: ``$\Lambda$-Str'' (southern $\Lambda$-shaped structure), ``NW-LSs'' (linear structures perpendicular to the NW jet), ``Fishhook'' (hook-shaped feature near the jet), and ``Ctr-LSs'' (linear structures extending NW/SE from the center).
}
\label{fig:comparison_niico}
\end{figure*}

\begin{figure*}[]
\begin{center}
\includegraphics[width=0.49\textwidth, bb=0 0 555 502]{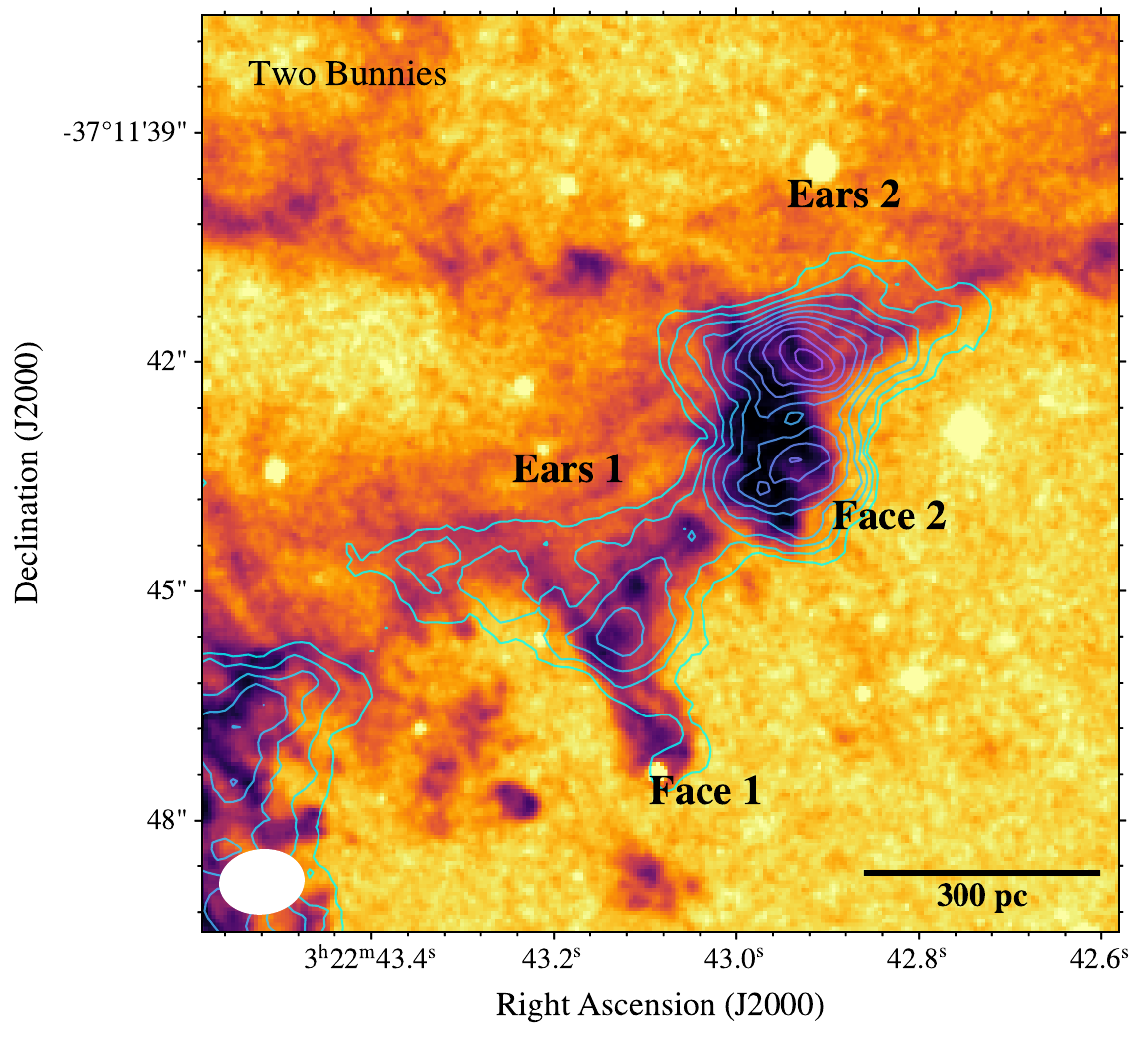}
\includegraphics[width=0.49\textwidth, bb=0 0 549 502]{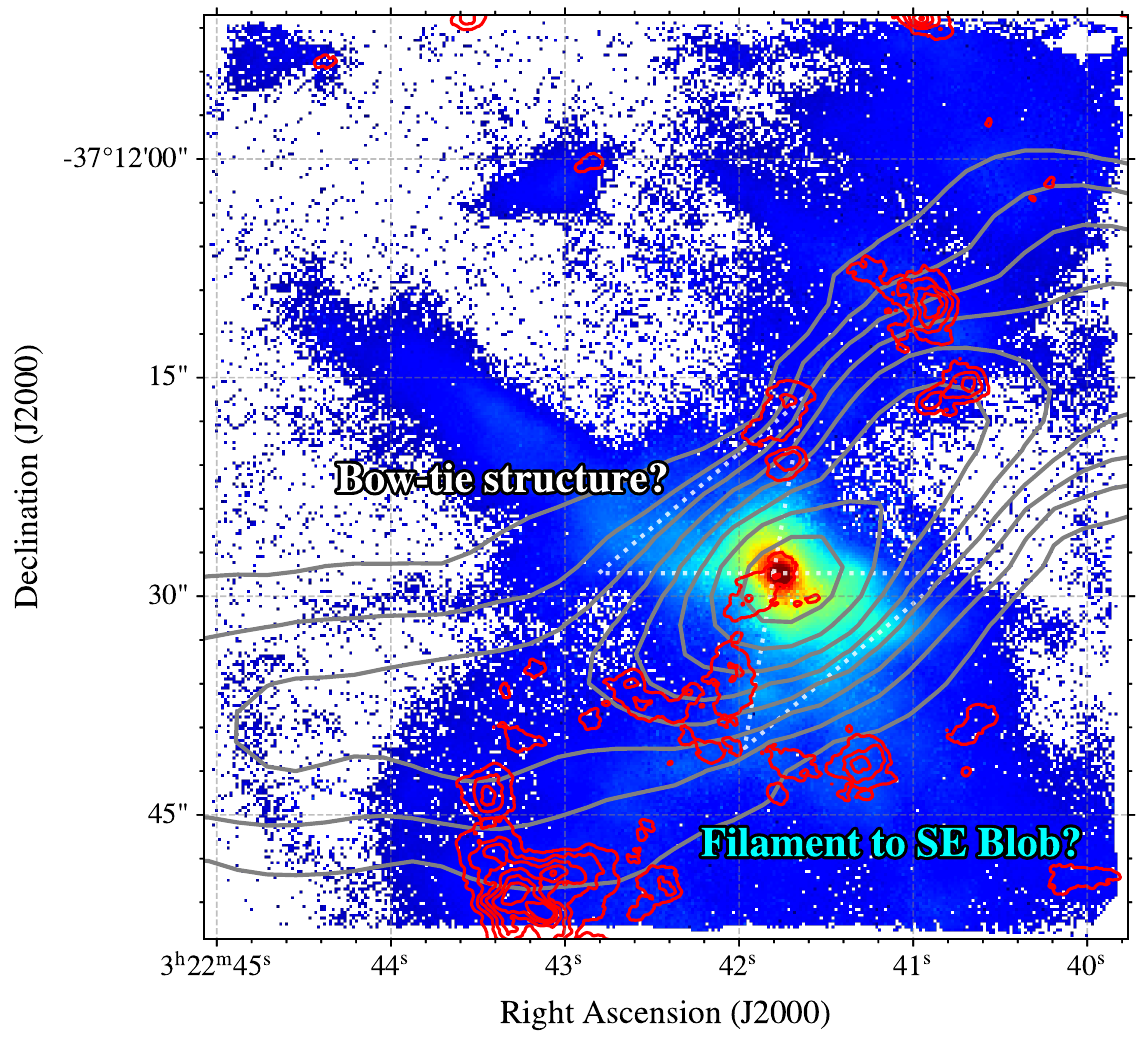}
\end{center}
\caption{
[Left] Example of the "bunny-face" structures observed in the east side of the NW Shell. The map shows the HST dust extinction (background) overlaid with CO($J=1-0$) emission contours. Two distinct ``bunny'' shapes are visible, with the CO emission peaks corresponding to the ``face'' of each structure. The white circle in the lower-left corner indicates the synthesized beam of the CO observations.
[Right] The \nii$\lambda6583$ emission map, with intensity levels adjusted to highlight bright central structures, overlaid with ALMA CO($J=1-0$) (red contours) and MeerKAT 1.4 GHz continuum (grey contours). The central emission exhibits a characteristic ``bow-tie'' shape.}
\label{fig:bunnies_and_bowtie}
\end{figure*}


Figure~\ref{fig:comparison_multiwavelengths} presents a comparison between the CO($J=1-0$) integrated intensity map (red solid line) and multi-wavelength data (greyscale). The S-shaped nuclear jet observed with MeerKAT at $1.4\ \text{GHz}$ is indicated by cyan solid contours.
The top row displays the Chandra X-ray \citep{Kim:2003yb,Lanz:2010ll}, VLT/MUSE H$\alpha$+\nii \citep{Maccagni:2021dz}, and MeerKAT \HI~images \citep{Serra:2019ph}. The bottom row shows the HST dust extinction map \citep{Duah-Asabere:2016ur} followed by the Spitzer 8.0~$\mu$m and 24~$\mu$m infrared images \citep{Lanz:2010ll}.
Note that the Spitzer data presented here have been processed to subtract the stellar contribution, following the methodology described in \cite{Lanz:2010ll}.

This figure demonstrates that the molecular gas shows good spatial correlation with the dust distribution. The $\text{Spitzer } 8.0\ \mu\text{m}$ emission is primarily attributed to the $7.7\ \mu\text{m}$ Polycyclic Aromatic Hydrocarbons (PAH) feature \citep{Lanz:2010ll}. If this assumption is correct, a strong spatial correspondence between the CO and PAH emission is also observed (Figure~\ref{fig:comparison_multiwavelengths}e).
However, it should be noted that the presence of PAH emission does not straightforwardly imply active star formation in this galaxy. Mid-infrared spectroscopic observations of NGC~1316 with {\it Spitzer} and {\it AKARI} have revealed that its PAH feature ratios deviate from those typically observed in normal star-forming galaxies, a characteristic shared with other radio galaxies \citep[e.g.,][]{Smith:2007pz,Kaneda:2007lh}. This indicates that the PAH may be excited by mechanisms other than star formation, such as interstellar shocks, AGN feedback, or the evolved stellar population, rather than UV photons from young massive stars. Therefore, caution is required when associating these PAH-emitting regions directly with ongoing star formation.

Furthermore, the molecular gas is adjacent to both hot/warm ionized gas and neutral atomic gas, which are traced by X-ray/\nii+H$\alpha$ and \HI~emission, respectively (Figure~\ref{fig:comparison_multiwavelengths}a, b, c).
Note that the \nii $\lambda$6583 emission surpasses the H$\alpha$ emission throughout this region displayed in Figures~\ref{fig:comparison_multiwavelengths} and \ref{fig:comparison_niico} (i.e., \nii$\lambda6583>$H$\alpha$), indicating a Low-Ionization Nuclear Emission-line Region (LINER)-type excitation \citep[e.g.,][]{Morokuma-Matsui:2019gv,Richtler:2020xr,Maccagni:2021dz} as is frequently observed in other early-type group-dominant galaxies \citep[e.g.,][]{Lagos:2022wv}.
Intriguingly, the region beyond the bend of the S-shaped nuclear jet appears devoid of X-ray gas (Figure~\ref{fig:comparison_multiwavelengths}a).
While \cite{Lanz:2010ll} proposed this feature as a small-scale cavity inflated by the jet, its identification as a sharp X-ray cavity is hindered by several factors. Specifically, the low X-ray photon counts limit the statistical significance of the feature. Furthermore, the inner jets have been inactive for a few~Myr \citep{Maccagni:2020pq}; this cessation of jet activity allows the plasma to spread out and diffuse, which tends to smear out the sharp cavity edges. These factors, potentially combined with the relatively faint or shallow hot atmosphere of a galactic group and projection along the line of sight, naturally explain why the region does not manifest as a well-defined X-ray hole compared to more prominent examples in massive clusters.

In the following sections, we characterize several prominent structures by combining the distribution of warm ionized gas (WIG), molecular gas, atomic gas, and dust, while also reviewing the properties of the nuclear jet in these regions. Our characterization is primarily based on the \nii $\lambda$6583 channel maps from \cite{Maccagni:2021dz} (Figure~\ref{fig:comparison_niico}), which provide a comprehensive view of the WIG and its kinematics over a wide spatial extent. It should be noted, however, that the gas affected by the jet or the jet-driven bubbles may not necessarily dominate over the secular kinematic patterns or merger-driven motions. For example, an infalling molecular gas complex may be halted by interaction with the jet, without necessarily leading to a net gas outflow. Therefore, identifying the specific mechanisms responsible for observed gas motions, particularly in distinguishing between inflows and outflows, remains challenging. A more extensive analysis of the gas kinematics can be found in \cite{Maccagni:2021dz}.

\subsubsection{Jet Geometry and X-ray Morphology}
The nuclear jet at 1.4 GHz appears relatively symmetric; however, higher-resolution VLA images reveal a slight dominance of the northern jet \citep{Geldzahler:1984rl}. This asymmetry likely arises from either relativistic beaming (suggesting the northern jet is inclined toward us) or higher ISM density in the north hindering jet propagation. Note that a key characteristic of the nuclear jet in NGC~1316 is its prominent S-shaped morphology. Such sinuous structures are typically attributed to either jet precession or interactions with the surrounding ISM. The diffuse X-ray emission predominantly occupies regions ``evading'' the jet path. In the northern region ($\sim$few kpc), the X-ray gas (or hot ionized gas, HIG) spatially correlates with WIG, \HI, and dust, yet it remains distinct from, and seemingly excluded by, the 5-kpc shell structure. In the south, the X-ray emission aligns with the WIG and \HI, but exhibits a spatial offset from the molecular gas and dust peaks toward the east.

\subsubsection{The 5-kpc Shell}
A shell-like feature with a radius of $\sim$5~kpc is identified in the \nii $\lambda$6583 emission (Figure~\ref{fig:comparison_multiwavelengths}b), based on the same dataset presented by \cite{Maccagni:2021dz}.
This feature is prominent on the eastern side at $V_{\text{LSR}} \sim 1238$~km~s$^{-1}$, although the \nii~line is partially blended with the adjacent H$\alpha$ emission. As the velocity increases toward 2400~km~s$^{-1}$, this 5-kpc shell splits from the east toward the north and south, gradually shifting westward. The velocity field of this structure appears to connect seamlessly with the central \HI~kinematics \citep[Figure 3 in][]{Maccagni:2021dz}. The kinematic major axis of this 5-kpc shell seems to align with the photometric major axis of the stellar component, yet its rotation is counter-directional; while the gas is blueshifted in the east and redshifted in the west relative to the systemic velocity, the stars exhibit the opposite gradient \citep{Arnaboldi:1998lp}.

\subsubsection{``Bunny-Face'' Morphologies}
Within the 5-kpc northwestern shell, both dust extinction maps and ALMA CO maps reveal clumps exhibiting a striking ``bunny-face'' structure. This morphology is characterized by a central dense clump (the ``face'') and two thin, bifurcated tails that extend like ``ears'' (Figure~\ref{fig:bunnies_and_bowtie} left). Similar ``bunny-face'' morphologies, each consisting of a central clump and two ear-like tails, are observed throughout the dust extinction map. Their orientations vary locally and are not strictly radial relative to the nucleus. While the southern portion of the ionized gas shell lacks prominent counterparts in dust or molecular gas, faint ``bunny-like'' features are visible in the extinction maps of that region.

\subsubsection{North-South WIG Associated with the Hot Ionized Gas}
A diffuse component of WIG is observed in association with the HIG detected in X-rays. While the radio jets extend along the northwest–southeast (NW–SE) axis, the diffuse HIG is distributed asymmetrically relative to these jets: it is located to the east of the northwestern jet and to the west of the southeastern jet (Figures~\ref{fig:comparison_multiwavelengths} and \ref{fig:comparison_niico}). In the \nii~channel maps, a faint emission feature extends toward the north-northwest from the nucleus. This feature is likely associated with the northern HIG located to the east of the northwestern jet and is visible within the velocity range of $V_{\text{LSR}} \sim 1174\text{--}1496$~km~s$^{-1}$. Conversely, a similar diffuse component extends toward the southwest relative to the nucleus, presumably associated with the southern HIG situated to the west of the southeastern jet, and is observed between $1625\text{--}2013$ km s$^{-1}$.

\subsubsection{Distinct WIG Structures between the Northern Shell and the Galactic Center}

Multiple features with complex kinematics are identified in the region between the northern 5-kpc shell and the galactic center. Within the velocity range of $V_{\text{LSR}} \sim 1690\text{--}2206$~km~s$^{-1}$, a structure extends southwest from the northern shell, exhibiting a progressive redshift with increasing distance from the shell boundary.
Near the systemic velocity ($1625\text{--}1754$~km~s$^{-1}$), two nearly parallel linear features are observed perpendicular to the jet axis, located approximately 2.5 kpc and 3 kpc north-northwest of the nucleus (``NW-LSs'' in Figure~\ref{fig:comparison_niico}). Between $1819\text{--}2013$~km~s$^{-1}$, a ``fishhook-shaped'' or J-shaped structure appears, curving away from the 2.5-kpc linear feature. These parallel structures exhibit subtle velocity gradients, potentially indicating a blueshift as the gas recedes from the jet axis. Notably, molecular gas has been detected within the J-shaped feature associated with the 2.5-kpc structure; this gas is characterized by elevated velocity dispersions and high line ratios, such as $R_{21}$ and $R_{31}$. 
Furthermore, a diffuse emission feature reappears to the north-northwest at approximately 2206 km s$^{-1}$, followed by the emergence of a clump further to the west in the $2271\text{--}2400$ km~s$^{-1}$ range.

\subsubsection{Distinct WIG Structures Surrounding the Galactic Center and the Inner Southern Shell}

The galactic center exhibits a broad velocity component spanning $V_{\text{LSR}} \sim 1174\text{--}2400$~km~s$^{-1}$, potentially partially affected by H$\alpha$ contamination at lower velocities. Beyond a central point source, an extended ``bow-tie'' structure oriented along the northeast-southwest axis becomes prominent around 1690~km~s$^{-1}$ (Figure~\ref{fig:bunnies_and_bowtie} right). The northeastern lobe of this structure is associated with molecular gas characterized by high velocity dispersion and elevated $R_{21}$ and $R_{31}$ line ratios. A linear feature extends further east-northeast from the edge of this bow-tie between $1690\text{--}2013$ km s$^{-1}$, exhibiting an increasing redshift with increasing galactocentric distance (``Cnt-LSs'' in Figure~\ref{fig:comparison_niico}). On the opposite side, the southwestern lobe appears more diffuse near the systemic velocity ($1690\text{--}1948$~km~s$^{-1}$) and gradually extends further southwest at higher velocities, with redshifted molecular gas detected at its tip.

In the region between the center and the southern shell, a complex network of filaments is observed. Between $1432\text{--}1690$~km~s$^{-1}$, a filament extends toward the south-southwest, with a secondary branch diverging toward the east in the $1432\text{--}1625$~km~s$^{-1}$ range. This branch terminates at the ``SE Blob'', a region rich in molecular gas and dust emission. The corresponding warm ionized gas in this blob is most distinct at approximately 1367~km~s$^{-1}$ and displays a clear velocity gradient, with velocities increasing from the northeast toward the southwest. Further south, between $1303\text{--}1432$~km~s$^{-1}$, two faint filaments emerge in a ``$\Lambda$-shape'' (``$\Lambda$-Str'' in Figure~\ref{fig:comparison_niico}). To the south of these filaments, components associated with the southern shell appear at $V_{\text{LSR}} \sim 1432$~km~s$^{-1}$ and rotate progressively westward with increasing velocity. By 2400~km~s$^{-1}$, this structure reaches the eastern side, completing the ring-like morphology consistent with the 5-kpc shell seen in the north.

\subsubsection{The Tortuous Filamentary Structure of Molecular Phase between the SE Blob and the Galactic Center}

Although not prominent in WIG, a tortuous, filamentary structure composed of molecular gas and dust appears to bridge the SE Blob and the galactic center. 
This feature resembles a ``sideways Y-shape'' tilted to the right, characterized by a primary filament that bifurcates toward the east. The kinematics of this structure, as revealed by ALMA CO data (Figure~\ref{fig:mom_co10}), show that the base of the Y-shape emerges from the east at $V_{\text{LSR}} \sim 1405$~km~s$^{-1}$. As the velocity increases, the emission follows a westward path, reaching the junction of the Y-shape at $\sim 1570$~km~s$^{-1}$. From this intersection, the structure splits; the northern branch reaches the galactic center by $\sim 1700$~km~s$^{-1}$. Near the tip of the southern branch, separate emission components appear at slightly offset spatial positions between $1725\text{--}1790$~km~s$^{-1}$.

The dust extinction map reveals a ``bunny-face''-like morphology along the main axis of this Y-shaped filament, suggesting the influence of external pressure originating from the galactic center or the southern edge of the nuclear jet. Notably, the outer (southwestern) terminus of the bifurcated Y-shape exhibits broad CO line widths and relatively high $R_{21}$ ratios (see Figures~\ref{fig:mom_co10}, \ref{fig:coratios} and \ref{fig:r21_v}). The CO spectrum of this broad component exhibits three distinct velocity peaks at approximately 1615, 1690, and 1750 km s$^{-1}$. While the spatial and kinematic proximity of these features suggests they may form a single, continuous structure, current limitations in spatial resolution, spectral resolution, and sensitivity preclude a definitive conclusion. Further high-resolution observations are required to confirm the physical connectivity of this tortuous complex.


\subsection{Possible Origins of the characteristic structures}

\subsubsection{The 5-kpc Shells and ``Bunny'' structures}
The 5-kpc shell may be interpreted as the limb-brightened edge of a bubble-driven outflow \citep{Maccagni:2021dz}. The ``bunny-face'' structures are indicative of hydrodynamic instabilities, specifically Rayleigh–Taylor (RT) instabilities at the bubble interface or shock compression combined with gas stripping driven by ram-pressure and Kelvin-Helmholtz (KH) instabilities. The latter effect is often seen in hydrodynamic simulations of young (up to a few Myr) jets interacting with a clumpy ISM \citep{Wagner:2012lz, Mukherjee:2016wi}. However, if the observed ring structure originated from the limb-brightening of a purely spherically symmetric expanding shell, it would not exhibit the systematic velocity gradient currently observed. Consequently, the distinct east-west velocity gradient suggests two primary geometric scenarios: first, a non-spherical expansion (e.g., a prolate, rugby-ball-shaped shell) with its major axis inclined relative to the line of sight; or second, a spherically symmetric expansion involving a high-density ring-like structure that is itself inclined toward the observer. 
While such a well-defined WIG shell encompassing a radio jet has not yet been reported in other radio galaxies or early-type group-dominant galaxies \citep[e.g.,][]{Lagos:2022wv}, similar spherical shell structures are frequently observed surrounding SNRs \citep[e.g.,][]{Milisavljevic:2024im}.
It is worth noting that a kinematic misalignment between stellar components and cold/warm gas is a well-known feature in galaxies that have recently acquired external gas and are still undergoing a settling process within the host potential \citep[e.g.,][]{van-de-Voort:2015uk,Davis:2019ez,Ruffa:2019hk,Baker:2025ii}. However, the highly structured morphology (e.g., the 5-kpc shell and the instabilities) and the specific systematic velocity gradient observed here cannot be fully explained by a simple settling process alone, reinforcing the necessity of dynamic jet-ISM interactions.
Future work should involve a detailed comparison with numerical simulations to elucidate the underlying physics and a systematic search for such structures in other radio-loud systems.

\subsubsection{Jet-ISM Interaction and Lateral Outflows}

The multi-phase gas dynamics in this region suggest an interaction between jet and gas, favouring the jet deflection hypothesis as an explanation for the overall morphology of the kpc-scale jet. This scenario is supported by several multi-phase observational clues, such as the spatial coincidence of the molecular/multi-phase gas clump at the deflection point, as well as the enhanced velocity dispersion and elevated $R_{21}$ and $R_{31}$ line ratios located directly along the pre-deflection jet axis. Linear ISM structures extending perpendicular to the jet axis are features frequently seen in numerical simulations of jet-ISM interactions \citep{Mukherjee:2018my, Mukherjee:2018pd}. These structures are thought to originate when the jet head impacts a dense molecular cloud; the impact site is heated to extreme temperatures and undergoes isotropic expansion. However, because the jet's forward path is still obstructed by the remaining molecular gas, the expansion is funneled into a potential signature of an outflow directed perpendicular to the main jet stream's direction of propagation \citep[``lateral outflow'',][]{Mukherjee:2018pd}. This scenario is consistent with our observations of the molecular gas detected at the base of these linear structures (i.e., the side intersecting with the jet), which exhibits high velocity dispersion and elevated $R_{21}$ and $R_{31}$ line ratios. We note, however, that the enhancements in the CO line ratios observed in NGC~1316 are relatively moderate compared to those found in some other active systems, such as IC~5063 \citep{Oosterloo:2017oc}, the Teacup galaxy \citep{Audibert:2023gv}, and NGC~3100 \citep{Ruffa:2022zo}, where the spatial correlation between the radio jets and the higher line ratios is more pronounced.

Furthermore, the kinematics of the ionized gas corroborates this interpretation. WIG associated with the HIG appears blueshifted to the east of the northern jet and redshifted to the west of the southern jet. Notably, the linear WIG structure extending perpendicular to the northern jet also exhibits a blueshift, consistent with the surrounding diffuse component.
This kinematic agreement suggests that these HIG components could potentially be related to gas heated and accelerated by the aforementioned perpendicular outflows. However, given the complexity of the multi-phase gas kinematics, this interpretation remains tentative, and dedicated numerical simulations specific to the geometry of NGC~1316 will be required in future works to rigorously test and validate this scenario.

\subsubsection{The Tortuous Filamentary Structure above the SE Blob}

The origin of the tortuous filamentary structures observed between the SE Blob and the Galactic Center remains overall elusive. While the SE Blob, centered at $\sim$1367 km s$^{-1}$, exhibits a continuous velocity gradient from the northeast (lower $v$) to the southwest (higher $v$), its physical nature is difficult to constrain due to conflicting observational evidence. On one hand, the SE Blob shows strong dust emission despite relatively weak dust extinction, which typically suggests that the gas is located on the far side of the galaxy. In this configuration, the blueshifted motion would represent an inflow toward the nucleus. On the other hand, it is possible that the jet interaction has locally enhanced the dust emissivity, meaning that the weak extinction does not necessarily imply a far-side location. If this is the case, the filamentary structure might instead trace an outflow along magnetic field lines surrounding the jet. Given these complexities, a detailed comparison with magnetohydrodynamical (MHD) simulations is required to determine whether these features represent infalling gas or jet-driven outflows.

\subsection{Molecular Clouds Destruction}



Our ALMA observations of NGC~1316 reveal a high fraction of extended molecular gas components near the galactic center, coupled with a notable absence of GMAs.
This paucity of massive, discrete cloud structures might indicate that the molecular gas in the central region is undergoing significant disruption due to jet-ISM interactions.
Furthermore, the shell-like morphology observed in the WIG bears a striking resemblance to the structures found around SNRs, implying a commonality in the underlying feedback mechanisms.

The state of the ISM in radio galaxies involving an expanding, energetic bubble shares fundamental similarities with environments surrounding SNRs and stellar winds, where the physical processes of molecular cloud destruction have been extensively studied \citep[e.g.,][]{Cowie:1977lh,Cowie:1977oe,McKee:1977yu,McKee:1977qr,Nakamura:2006dj}. Bearing the potential effects of the past galaxy merger in mind, in this section, we first evaluate whether the evolution of the hot bubble in NGC~1316 aligns with theoretical models established for SNRs and massive stellar winds. Subsequently, we assess the feasibility of molecular cloud destruction through a simplified energetics analysis and discuss the governing physical processes by comparing several characteristic timescales.

\subsubsection{Bubble Evolution}

\begin{figure*}[]
\begin{center}
\includegraphics[width=0.49\textwidth, bb=0 0 576 576]{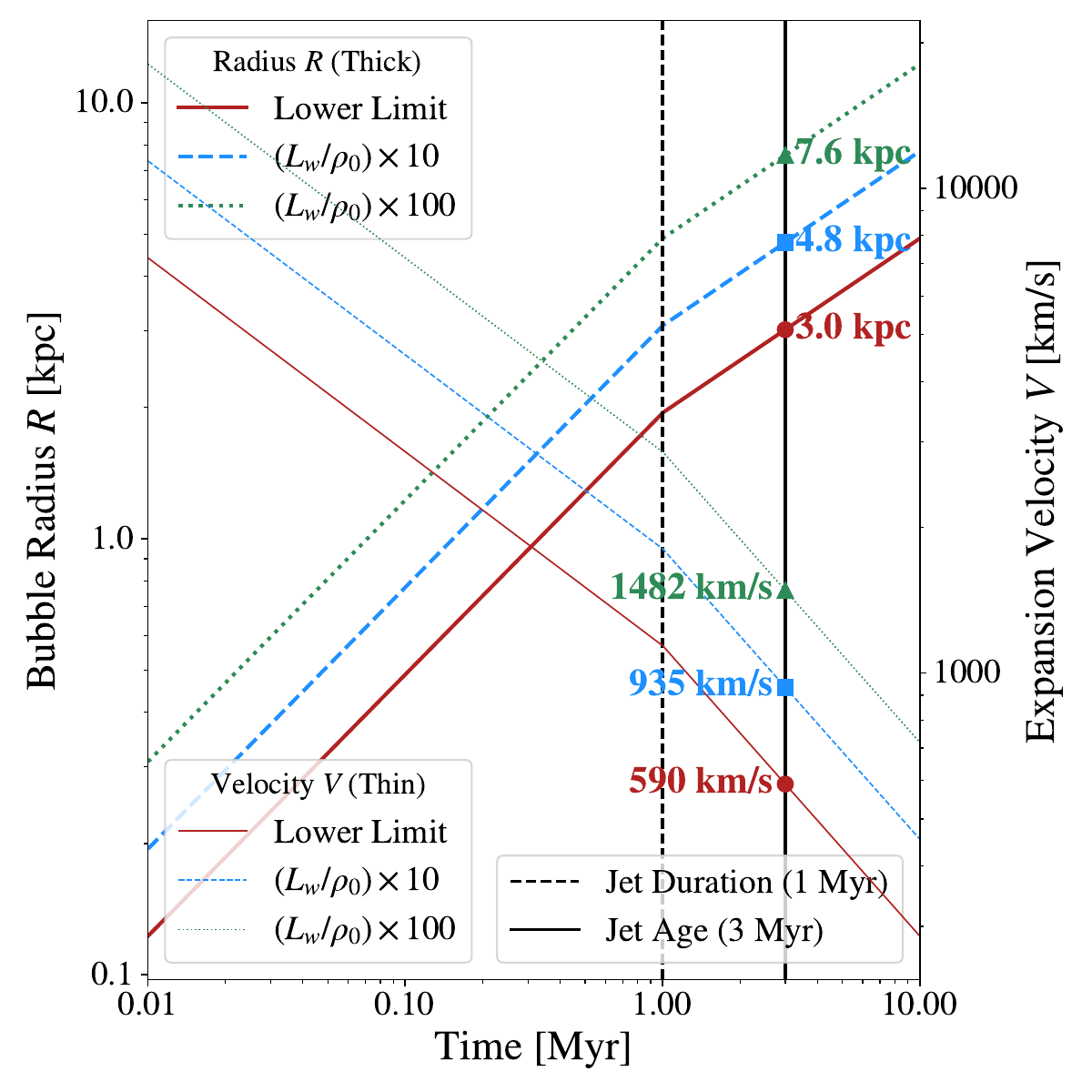}
\includegraphics[width=0.49\textwidth, bb=0 0 556 552]{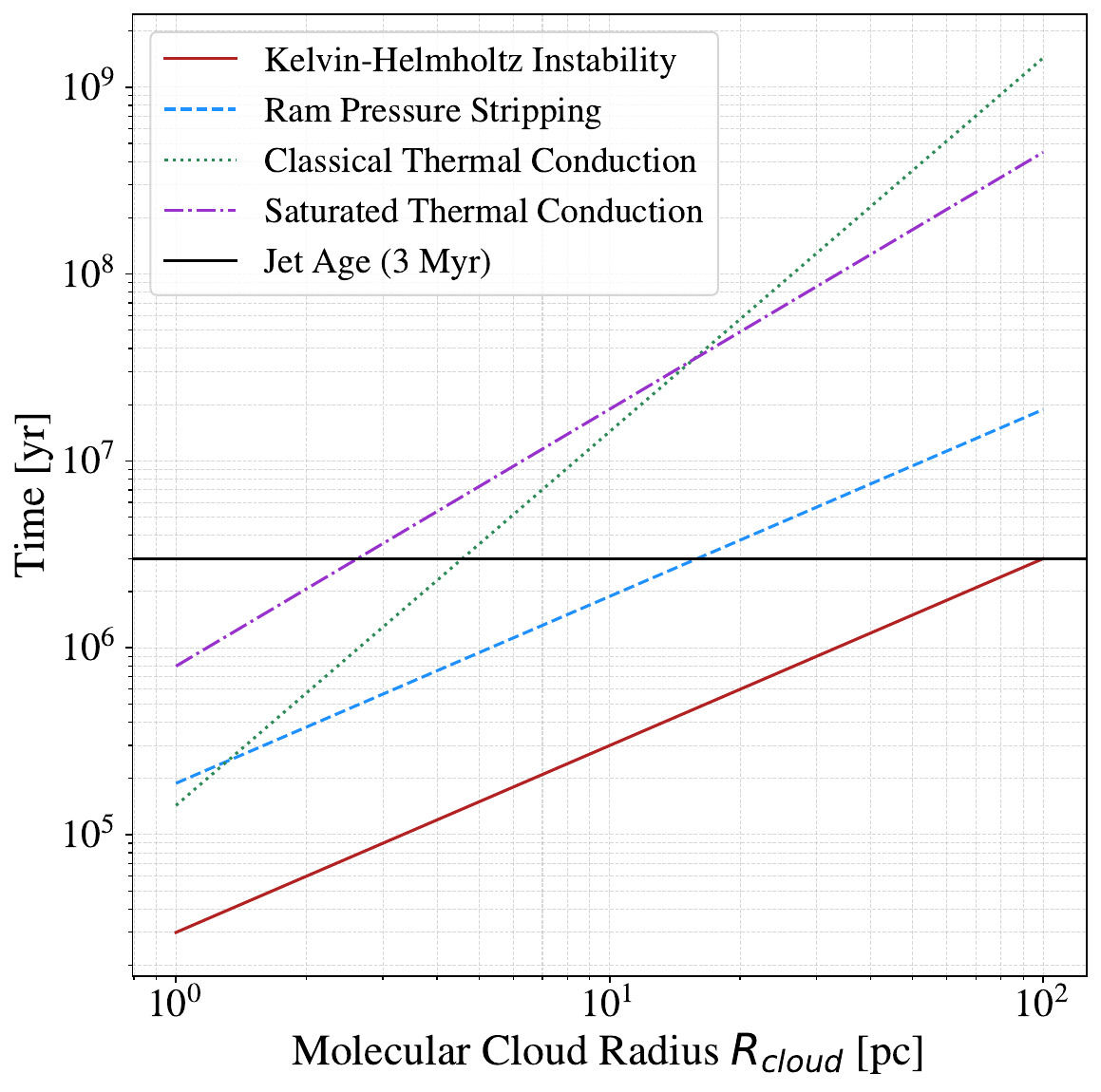}
\end{center}
\caption{
[Left] Evolution of bubble radius $R(t)$ (left axis) and expansion velocity $V(t)$ (right axis) based on Equation~\ref{eq:weaver77}. The red solid, blue dashed, and green dotted lines represent the lower limit ($L_w = 1.6 \times 10^{43}$ erg s$^{-1}$, $n_{\rm e} = 0.01$ cm$^{-3}$) and cases where $L_w/\rho_0$ is increased by factors of 10 and 100, respectively. Symbols indicate the expected values at a jet age of 3 Myr, as denoted by the vertical line.
[Right] Destruction timescales of molecular clouds as a function of cloud radius. The red solid and blue dashed lines represent the timescales for KH instability ($t_{\rm KHI}$) and cloud ablation ($t_{\rm abl}$), respectively. The green dotted and purple dot-dashed lines correspond to classical and saturated thermal conduction timescales. The horizontal black line indicates the jet age of 3~Myr. The intersection of this reference line with each timescale defines the maximum cloud size that can be fully disrupted within the jet's lifetime.
}
\label{fig:models}
\end{figure*}

Observations with MeerKAT at 1.4~GHz indicate that the nuclear jet of NGC~1316 emits a power of $1.1\times10^{23}$~W~Hz$^{-1}$ \citep{Maccagni:2020pq}.
Given the observation frequency of 1.4~GHz, this corresponds to radio luminosity of $P_{\rm radio}=\nu L_\nu=1.5\times10^{39}$~erg~s$^{-1}$. We can convert this radio luminosity to jet kinetic power ($P_{\rm jet}$) using the following empirical relation from \cite{Cavagnolo:2010rl}:
\begin{equation}
P_{\rm jet} \sim 5.8\times10^{43} \left(\frac{P_{\rm radio}}{[10^{40}~{\rm erg~s^{-1}}]} \right)^{0.70},
\end{equation}
which yields an estimated jet power of $P_{\rm jet} \sim 1.6\times10^{43}$ erg s$^{-1}$ for NGC~1316.

Regarding the evolution of the bubble associated with the jet, we assume an active jet duration of 1 Myr and a total age of 3 Myr, based on the observational constraints for NGC~1316 \citep{Maccagni:2020pq}. It is important to consider that the empirical scaling relations used to estimate $P_{\rm jet}$ may not fully account for jets confined within the ISM. As suggested by \cite{Mukherjee:2018pd}, the derived jet power could be underestimated by a factor of a few. This potential underestimation of the energy input provides a primary motivation for exploring models with higher $P_{\rm jet}$ in the subsequent analysis.

To assess whether this jet could have created the observed X-ray bubble, we model its expansion in two distinct phases: an initial active injection phase followed by a passive energy-conserving phase.
For the first $1$~Myr, we assume the jet provides a constant power, creating a bubble whose radius $R(t)$ evolves according to the model for stellar wind bubbles
\begin{equation}
R(t) = \left(\frac{250}{308\pi}\right)^{1/5} \left(\frac{P_{\rm jet}}{\rho_{\rm 0}}\right)^{1/5} t^{3/5}
\label{eq:weaver77}
\end{equation}
where $\rho_0$ is the mass density of the ambient gas. For our calculation representing the lower end of the expected bubble radius, we adopt $\rho_0 = m_p n_e$ with $n_e = 0.01$ cm$^{-3}$, noting that this density is derived from X-ray observations of the innermost galaxy core. Since the density is expected to decrease significantly as one moves away from the center, this value likely represents an upper limit for the bulk of the region through which the bubble expands. Furthermore, as noted above, $P_{\rm jet}$ itself might be underestimated for jets confined within the ISM. These factors suggest that the ratio $P_{\rm jet}/\rho_0$ could be much higher across most of the volume, so we also consider cases for which this ratio is increased by factors of 10 and 100.

After the jet activity ceases at $t = 1$~Myr, we assume the bubble continues to expand adiabatically without further energy injection, following the Sedov-Taylor solution. For this phase ($1 < t \le 3$ Myr), we evolve the radius and expansion velocity from their values at $t = 1$~Myr using the scaling relations $R \propto t^{2/5}$ and $V \propto t^{-3/5}$.


The resulting evolution of size, $R$ and expansion velocity, $V$ is shown in Figure~\ref{fig:models} (left). At the current estimated jet age of $t = 3$ Myr, the baseline model yields $R \sim 3.0$ kpc and $V \sim 590$ km s$^{-1}$. However, in the cases in which the $P_{\rm jet}/\rho_0$ ratio is increased by factors of 10 and 100 to account for lower $\rho_0$ and higher $P_{\rm jet}$, the predicted radius reaches $R \sim 4.8$ kpc and $7.6$ kpc, with corresponding velocities of $V \sim 935$ km s$^{-1}$ and $1482$ km s$^{-1}$, respectively. Given that the empirical scaling relation of \cite{Cavagnolo:2010rl} exhibits a well-known scatter spanning a few orders of magnitude, the baseline $P_{\rm jet}$ carries inherent systematic uncertainties. Nevertheless, our multi-case analysis explicitly encompasses this substantial empirical scatter. The fact that the observed shell radius ($\sim 5$~kpc) is well reproduced within these modified $P_{\rm jet}/\rho_0$ models strongly reinforces that the notion that the observed bubble was inflated by recent jet activity is physically plausible.

Although the expansion velocity decelerates over time, we adopt a representative speed of $V \sim 1000$ km s$^{-1}$ for our subsequent analysis. This value is consistent with the velocity range derived from our modified $P_{\rm jet}/\rho_0$ models, even though a constant expansion to $5$ kpc over $3$ Myr would require a higher average velocity of $V_{\text{const}} \sim 1630$ km s$^{-1}$. Given that the sound speed ($c_s$) in fully ionized hydrogen gas at $T = 6 \times 10^6$ K is approximately $287$ km s$^{-1}$, the bubble expansion at this representative speed is expected to drive a shock wave with a Mach number $M = V/c_s \sim 3.5$.

\subsubsection{Energetics}


\cite{Maccagni:2020pq} has constrained the AGN jet activity that formed the bubble to have started approximately $3 \text{ Myr}$ ago and lasted for less than $1 \text{ Myr}$. Assuming the current nuclear-jet power ($P_{\text{jet}}$) was sustained for $t_{\text{jet}} = 1 \text{ Myr}$, the total injected energy is estimated as $E_{\text{jet,total}} = P_{\text{jet}} \times t_{\text{jet}} \sim 5 \times 10^{56} \text{ erg}$.

To evaluate how much of this injected energy is lost through radiative cooling, we estimate the cooling time ($t_{\rm cool}$) of the X-ray emitting gas. Using the cooling function $\Lambda$ from \cite{Sutherland:1993tk} for solar metallicity ($\log \Lambda \sim -22.4$ at $T \sim 6 \times 10^6$ K), the cooling time for a bubble with a radius of $5$ kpc and an ambient density of $n_{\rm H} = 0.01$ cm$^{-3}$ is approximately $208$ Myr. This is nearly two orders of magnitude longer than the jet age of $3$ Myr. If the bubble interior is more dilute, such as $n_{\rm H} = 0.001$ cm$^{-3}$, $t_{\rm cool}$ extends to $\sim 2.08$ Gyr. These results indicate that radiative energy loss is negligible, and the energy injected by the jet remains largely stored within the hot bubble.

Next, we evaluate whether this energy is sufficient to (self-)gravitationally unbind the molecular gas reservoir. We adopt several assumptions: (1) The entire currently observed molecular gas mass ($M_{\text{mol}} \sim 6 \times 10^8 M_{\odot}$) existed as molecular clouds prior to interacting with the bubble. (2) The molecular clouds followed a truncated power-law mass function typical of star-forming galaxies, with a minimum cloud mass of $M_{\text{min}} = 10^4 M_{\odot}$. (3) The energy required is purely that needed to overcome the gravitational binding of the clouds (molecular dissociation and ionization energies are not considered). (4) Each cloud is approximated as a sphere with a constant mean density of $n_{\text{cloud}} = 100 \text{ cm}^{-3}$. 

The gravitational binding energy ($E_{\text{bind}}$) of a single uniform-density spherical cloud is $E_{\text{bind}} = \frac{3}{5} \frac{G M_{\text{cloud}}^2}{r_{\text{cloud}}}$, which, for a fixed density, scales only with the cloud's radius ($r_{\text{cloud}}$) or mass ($M_{\text{cloud}} \propto r_{\text{cloud}}^3$). The molecular cloud mass functions (MCMFs) have been well-studied in both the Milky Way and nearby spiral galaxies \citep[e.g.,][]{Heyer:2001km,Fukui:2001et,Rosolowsky:2003mw,Rosolowsky:2007ta,Colombo:2014sw,Rice:2016oq,Tosaki:2017cr,Hirota:2018hs,Hirota:2024ij}.
For this study, we specifically employ the MCMF of M~83 \citep{Hirota:2024ij}. The cumulative MCMF, $N(>M_{\rm cloud})$, is modeled as follows:
\begin{equation}
N(>M_{\rm cloud}) = A \left[ \left(\frac{M_{\rm cloud}}{M_{\text{lim}}}\right)^{\gamma_{\text{PL}}+1} - 1 \right],
\end{equation}
where $N(>M_{\rm cloud})$ is the cumulative number of clouds with mass greater than $M_{\rm cloud}$. We adopt the parameters $\gamma_{\text{PL}}=-1.85$ and $M_{\text{lim}}=8.58 \times 10^6 M_{\odot}$ from M~83 observations \citep{Hirota:2024ij}, and the constant $A$ is determined by normalizing the function to the total molecular gas mass $M_{\text{mol}}$ of NGC~1316. 
Using these constraints, we calculated the total energy required to unbind all molecular clouds across various mass ranges (e.g., $10^4-10^5 M_{\odot}$, $10^4-10^6 M_{\odot}$, $10^4-10^7 M_{\odot}$ as summarized in Table~\ref{tab:results_05dex}). The total energy required to unbind nearly all molecular clouds is found to be $E_{\text{dest}} \sim 1 \times 10^{54} \text{ erg}$.



We evaluate the feasibility of cloud destruction by estimating the fraction of jet energy intercepted by the molecular clouds and the resulting efficiency required for their disruption. We assume a total jet energy of $E_{\text{jet,total}} \sim 5 \times 10^{56}$~erg, representing the available energy budget over the active phase. The total molecular gas mass is taken as $M_{\text{mol}} = 5.5 \times 10^8 M_{\odot}$. For simplicity, we model this mass as a population of $N_{\text{total}} = 5500$ discrete, uniform clouds, each with a typical radius of $r_{\text{cl}} = 20$ pc and a mass of $M_{\text{cl}} = 10^5 M_{\odot}$. Under this assumption, the gravitational binding energy required to disrupt a single cloud is $E_{\text{bind}} \sim 1.3 \times 10^{50}$~erg, leading to a total required energy of $E_{\text{dest, total}} \sim 7.1 \times 10^{53}$~erg for the entire population. This value is in agreement with $E_{\text{dest}} \sim 1 \times 10^{54}$ erg derived in the previous section, where we assumed a power-law size distribution of molecular clouds.

To calculate the total covering factor, $f_{\text{cover}}$, which represents the fraction of the sky covered by clouds as seen from the nucleus, we consider the following two spatial distributions:

\noindent
\textbf{Case 1: Uniform Spherical Distribution.}
Assuming the clouds are distributed uniformly within a sphere of $R_{\text{max}} = 5$ kpc, the covering factor is calculated as $f_{\text{cover}} = \frac{3}{4} N_{\text{total}} (r_{\text{cloud}} / R_{\text{max}})^2 \sim 0.066$. It is worth noting that in this uniform integration, the $R^2$ term in the denominator (from the solid angle of each cloud) and the $R^2$ term in the numerator (from the volume element's surface area) cancel each other out. In this scenario, the clouds intercept approximately $E_{\text{rec}} = f_{\text{cover}} \times E_{\text{jet, total}} \sim 3.3 \times 10^{55}$ erg. The destruction efficiency, defined as the ratio of the required disruption energy to the intercepted energy ($\epsilon = E_{\text{dest, total}} / E_{\text{rec}}$), is approximately $2.2\%$.

\noindent
\textbf{Case 2: Multi-shell Discrete Distribution.} 
Alternatively, if the clouds are distributed equally across five shells at radii of 1, 2, 3, 4, and 5 kpc (1100 clouds per shell), the total covering factor is $f_{\text{cover}} = \sum_{i=1}^{5} \frac{N_i}{4} (\frac{r_{\text{cl}}}{R_i})^2 \sim 0.155$. This distribution results in a higher intercepted energy of $E_{\text{rec}} \sim 7.7 \times 10^{55}$ erg due to the larger solid angles subtended by clouds at smaller radii. Consequently, the required destruction efficiency decreases to approximately $0.9\%$.

It should be noted that we have assumed a single cloud size and mass for these spatial distribution models. In reality, accounting for a realistic size distribution would also require considering the specific spatial arrangement of clouds of different sizes, which significantly increases the complexity of the calculation. To fully capture the interplay between the cloud size distribution and their spatial locations, more detailed numerical simulations would be necessary.

Furthermore, clouds located at smaller radii are expected to be more susceptible to disruption, as they are exposed to the jet-driven bubble for a longer duration and intercept a larger energy. This physical expectation is consistent with our observations of NGC~1316, where the innermost regions show a larger fraction of extended molecular gas and a deficit of GMAs (Figures~\ref{fig:dfiffuse_totalspec} and ~\ref{fig:pycprops}). This suggests that the intense energy injection from the nuclear jet has already driven the disruption of molecular clouds in the central parts of the galaxy. Nevertheless, in both cases considered here, the required efficiency $\epsilon$ remains well below 10\%, suggesting that the jet provides sufficient energy to drive the observed cloud destruction across the galaxy.




In support of this energy budget, simulations of jet-driven feedback onto dense gas indicate that a significant fraction, often more than 10\%, of the jet energy can be converted into the kinetic energy of the warm and cold phases of the ISM \citep{Wagner:2012lz}. This implies that our required efficiency, $\epsilon < 10\%$, is well within physically plausible limits, although such kinetic energy does not necessarily lead to the immediate or full destruction of all clouds. While the energetics provide a necessary criterion for substantial feedback, the actual destruction of molecular clouds proceeds over an extended period of hydrodynamic and thermal ablation, whose efficiencies remain poorly constrained.

\begin{table*}[]
    \centering
    \caption{Total energy required to disrupt molecular clouds}
    \label{tab:results_05dex}
    \begin{tabular}{llll}
        \hline
        Mass Upper Limit ($M_{\odot}$) & $N_{\rm clouds,in-bin}$ & $M_{\rm total}$ ($M_{\odot}$) & $E_{\text{bin,total}}$ (erg) \\
        \hline
        $10^5$ & $4.84\times10^3$ & $1.32\times10^8$ & $2.26\times10^{52}$ \\
        $10^6$ & $5.52\times10^3$ & $3.18\times10^8$ & $1.71\times10^{53}$ \\
        $10^7$ & $5.61\times10^3$ & $5.60\times10^8$ & $1.01\times10^{54}$ \\
        \hline
    \end{tabular}
\end{table*}

Finally, we estimate the energy required to increase the velocity dispersion of the surviving large molecular clouds. We continue to adopt the same MCMF as discussed above, focusing on the surviving large clouds.
It should be noted that our 100~pc resolution data primarily trace GMAs rather than individual clouds. Since the actual sizes of the constituent clouds remain unresolved, we focus on the surviving large clouds with a radius larger than 50~pc, which corresponds to half of the beam size of our observations.
We assume that prior to the jet-ISM interaction, these clouds followed the size-linewidth relationship of $r_{\rm cloud}~\text{[pc]} = (\sigma~\text{[km s$^{-1}$]})^{0.5}$ \citep{Solomon:1987gx}. Given that the velocity dispersion of GMAs detected in NGC~1316 is roughly 2--3 times that of M83's GMAs, we calculate the energy required to double the velocity dispersion of these surviving clouds. 



The energy needed to double the dispersion for clouds larger than 50~pc is $5.1 \times 10^{58}$~erg by adopting a sufficiently fine mass-binning width of 0.01~dex to closely approximate the continuous integration of the mass function. This value is two orders of magnitude larger than the estimated energy of the expanding hot bubble ($\sim 5 \times 10^{56}$~erg). These findings are broadly consistent with numerical simulations of jet-ISM interactions. For instance, \cite{Mukherjee:2016wi} demonstrated that powerful jets with $P_{\text{jet}} \gtrsim 10^{44\text{--}45} \text{ erg s}^{-1}$ are capable of enhancing the velocity dispersion by more than a factor of two, although these simulations were designed with high cloud filling factors that enhance the energy transfer. Even with such high jet powers, the impact on velocity dispersion remains limited if the interaction geometry is suboptimal, such as when the jet is inclined away from the galactic disk \citep{Mukherjee:2018my}.

Furthermore, the elevated velocity dispersion observed at the GMA scale could potentially suggest high virial parameters within the constituent molecular gas. If these high dispersions reflect the internal kinematics of the gas, the resulting virial ratio ($\alpha_{\rm vir} \sim 5 \sigma^2 r_{\rm cloud} / GM$) might exceed unity. Such a state could potentially contribute to the prevention of gravitational collapse and the suppression of star formation, although higher-resolution observations are needed to confirm the virial state of individual clouds.

This analysis, alongside the situational constraints of jet-cloud coupling, may indicate the difficulty in doubling the velocity dispersion of large molecular clouds through the jet alone. The complex dynamical history and unique structure of NGC 1316 likely play significant roles. The stellar body of NGC 1316 exhibits numerous tidal tails and loops, evidence of one or more merger events occurring between 1 and 3 Gyr ago \citep[e.g.,][]{Schweizer:1980gl, Goudfrooij:2001ya,Serra:2019ph}. The molecular gas in such systems is often interpreted as having an external origin, having been transported into the galaxy during these past interactions \citep[e.g.,][]{Xilouris:2004bk,Lanz:2010ll}. It is possible that this accreted gas remains in a dynamically unsettled or naturally turbulent state, which could contribute to the enhanced velocity dispersions observed in our data.

Additionally, the distinct gravitational potential of the elliptical host galaxy may further influence the gas kinematics. Unlike the exponential disk of M~83, NGC~1316 follows a de Vaucouleurs ($r^{1/4}$) mass distribution, which can impose higher galactic shear on gas structures of comparable size. Consequently, the high velocity dispersions in NGC~1316 may result from a combination of jet-induced turbulence, the inherent dynamical instability of accreted gas from past mergers, and galactic shear. To precisely isolate the impact of the AGN jet, a comparison with GMAs in quiescent elliptical galaxies (specifically those with similar merger histories but lacking active jets) is ideally required to disentangle these environmental effects from jet-induced turbulence.

\subsubsection{Molecular Clouds Destruction Processes}


Previous studies on the disruption of molecular gas surrounding SNRs and stellar winds have primarily identified following key mechanisms: thermal conduction and the dynamical effects associated with high-velocity expanding environments. 
In this study, we argue that the destruction of molecular clouds in jet-ISM interaction regions is primarily driven by Kelvin-Helmholtz (KH) instability, while ram pressure plays a secondary but complementary role.

We consider a molecular cloud of density $\rho_{\rm c}$ embedded in a hot diffuse medium of density $\rho_{\rm h}$, subject to a relative velocity $\Delta v$ induced by an AGN jet or a jet-driven expanding bubble. The density contrast is defined as
\begin{equation}
    \chi \equiv \frac{\rho_{\rm c}}{\rho_{\rm h}}.
\end{equation}
The growth rate of KH instability at a shear interface is given by
\begin{equation}
    \omega = k \frac{\sqrt{\rho_{\rm c}\rho_{\rm h}}}{\rho_{\rm c}+\rho_{\rm h}} \Delta v,
\end{equation}
where $k$ is the wavenumber of the perturbation. Adopting pertubations on the scale of cloud radius $r_{\rm cloud}$, $k \simeq 2\pi/r_{\rm cloud}$, and assuming $\rho_{\rm c}\gg \rho_{\rm h}$, the characteristic KH growth time becomes
\begin{equation}
    t_{\rm KHI} \simeq \frac{r_{\rm cloud}}{2 \pi \Delta v} \sqrt{\chi}.
\end{equation}
For the fiducial parameters adopted in this work ($\chi\sim10^4$ and $\Delta v\sim10^3$~km s$^{-1}$), molecular clouds with radii up to several tens of parsecs are expected to be disrupted within $\sim3$~Myr, comparable to the estimated jet age of NGC~1316. This strongly suggests that KH instability is efficient enough to fragment molecular clouds during the active phase of the jet.

The ram pressure exerted by the surrounding hot medium is
\begin{equation}
    P_{\rm ram}=\rho_{\rm h} \Delta v^2.
\end{equation}
A characteristic timescale for cloud ablation due to ram pressure stripping can be written as
\begin{equation}
    t_{\rm abl} \sim \frac{r_{\rm cloud}}{\Delta v}\sqrt{ \frac{\rho_{\rm c}}{\rho_{\rm h}}} = \frac{r_{\rm cloud}}{\Delta v} \sqrt{\chi}.
\end{equation}
For the same fiducial parameters, $t_{\rm abl}$ typically exceeds $10^8$~yr for $r_{\rm cloud}\sim10$~pc, which is much longer than the jet lifetime. Ram pressure alone is therefore insufficient to ablate massive molecular clouds on Myr timescales.

However, for clouds embedded in fast flows, the mixing layers created by the KH instability are carried away by ram pressure, forming cometary tails, and, presumably the bunny structures, if they remained cone-shaped. Furthermore, once the KH instability fragments a clouds into smaller substructures, the effective cloud radius $r_{\rm cloud}$ is reduced, and ram-pressure stripping and turbulent mixing become increasingly efficient. In this sense, ram pressure acts in concert with the KH instability, accelerating the conversion of bound molecular clouds into a diffuse molecular phase.

Another potential cloud destruction mechanism is evaporation driven by thermal conduction. Following the formulation by \cite{Cowie:1977lh}, we calculated the mass evaporation rates ($\dot{m}$) and the corresponding evaporation timescales ($t_{\text{evap}}$) for spherical clouds in both the classical and saturated thermal conduction regimes.
For classical thermal conduction (Eq. 22 in their paper), the mass evaporation rate is expressed as:
\begin{equation}
\dot{m}_{\text{clas}} \sim 2.75 \times 10^4 \, T_{\text{hot}}^{5/2} \, r_{\text{cloud,pc}} \left( \frac{30}{\ln\Lambda} \right) \quad [\text{g s}^{-1}],
\end{equation}
where $r_{\text{cloud,pc}}$ is the cloud radius in parsecs and $\ln \Lambda$ is the Coulomb logarithm. Since the evaporation rate scales linearly with $r$ while the cloud mass $m_c$ scales as $r^3$, the resulting evaporation timescale ($t_{\text{evap}} = m_c / \dot{m}$) depends on $r_{\text{cloud,pc}}^2$ as follows:
\begin{equation}
t_{\text{evap, clas}} \sim 3.3 \times 10^{20} \, n_{\text{cloud}} \, r_{\text{cloud,pc}}^2 \, T_{\text{hot}}^{-5/2} \, \left( \frac{\ln\Lambda}{30} \right) \quad [\text{yr}].
\end{equation}
In the case of saturated thermal conduction (Eq. 64), the evaporation rate is determined by:
\begin{equation}
\dot{m}_{\text{sat}} = 3.25 \times 10^{18} \, n_{\text{hot}} \, T_{\text{hot}}^{1/2} \, r_{\text{cloud,pc}}^2 \, \phi_s \, F(\sigma_0) \quad [\text{g s}^{-1}].
\end{equation}
Here, $n_{\text{hot}}$ and $T_{\text{hot}}$ represent the density and temperature of the surrounding hot gas, and $\phi_s$ is a dimensionless constant of order unity. The saturation parameter $\sigma_0$, assuming $\ln\Lambda=30$, is defined as:
\begin{equation}
\sigma_0 = \left[ \frac{T_{\text{hot}}}{1.54 \times 10^7 \text{ K}} \right]^2 \frac{1}{n_{\text{hot}} r_{\text{cloud,pc}} \phi_s}.
\end{equation}
For $\phi_s=1$, the function $F(\sigma_0)$ is given by $2.73 \, \sigma_0^{3/8}$, and the evaporation timescale in the saturated regime is calculated as:
\begin{equation}
t_{\text{evap, sat}} \sim 2.8 \times 10^6 \left( \frac{n_{\text{cloud}}}{n_{\text{hot}}} \right) \frac{r_{\text{cloud,pc}}}{T_{\text{hot}}^{1/2} \, \phi_s \, F(\sigma_0)} \quad [\text{yr}].
\end{equation}

Figure~\ref{fig:models} (right) illustrates the molecular cloud radius dependence of four different timescales. Crucially, the figure demonstrates that all considered timescales decrease with decreasing cloud radius, indicating that smaller molecular clouds are more susceptible to destruction. The timescales required to destroy a molecular cloud with a radius of 10~pc are $t_{\rm KH,10pc}=3.0 \times 10^{5}~\text{yr}$, $t_{\rm abl,10pc}=1.9 \times 10^6~\text{yr}$, $t_{\rm evap,10pc}=1.4 \times 10^7~\text{yr}$, and $t_{\rm evap,sat,10pc}=1.9 \times 10^7~\text{yr}$, respectively. Furthermore, the maximum radii of molecular clouds that can be destroyed within the 3~Myr age of NGC~1316's jet are approximately $r_{\rm KH,3Myr}=100~\text{pc}$, $r_{\rm abl,3Myr}=16~\text{pc}$, $r_{\rm evap,3Myr}=4.6~\text{pc}$, and $r_{\rm evap,clas,3Myr}=2.6~\text{pc}$, respectively, for each of these timescales. This suggests that KHI plays a crucial role in molecular cloud destruction within the expanding hot gas. However, in reality, these effects occur simultaneously, and thus 100~pc might represent an upper limit. 
Simulations, for example, suggest that thermal conduction inhibits the KH instability and tends to protect cloud cores from ablation, although it enhances ablation from cometary tails \citep[see][for a recent review]{Gronke:2026ry}.


\subsubsection{Caveats}

The scenario presented above is a highly simplified model and the actual physical interactions are considerably more complex. The primary simplifications involve the assumption of uniform density for both the ambient gas and the molecular clouds. In reality, both the ambient gas and the clouds exhibit significant density structures (e.g., density gradients in the ambient medium and clumpy substructure within the molecular clouds). The density and temperature contrast adopted in our simplified energetics calculation represents an extreme case, assuming that the moderate-density transitional layers have been rapidly stripped away, maintaining a persistently sharp contrast between the hot bubble gas and the dense molecular material.


Furthermore, the thermal response of the clouds is a critical factor. Based on the internal energy density and the H$_2$ radiative cooling rates from \cite{Shull:1978kc}, we estimate the post-shock cooling time for clouds with $n_{\text{H}_2} = 100 \text{ cm}^{-3}$ to be as short as $\sim 150$ yr (assuming a Mach 10 shock, $T \approx 312 \text{ K}$). This timescale is several orders of magnitude shorter than the jet activity duration ($\sim 1 \text{ Myr}$), indicating that the shock-injected energy is almost immediately radiated away. Since clouds tend to be longer-lived when they cool efficiently, the time-scales of the disruption processes considered in our simplified mechanical models should be considered as conservative lower limits \citep[e.g.,][]{Cooper:2009lh, Banda-Barragan:2021fu}.

Lastly, a complete physical description must account for the influence of magnetic fields. The presence and geometry of magnetic fields can significantly affect several critical processes; for instance, they can suppress fluid instabilities (e.g., RT or KH instabilities) that would otherwise shred the molecular clouds rapidly. They can also inhibit thermal conduction from the hot bubble interior to the cold molecular clouds. Accurately modeling these detailed processes, including non-uniform density distributions and MHD-related transport phenomena, is beyond the scope of this paper and would necessitate dedicated numerical simulations.

\section{Summary}

We performed ALMA CO($J=1-0$) observations of a nearby radio galaxy NGC~1316 at a 100-pc scale. The observations reveal a complex spatial and kinematic distribution of molecular gas, characterized by broad line widths exceeding 50 km~s$^{-1}$ in several regions. The interferometric CO flux recovery ranges from 34\% to 38\% of the single-dish values. The majority of the gas near the galactic center is not detected with the interferometers. In contrast, flux recovery reaches approximately 50\% in the NW Shell and SE Blob, where 24 GMAs were identified using {\tt PYCPROPS}. While these GMAs are similar in size to those in typical star-forming galaxies, their velocity dispersions are approximately twice as high.

CO line ratio analysis using archival CO($J=2-1$) and CO($J=3-2$) data shows typical star-forming values ($R_{21} \sim 0.7, R_{31} \sim 0.3$) in most regions, but these ratios reach unity near the jet where velocity dispersion is elevated. Multiwavelength comparisons show that molecular gas coexists with hot, warm, and neutral phases. Notably, the warm ionized gas forms a shell structure with a $\sim5$ kpc radius that encompasses the molecular NW Shell.

These findings (high diffuse gas fractions, elevated GMA velocity dispersions, and high CO line ratios) strongly suggest active interaction between the jet and the surrounding ISM. An expanding bubble model based on stellar wind theory, incorporating jet age, power, and hot-gas density, successfully reproduces the observed bubble size. Given that the jet energy exceeds the binding energy of the molecular clouds, we propose a hypothesis where the high extended gas fraction results from small cloud destruction. Our analysis suggests that KH instabilities and ram pressure stripping preferentially disrupt smaller molecular clouds. This mechanism may explain the suppression of star formation observed in distant, gas-rich radio galaxies. 

Future high-resolution and high-sensitivity ALMA observations focusing on the molecular cloud mass function are required to confirm the selective destruction of small-scale clouds. By conducting such observations, we will be able to clarify the origin of the discrepancy in the GMA size distribution between NGC~1316 and M~83, where the former potentially exhibits a steeper truncation at the larger end (Figure~\ref{fig:gma_comparison}). Specifically, this approach will allow us to distinguish whether this trend arises from physical mechanisms, such as compression by the expanding bubble and the selective stripping of diffuse GMC envelopes as suggested by numerical simulations, or is instead attributed to observational limitations, such as the higher sensitivity of the M~83 data which may facilitate the detection of more extended cloud outskirts.

On the theoretical side, further advancements in numerical simulations are essential to fully capture the complexity of the jet-ISM interaction observed in this system. While most existing idealized simulations modeling jet feedback assume pre-existing, stable gaseous disks, the unique environment of NGC~1316 suggests a far more dynamic scenario. 
The multiphase gas has been first supplied by a major merger 1~Gyr ago, since then several nuclear activities have shaped the inner regions of this source. The AGN likely self-sustained this flickering \citep{Maccagni:2021dz}.
To systematically investigate such complex setups, future studies will benefit from exploiting high-resolution cosmological galaxy formation simulations that naturally reproduce these unsettled gas configurations.

\begin{acknowledgments}
We thank the anonymous referee for his/her thorough review and helpful comments that improved the manuscript.
KMM is grateful to Prof. Cathy Horellou and Prof. Paolo Serra for providing the HST data used in \cite{Duah-Asabere:2016ur} and MeerKAT \HI~data used in \cite{Serra:2019ph}. KMM also would like to thank Prof. Ronald D. Ekers and Prof. Arif Babul for the insightful comments regarding the jet-ISM interaction.
This paper makes use of the following ALMA data: ADS/JAO.ALMA\#2017.0.00129.S; 2017.0.01140.S; 2019.1.01845.S. ALMA is a partnership of ESO (representing its member states), NSF (USA) and NINS (Japan), together with NRC (Canada), NSTC and ASIAA (Taiwan), and KASI (Republic of Korea), in cooperation with the Republic of Chile. The Joint ALMA Observatory is operated by ESO, AUI/NRAO and NAOJ.
This work was supported by JSPS Core-to-Core Program (grant number: JPJSCCA20210003) and JSPS Grants-in-Aid for Scientific Research (21H01128, JP23K13142, JP23K20035, 23H04899, 24H00247, and 25H00672).
This work has also been supported in part by the Collaboration Funding of the Institute of Statistical Mathematics ``New Perspective of the Cosmology Pioneered by the Fusion of Data Science and Physics'' and ``Machine-Learning-Based Cosmogony: From Structure Formation to Galaxy Evolution''. JK acknowledges support from NSF through grants AST-2006600 and AST-2406608.
FMM carried out part of the research activities described in this paper with contribution of the Next Generation EU funds within the National Recovery and Resilience Plan (PNRR), Mission 4 - Education and Research, Component 2 - From Research to Business (M4C2), Investment Line 3.1 - Strengthening and creation of Research Infrastructures, Project IR0000034 – ``STILES - Strengthening the Italian Leadership in ELT and SKA''.

\end{acknowledgments}





%
\facilities{ALMA, VLT:Yepun, CXO, MeerKAT, Spitzer, HST}

\software{CASA \citep{McMullin:2007bf},
          astropy \citep{Astropy-Collaboration:2013ef,Astropy-Collaboration:2018qf,Astropy-Collaboration:2022ee},
          APLpy \citep{Robitaille:2012mi},
          Matplotlib,
          numpy,
          SoFiA \citep{Serra:2015ix,Westmeier:2021kk}, 
          PYCPROPS \citep{Rosolowsky:2021qo}
          }


\appendix

\section{Appendix}

This appendix provides supplementary information and data products that support the main analysis presented in this paper. Specifically, we include the detailed input parameters used for source finding, as well as channel-by-channel ratio maps to evaluate the consistency of the observed gas properties across different velocity ranges.

\begin{table}[htbp]
    \centering
    \caption{SoFiA Parameters for Mask Generation}
    \label{tab:sofia_parameters}
    \begin{tabular}{llc}
        \toprule
        \textbf{Module} & \textbf{Parameter} & \textbf{Value} \\
        \midrule
        \multirow{10}{*}{Noise Scaling}
        & \texttt{scaleNoise.enable} & \texttt{true} \\
        & \texttt{scaleNoise.mode} & \texttt{local} \\
        & \texttt{scaleNoise.statistic} & \texttt{mad} \\
        & \texttt{scaleNoise.fluxRange} & \texttt{negative} \\
        & \texttt{scaleNoise.windowXY} & \texttt{300} \\
        & \texttt{scaleNoise.windowZ} & \texttt{10} \\
        & \texttt{scaleNoise.gridXY} & \texttt{0} \\
        & \texttt{scaleNoise.gridZ} & \texttt{0} \\
        & \texttt{scaleNoise.interpolate} & \texttt{true} \\
        & \texttt{scaleNoise.scfind} & \texttt{false} \\
        \midrule
        \multirow{7}{*}{S+C Finder}
        & \texttt{scfind.enable} & \texttt{true} \\
        & \texttt{scfind.kernelsXY} & \texttt{0, 4, 8} \\
        & \texttt{scfind.kernelsZ} & \texttt{0, 3, 6, 9} \\
        & \texttt{scfind.threshold} & \texttt{4} \\
        & \texttt{scfind.replacement} & \texttt{2.0} \\
        & \texttt{scfind.statistic} & \texttt{mad} \\
        & \texttt{scfind.fluxRange} & \texttt{negative} \\
        \midrule
        \multirow{12}{*}{Linker}
        & \texttt{linker.enable} & \texttt{true} \\
        & \texttt{linker.radiusXY} & \texttt{4, 8} \\
        & \texttt{linker.radiusZ} & \texttt{3, 6, 9} \\
        & \texttt{linker.minSizeXY} & \texttt{4} \\
        & \texttt{linker.minSizeZ} & \texttt{3} \\
        & \texttt{linker.maxSizeXY} & \texttt{0} \\
        & \texttt{linker.maxSizeZ} & \texttt{0} \\
        & \texttt{linker.minPixels} & \texttt{0} \\
        & \texttt{linker.maxPixels} & \texttt{0} \\
        & \texttt{linker.minFill} & \texttt{0.0} \\
        & \texttt{linker.maxFill} & \texttt{0.0} \\
        & \texttt{linker.positivity} & \texttt{false} \\
        & \texttt{linker.keepNegative} & \texttt{false} \\
        \bottomrule
    \end{tabular}
\end{table}

\begin{figure*}[]
\begin{center}
\includegraphics[width=\textwidth, bb=0 0 1749 1495]{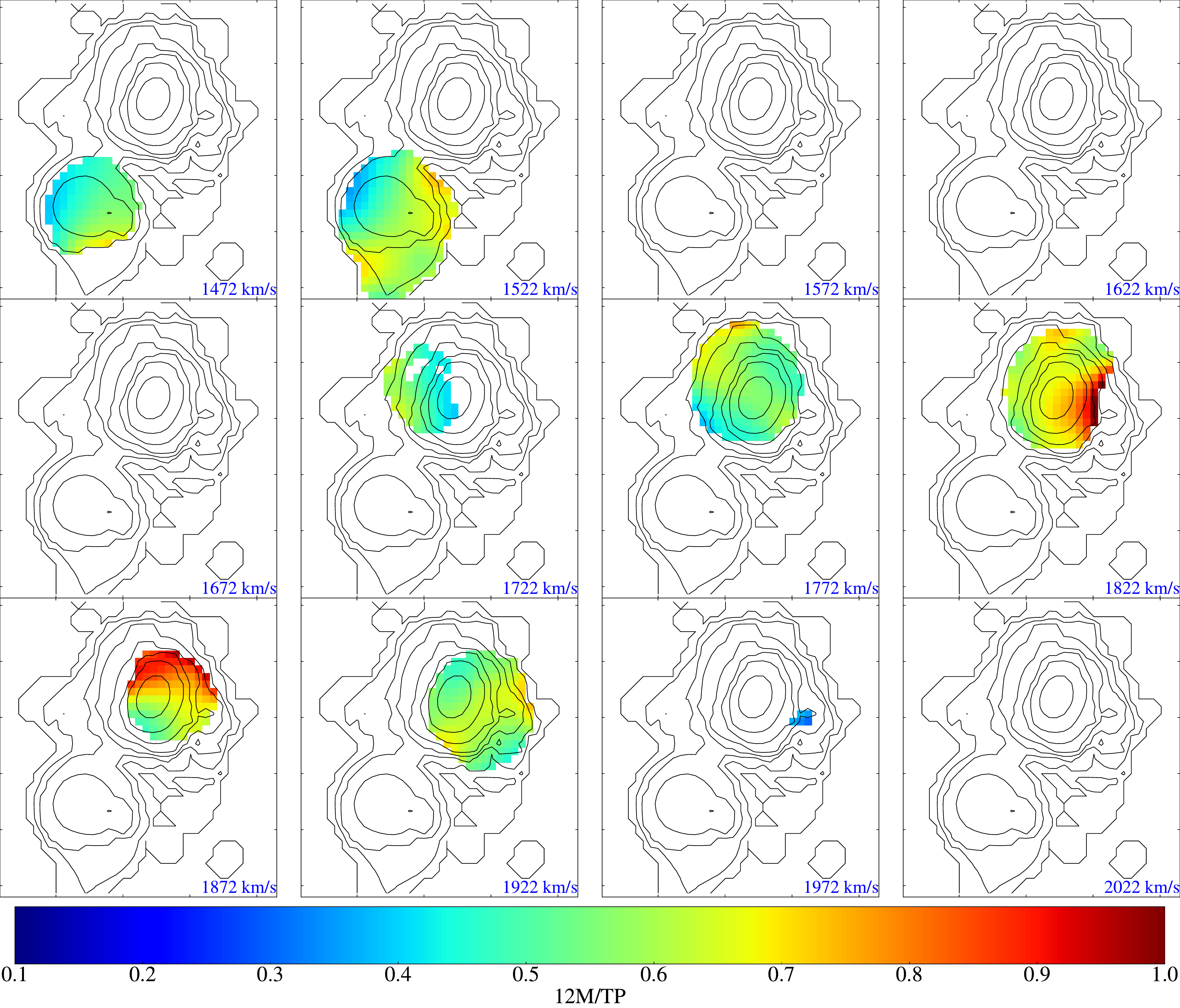}
\end{center}
\caption{
12-m/TP ratio map at different velocity channels.
Contours indicate the TP integrated intensity.
}
\label{fig:12mtp_v}
\end{figure*}

\begin{figure*}[]
\begin{center}
\includegraphics[width=\textwidth, bb=0 0 1749 1495]{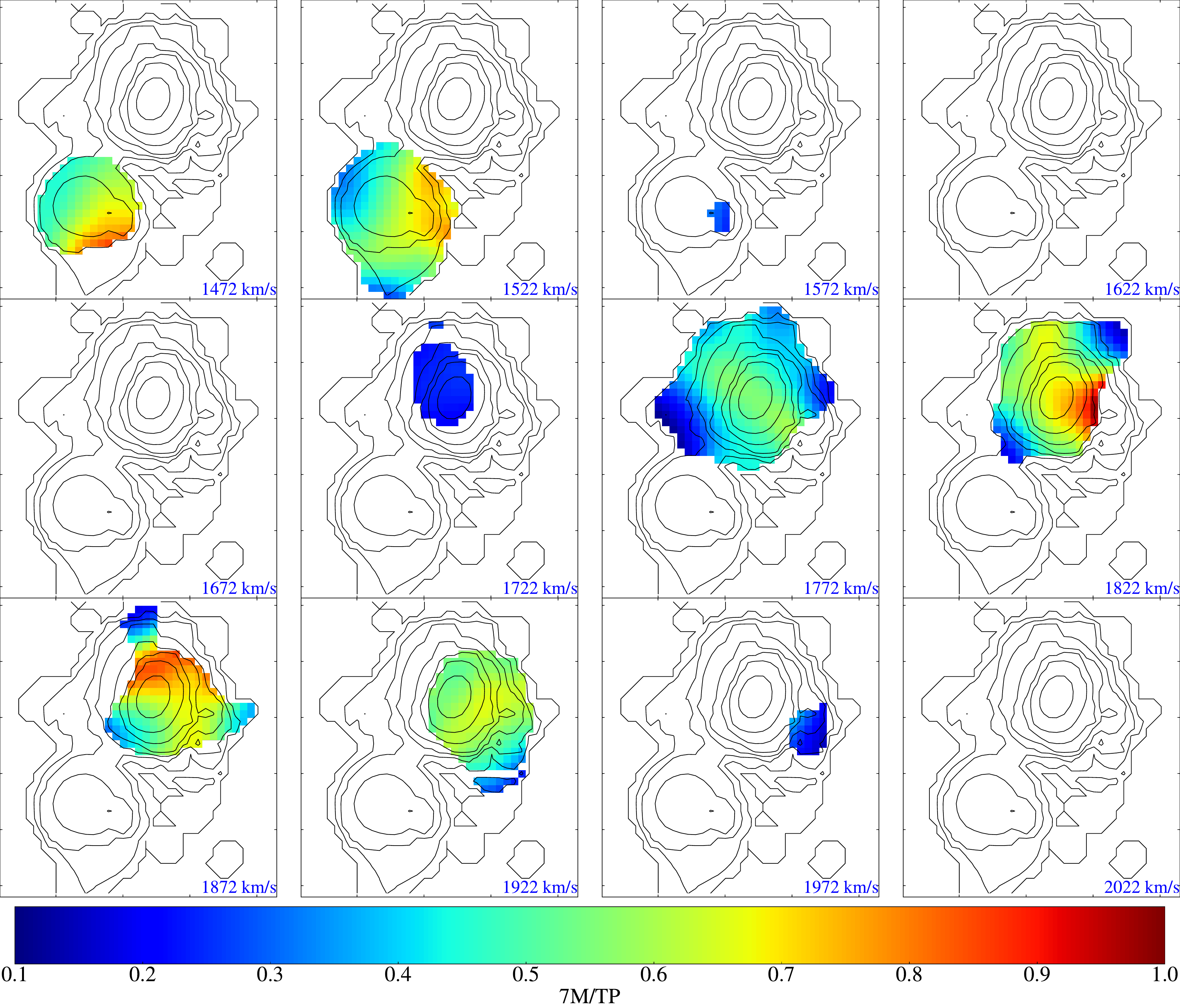}
\end{center}
\caption{
7-m/TP ratio map at different velocity channels.
Contours indicate the TP integrated intensity.
}
\label{fig:7mtp_v}
\end{figure*}


Table~\ref{tab:sofia_parameters} summarizes the specific input parameters adopted for the {\tt SoFiA2} source-finding pipeline. These parameters were optimized to ensure a robust extraction of the molecular gas emission while minimizing the inclusion of noise artifacts.


Figures~\ref{fig:12mtp_v} and \ref{fig:7mtp_v} present velocity-channel maps of the intensity ratios for 12-m/TP and 7-m/TP, respectively. These maps are used to evaluate the flux recovery fraction of the interferometric data relative to the TP observations across different spatial scales and velocity components. In each panel, the integrated intensity (moment 0) of the TP data is overlaid as contours to indicate the overall distribution of the molecular gas.

Our channel-by-channel analysis reveals that the interferometric data hardly recover any emission in the velocity range of 1,572-–1,722~km~s$^{-1}$. In other velocity ranges, including the NW Shell and SE Blob, the flux recovery fraction is found to be typically around 0.5. These trends are consistent with the results observed in the integrated intensity ratio maps (Figure~\ref{fig:dfiffuse}), confirming that the missing flux of extended structures is a general characteristic of the interferometric data in this study.


\bibliography{myref_ngc1316}{}
\bibliographystyle{aasjournalv7}



\end{document}